\documentclass[11pt]{cernyrep}

\usepackage{axodraw}
\usepackage{pst-coil}
\usepackage{pstricks}
\DeclareMathOperator{\IIm}{Im}
\let\vec\mathbf
\providecommand{\Rule}[2][0mm]{\rule[#1]{0mm}{#2}} 
\begin{document}

\title{Neutrino physics}
\author{P.~Hern\'andez}
\institute{IFIC, 
Universidad de Val\`encia and CSIC, 
E-46071 Valencia, Spain}
\maketitle

\begin{abstract}
The topics discussed in this lecture include: general properties of
neutrinos in the SM, the theory of neutrino masses and mixings (Dirac
and Majorana), neutrino oscillations both in vacuum and in matter, an
overview of the experimental evidence for neutrino masses and of the
prospects in neutrino oscillation physics. We also briefly review the
relevance of neutrinos in leptogenesis and in
beyond-the-Standard-Model physics.
\end{abstract}

\section{Neutrinos in the Standard Model}

LEP era  established the validity of the Standard Model (SM) with
an accuracy below the per cent level. The SM is based on the gauge
group $SU(3)\times SU(2) \times U_Y(1)$ that is spontaneously broken
to the subgroup $SU(3)_{color} \times U(1)_{em}$.  All the fermions of
the SM fall into irreducible representations of this group with the
quantum numbers summarized in \Tref{tab:reps} \cite{kleiss}.

Neutrinos are the most elusive particles of this table. They do not carry 
electromagnetic or colour charge, but only the weak charge under the
spontaneously broken subgroup.  For this reason they are extremely
weakly interacting, since their interactions are mediated by massive
gauge bosons.

The history of neutrinos goes back to W.~Pauli who postulated the
existence of the electron neutrino in an attempt to restore
energy--momentum conservation in $\beta$ decay, but he did so with
great regret: \emph{I have done a terrible thing, I have postulated a
particle that cannot be detected}.  Fortunately Pauli was wrong, not
only have neutrinos been detected but they have been extremely useful
in establishing the two most striking features of \Tref{tab:reps}: the
left--handedness of the weak interactions (the left--right asymmetry of
the table) and the family structure (the three--fold repetition of the
same representations).

In the SM only the left-handed fields carry the $SU(2)$ charge, where 
by left-handed we denote the negative chirality component (\ie eigenstate
of $\gamma_5$ with eigenvalue minus one) of the fermion field \cite{kleiss}:
\begin{eqnarray}
\Psi = \Psi_R + \Psi_L = \underbrace{\left(\frac{1 + \gamma_5}{2}\right)}_{P_R}  
       \Psi + \underbrace{\left(\frac{ 1 - \gamma_5}{2}\right)}_{P_L}  \Psi\SPp.
\end{eqnarray}
For relativistic fermions (\ie massless), it is easy to see that the
chiral projectors are equivalent to the projectors
on helicity components:
\begin{eqnarray}
P_{R,L} = \frac{1}{2} \left( 1 \pm \frac{\vec{s}\cdot\vec{p}}{|p|} \right) 
        + O\left(\frac{m_i}{E}\right),
\end{eqnarray}
where the helicity operator $\vec\Sigma = \frac{\vec{s}\cdot\vec{p}}{|p|}$ measures
the component of the spin in the direction of the spatial momentum. Therefore
for massless fermions only the left-handed states (with the spin pointing 
in the opposite direction to the momentum) carry $SU(2)$ charge. This is
not inconsistent with Lorenzt invariance, since for a fermion travelling 
at the speed of light, the helicity is the same in any reference frame. In 
other words, the helicity operator commutes with the Hamiltonian for a massless
fermion and is thus a good quantum number. 

The discrete symmetry under CPT (charge conjugation, parity, and time
reversal), which is a basic building block of any Lorenzt invariant
and unitary field theory, requires that for any left-handed fermion,
there exists a right-handed antiparticle, with opposite charge, but
the right-handed particle state may not exist. This is precisely what
happens with neutrinos in the SM. Since only the left-handed states
carry charge and their masses were compatible with zero when the SM
was established, they were postulated to be Weyl fermions: \ie a
left-handed particle and a right-handed antiparticle.

\begin{table}
\caption[]{Irreducible fermionic representations in the Standard Model: ($I_{SU(3)}, I_{SU(2)})_{Y}$}
\label{tab:reps}
\[
\begin{array}{@{}cc|ccc@{}}                          \hline\hline
\Rule[-1em]{2.5em}
(\vec{1},\vec{2})_ {-\frac{1}{2}} 
 & (\vec{3},\vec{2})_{-\frac{1}{6}} 
  & (\vec{1},\vec{1})_{-1} 
   & (\vec{3},\vec{1})_{-\frac{2}{3}} 
    & (\vec{3},\vec{1})_{-\frac{1}{3}}             \\\hline  
\Rule[-2em]{4em}
\begin{pmatrix}\nu_e  \\ e \end{pmatrix}_{_L} 
 & \begin{pmatrix}u^i \\ d^i \end{pmatrix}_{_L} 
  & e_R 
   & u^i_R 
    & d^i_R                                        \\ 
\Rule[-2em]{4em}
\begin{pmatrix}\nu_\mu \\ \mu\end{pmatrix}_{_L} 
 & \begin{pmatrix}c^i  \\ s^i\end{pmatrix}_{_L} 
  & \mu_R 
   & c^i_R 
    & s^i_R                                        \\ 
\Rule[-2em]{4em}
\begin{pmatrix} \nu_\tau \\ \tau\end{pmatrix}_{_L} 
 & \begin{pmatrix}t^i \\ b^i\end{pmatrix}_{_L} 
  & \tau_R 
   & t^i_R 
    & b^i_R                                        \\\hline\hline
\end{array}
\]
\end{table}

Under parity, a left-handed particle state transforms into a
right-handed particle state, thus the left-handedness of the weak
interactions implies a maximal violation of parity, which is nowhere
more obvious than in the neutrino sector, where the reflection of a SM
neutrino in a mirror is nothing.

\begin{figure}
\centering
\includegraphics[width=.8\linewidth]{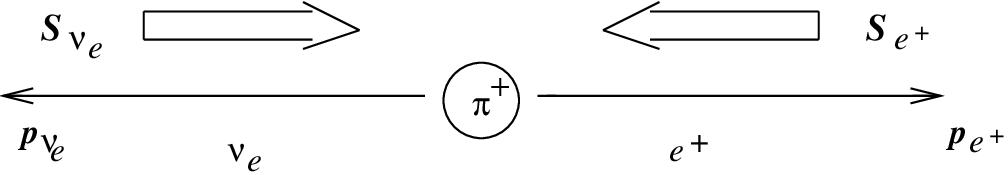}
\caption[]{Kinematics of pion decay}
\label{fig:pidecay}
\end{figure}

The weak current is therefore $V-A$ since it only couples to the left
fields: $\bar{\Psi}_L \gamma_\mu \Psi_L = \bar{\Psi} \gamma_\mu (1
-\gamma_5)/2 \Psi$. This structure is clearly seen in the kinematics
of weak decays involving neutrinos, such as the classic example of
pion decay to $e \nu_e$ or $\mu \nu_\mu$. In the limit of vanishing
electron or muon mass, this decay is forbidden, because the spin of
the initial state is zero and thus it is impossible to conserve
simultaneously momentum and angular momentum if the two recoiling
particles must have opposite helicities, as shown in
\Fref{fig:pidecay}. Thus the ratio of the decay rates to electrons and
muons, in spite of the larger phase space in the former, is strongly
suppressed by the factor $\left(\frac{m_e}{m_\mu}\right)^2 \sim
2\times 10^{-5}$.

Another profound consequence of the chiral nature of the weak
interaction is anomaly cancellation. The chiral coupling of fermions
to gauge fields leads generically to inconsistent gauge theories due
to chiral anomalies: if any of the diagrams depicted in \Fref{fig:ano}
is non-vanishing, the weak current is conserved at tree level but not
at one loop, implying a catastrophic breaking of gauge
invariance. Anomaly cancellation is the requirement that all these
diagrams vanish, which imposes strong constraints on the hypercharge
assignments of the fermions in the SM, which are \emph{miraculously}
satisfied:
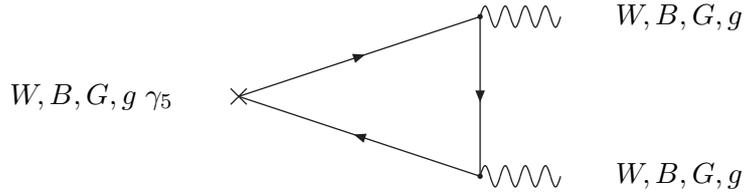
\begin{figure}
\begin{center}
\begin{picture}(280,70)(-55,15)
  \SetOffset(0,20)
  \Line(27,33)(33,27) \Line(27,27)(33,33)
  \ArrowLine(30,30)(120,60) \ArrowLine(120,60)(120,0)
  \ArrowLine(120,0)(30,30)
  \Text(-25,30)[]{$W,B,G,g \;\gamma_5$}
  \Photon(120,60)(150,60){4}{4} \Text(195,60)[]{$W,B,G,g$}
  \Photon(120,0)(150,0){4}{4} \Text(195,0)[]{$W,B,G,g$}
  \Vertex(120,60)1 \Vertex(120,0)1
\end{picture}
\end{center}
\caption[]{Triangle diagrams that can give rise to anomalies. $W, B,
           G$ are the gauge bosons associated to the $SU(2), U_Y(1),
           SU(3)$ gauge groups, respectively, and $g$ is the
           graviton}
\label{fig:ano}
\end{figure}
\begin{eqnarray}
\overbrace{\sum_{i=\text{quarks}} Y^L_i- Y^R_i}^{GGB} = 
\overbrace{\sum_{i=\text{doublets}} Y^L_i}^{WWB} = 
\overbrace{\sum_{i} Y^L_i-Y^R_i}^{Bgg} = 
\overbrace{\sum_i (Y_i^L)^3-(Y_i^R)^3}^{B^3} = 0, 
\label{eq:ano}
\end{eqnarray}
where $Y^{L/R}_i$ are the hypercharges of the left/right components of the 
fermionic field $i$, and the triangle diagram corresponding to each of the 
sums is indicated above the bracket. 

\begin{figure}
\centering
\includegraphics[angle=270,width=.7\linewidth]{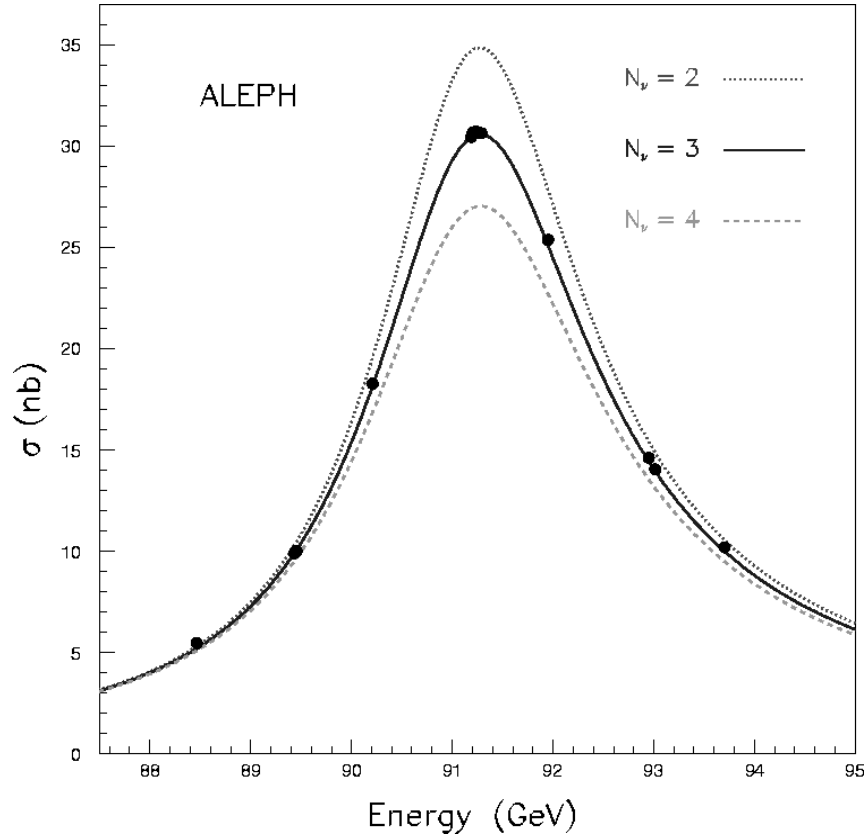}
\caption[]{$Z^0$ resonance from the ALEPH experiment at LEP. Data are
           compared to the case of $N_\nu=2,3$ and 4}
\label{fig:z0} 
\end{figure}

Concerning the family structure, we know, thanks to neutrinos, that
there are exactly three families in the SM. An extra SM family with quarks
and charged leptons so heavy that they remain unobserved, would also
have massless neutrinos that would have been produced in $Z^0$ decay,
modifying its width, which has been measured at LEP with an impressive
precision, as shown in \Fref{fig:z0}. This measurement excludes any
number of standard neutrino families different from three \cite{pdg}:
\begin{eqnarray}
N_\nu = 2.984 \pm 0.008.
\end{eqnarray}

\section{Neutrino masses and mixings}

When the SM was invented, there were only upper limits on the neutrino
masses so these were conjectured to be zero. The direct limit on
neutrino masses comes from the precise measurement of the end-point of
the lepton energy spectrum in weak decays, which gets modified if
neutrinos are massive.  In particular the most stringent limit is
obtained from tritium $\beta$-decay for the electron neutrino:
\begin{eqnarray}
H^3 \rightarrow ^3He + e^- + \bar{\nu}_e .
\end{eqnarray}
\Fref[b]{fig:kurie} shows the effect of a neutrino mass in the
end-point electron energy spectrum in this decay. The functional form
of this curve is $K(E_e) \propto \sqrt{(E_0 - E_e)((E_0 - E_e)^2 -
m_\nu^2)^{1/2}}$.
\begin{figure}
\centering
\includegraphics[width=.5\linewidth]{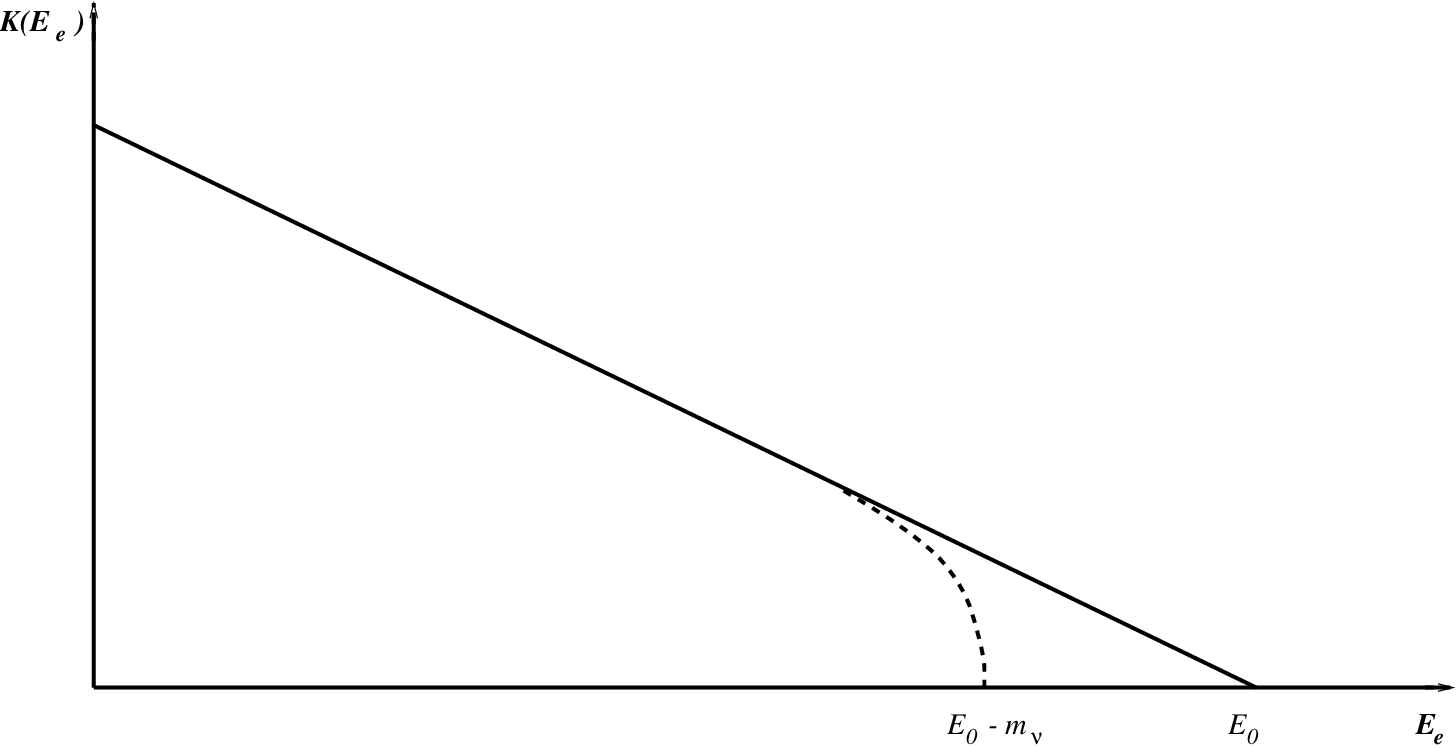}
\caption[]{Effect of a neutrino mass in the end-point of the lepton
           energy spectrum in $\beta$ decay}
\label{fig:kurie}
\end{figure}
The best limit has been obtained by the Mainz and Troitsk experiments
\cite{mainz-troitsk}:
\begin{eqnarray}
m_{\nu_e} < 2.2\UeV \,\text{(Mainz)},\quad
m_{\nu_e} < 2.1\UeV \,\text{(Troitsk)}\SPp, 
\label{mnue}
\end{eqnarray}
both at $95\%$ CL.  The direct limits on the other two neutrino
masses are much weaker. The best limit on the $\nu_\mu$ mass
($m_{\nu_\mu} < 170\UkeV$ \cite{psi}) was obtained from the
end-point spectrum of the decay $\pi^+ \rightarrow \mu^+ \nu_\mu$,
while that on the $\nu_\tau$ mass was obtained at LEP
($m_{\nu_\tau} < 18.2\UMeV$ \cite{lep_nutau}) from the decay $\tau
\rightarrow 5 \pi \nu_\tau$.

As we shall see, there is now  strong evidence that neutrinos are
indeed massive, although extremely light, below the stringent bound of
\Eref{mnue}.

Neutrino masses can be easily accommodated in the SM.  A massive
fermion necessarily has two states of helicity, since it is always
possible to reverse the helicity of a state that moves at a slower
speed than light by looking at it from a boosted reference frame.  In
fact a mass can be thought of as the strength of the coupling between
the two helicity states:
\begin{eqnarray}
m \;\overline{\psi_L} \psi_R + \text{h.c.}
\end{eqnarray}
In order to include such a coupling in the SM for the neutrinos we need
to identify the neutrino right-handed state, which in the SM is
absent. It turns out there are two ways to proceed:

\subsubsection*{Dirac massive neutrinos}

\begin{figure}
\centering
\includegraphics[width=\linewidth]{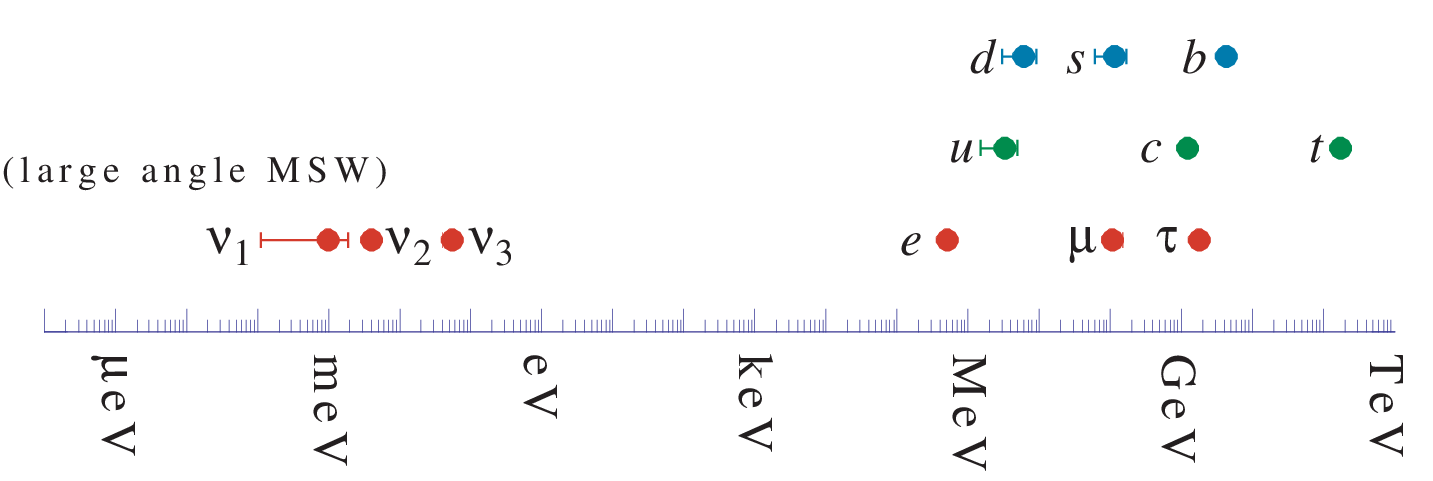}
\caption[]{Fermion spectrum in the Standard Model}
\label{fig:hierar}
\end{figure}

We can enlarge the SM by adding a set of three right-handed neutrino
states, which would be singlets under $SU(3)\times SU(2) \times
U_Y(1)$, but coupled to matter just through the neutrino masses.  This
coupling has to be of the Yukawa type to preserve the gauge symmetry
in such a way that the masses are proportional to the vacuum
expectation value of the Higgs field, $v$, exactly like for the
remaining fermions \cite{kleiss}:
\begin{eqnarray}
\lambda_\nu \;\overline{L_L}\; \tilde\Phi\; \nu_R + \text{ h.c. }  
  \rightarrow m_\nu = \lambda_\nu\; v,
\end{eqnarray}
where $L_L= (\nu_L \,\, l_L)$ is the lepton doublet and $\tilde\Phi$
is the scalar doublet that gets a vacuum expectation value $\langle
\tilde\Phi \rangle = (v \,\,0)$.  There are two important consequences
of proceeding in this way.  Firstly there is a new hierarchy problem
in the SM to be explained: why neutrinos are much lighter than the
remaining leptons, even those in the same family (see
\Fref{fig:hierar}).
Secondly, lepton number, $L$, which counts the number of leptons minus
that of antileptons, remains an exactly conserved global symmetry at
the classical level \footnote{As usual $B+L$ is broken by the anomaly
and only $B-L$ remains exact at all orders.}, just as baryon number,
$B$, is.

\subsubsection*{Majorana massive neutrinos}

For neutral particles, Majorana realized that one can get rid of half
of the degrees of freedom in a massive Dirac spinor in a
Lorenzt-invariant way by identifying the right-handed state with the
antiparticle of the left-handed state:
\begin{eqnarray}
\nu_R \rightarrow (\nu_L)^c = C \bar{\nu}^T_L = C \gamma_0 \nu_L^*,
\end{eqnarray}
where $C$ is the operator of charge conjugation in spinor space. 

\begin{figure}
\centering
\includegraphics[angle=270,width=.5\linewidth]{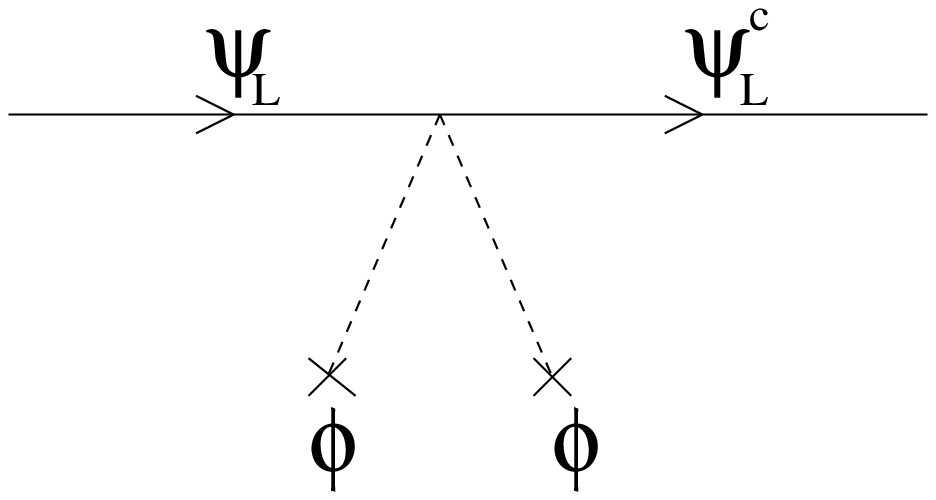}
\caption[]{Majorana coupling of the light neutrinos to the Higgs
           field}
\label{fig:majo}
\end{figure}

Neutrinos are the only particles for which this possibility is
compatible with charge conservation, because they are  charged only
under the spontaneously broken subgroup of the SM and thus a Majorana
mass term can be written in a gauge invariant way by including two
Higgs fields, as shown in \Fref{fig:majo}:
 \begin{eqnarray}
\frac{1}{M} L_L^T C\; \alpha_\nu {\tilde\Phi}^T\; {\tilde\Phi} \; L_L + \text{ h.c. }  ,
\label{majo}
\end{eqnarray}
where an energy scale, $M$, has been introduced for dimensional reasons, so that
the coupling $\alpha_\nu$ is adimensional. 
Upon spontaneous symmetry breaking, these couplings become Majorana
neutrino masses of the form
\begin{eqnarray}
m_\nu = \alpha_\nu \frac{v^2}{M}. 
\end{eqnarray}
If the scale $M$ is much higher than the electroweak scale $v$, a
strong hierarchy between the neutrino and the charged lepton masses
arises naturally.

\subsection{See-saw models}
\label{sec:seesaw}

\begin{figure}
\centering
\includegraphics[width=.4\linewidth]{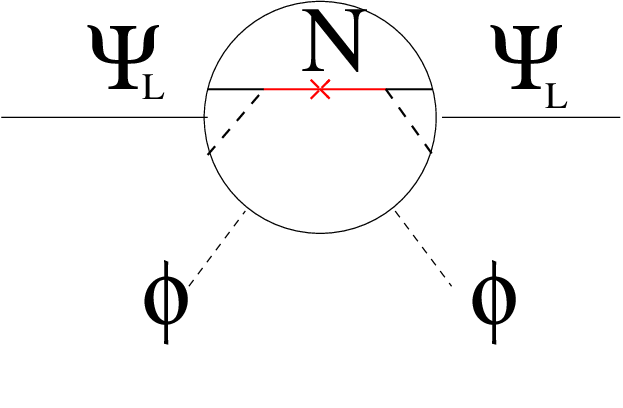}
\caption[]{Neutrino masses in the see-saw model}
\label{fig:ss}
\end{figure}

It is interesting to consider the simplest example to explain the
origin of the scale $M$ in the Majorana masses. This is the famous
see-saw model of Gell-Mann, Ramond, Slansky, and Yanagida \cite{typeI}.
In this model, the Majorana effective interaction of \Eref{majo}
results from the interchange of very heavy right-handed Majorana
neutrinos, as depicted in \Fref{fig:ss}.
The SM Lagrangian is enlarged with the terms
\begin{eqnarray}
\delta \mathcal{L}^\nu_Y &=& \; \bar{L}_L \tilde{\lambda}_\nu\;\tilde{\Phi}\; N_R + \frac{1}{2} N_R^T C \;M_R \;N_R  + \text{ h.c. } ,
\end{eqnarray}
that is a Yukawa coupling of the lepton doublet and the heavy singlets
plus a Majorana mass term for the singlets. Upon spontaneous symmetry
breaking these couplings become mass terms:
\begin{eqnarray}
\delta \mathcal{L}^\nu_Y &\rightarrow
  &\frac{1}{2} \begin{pmatrix}\nu^T_L & {N_R}^T\end{pmatrix} C 
               \begin{pmatrix}0 & {\tilde \lambda}_\nu v \\
                              {\tilde \lambda}^T_\nu v & M_R
               \end{pmatrix} 
               \begin{pmatrix}\nu_L \cr N_R\end{pmatrix}\SPp.
\end{eqnarray}
When $v \ll M_R$, the diagonalization of the mass matrix can be done
in perturbation theory:
\begin{eqnarray}
\mathcal{M} = \mathcal{M}^{(0)} + \mathcal{M}^{(1)}\equiv 
              \begin{pmatrix}0  & 0 \\ 0 & M_R \end{pmatrix} 
            + \begin{pmatrix}0  & {\tilde \lambda}_\nu v \\
                             {\tilde \lambda}^T_\nu v & 0 
              \end{pmatrix}\SPp.
\end{eqnarray}
To second order we find: 
\begin{eqnarray}
U^T \mathcal{M} U = \begin{pmatrix} 
                      - v^2 {\tilde \lambda}_\nu \frac{1}{M_R} {\tilde \lambda}^T_\nu 
                         & 0 \\
                      0  & M_R
                    \end{pmatrix} \;\;\; 
                U = \begin{pmatrix} 
                      1 & \tilde\lambda_\nu \frac{v}{M_R} \\
                      - \frac{v}{M_R} \tilde\lambda^T_\nu & 1
                    \end{pmatrix}\SPp.
\end{eqnarray}
There are three light Majorana neutrinos 
($\nu'_L \simeq \nu_L + \tilde{\lambda}_\nu \frac{v}{M_R} N_R$) with a mass matrix:
\begin{eqnarray}
v^2 {\tilde \lambda}_\nu \frac{1}{M_R} {\tilde \lambda}^T_\nu, 
\end{eqnarray}
and three heavy ones ($N'_R \simeq N_R - \frac{v}{M_R}
\tilde{\lambda}^T_\nu \nu_L$) with the mass matrix $M_R$.

Equivalently we say that the heavy Majoranas can be integrated out
leaving a trace of higher dimensional operators:
\begin{eqnarray}
\mathcal{L}^{d=5}_{eff} = \frac{1}{2} {L_L^T} C\;  \;{\tilde \Phi}^T\; \left({\tilde \lambda}_\nu \frac{1}{M_R} {\tilde \lambda}^T_\nu \right) \;  {\tilde \Phi}\; L_L \\
\mathcal{L}^{d=6}_{eff} = \mathcal{O}\left(\frac{1}{M_R^2}\right)...
\end{eqnarray}
The one with lowest dimension is the one we obtained from symmetry
arguments in \Eref{majo}.

A few observations are in place:
\begin{itemize}
\item The new physics scale $M$ in \Eref{majo} is simply related
      to the masses of the heavy Majorana neutrinos and the Yukawa
      couplings:
\begin{eqnarray}
\frac{\alpha_\nu}{M} \rightarrow 
  {\tilde \lambda}_\nu \frac{1}{M_R} {\tilde \lambda}^T_\nu . 
\end{eqnarray}
      As we shall see, data imply there is at least one $m_\nu \geq
      0.05\UeV$. If $\tilde{\lambda}_\nu \sim O(1)$ then:
\begin{eqnarray}
v \; < \; M_R \sim 10^{15}\UGeV < M_\text{Planck}, 
\end{eqnarray}
    and the masses are close to the typical Grand Unification (GUT)
    scale.

\item In order to give non-vanishing masses to all the three
      left-handed neutrinos, the number of Majorana singlets must
      satisfy $N_R \geq N_L =3$. The reason is that the matrix
      $\underbrace{{\tilde \lambda}_\nu}_{N_L \times N_R}
      \underbrace{\frac{1}{M_R}}_{N_R\times N_R} \underbrace{{\tilde
      \lambda}^T_\nu}_{N_R\times N_L}$ has $N_L -N_R$ zero modes.
\end{itemize}

\subsection{Majorana versus Dirac}

The consequences of the SM neutrinos being massive Majorana particles
are profound:
\begin{itemize}
\item A new physics scale $M$ must exist and is accessible in an
      indirect way through neutrino masses.

\item Lepton number is not conserved: a Majorana mass violates the
      conservation of all the charges carried by the fermion,
      including the global charges such as lepton number. As we shall
      see in \Sref{sec:baryo}, the dynamics associated to the
      scale $M$ could be responsible for the generation of the baryon
      asymmetry in the Universe.

\item The anomaly cancellation conditions fix all the hypercharges
      (\ie there is only one possible choice for the hypercharges that
      satisfies \Eref{eq:ano}), which implies that
      electromagnetic charge quantization is the only possibility in a
      field theory with the same matter content as the SM.
\end{itemize}

\begin{figure}
\centering
\includegraphics[angle=270,width=.75\linewidth]{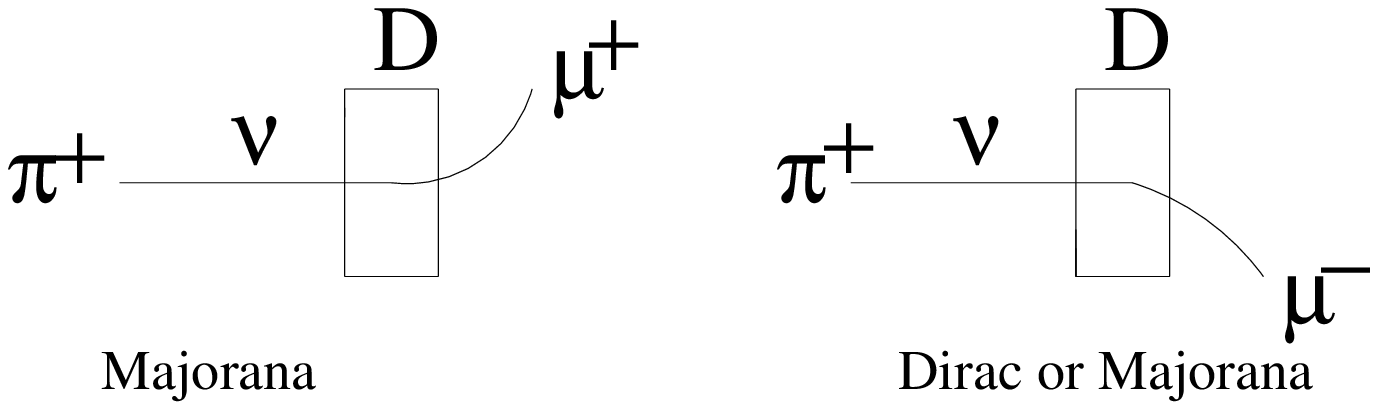}
\caption[]{A neutrino beam from $\pi^+$ decay ($\nu_\mu$) could
           interact in the magnetized detector producing a $\mu^+$
           only if neutrinos are Majorana.}
\label{fig:signature}
\end{figure}

It is clear that establishing the Majorana nature of neutrinos is of
great importance.  In principle there are very clear signatures, such
as the one depicted in \Fref{fig:signature}, where a $\nu_\mu$ beam
from $\pi^+$ decay is intercepted by a detector. In the Dirac case,
the interaction of neutrinos on the detector via a charged current
interaction will produce a $\mu^-$ in the final state. If neutrinos
are Majorana, a wrong-sign muon in the final state is also possible.
Unfortunately the rate for $\mu^+$ production is suppressed by
$m_\nu/E$ in amplitude with respect to the $\mu^-$. For example, for
$E_\nu = \mathcal{O}(1)\UGeV$ and $m_\nu \sim \mathcal{O}(1)\UeV$ the cross--section for this process will be roughly $10^{-18}$ times the usual CC
neutrino cross-section, which means it is impossible to detect.

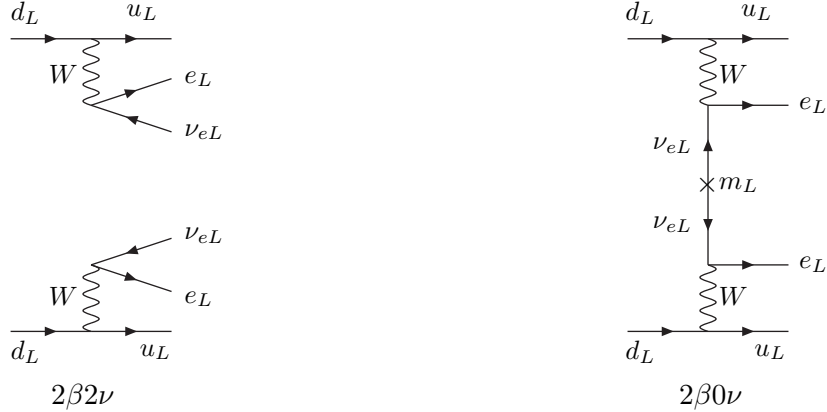
\begin{figure}[h]
\centering
\begin{picture}(330,160)(-15,-120)
\ArrowLine(-10,25)(20,25)
\ArrowLine(-10,-85)(20,-85)
\Photon(20,25)(20, 0)3 4
\Text(15,12)[r]{\small $W$}
\Text(17,-110)[c]{$2\beta2\nu$}
\ArrowLine(20,0)(50,10)
\ArrowLine(50,-10)(20,0)
\ArrowLine(50,-50)(20,-60)
\ArrowLine(20,-60)(50,-70)
\Photon(20,-60)(20,-85)3 4
\Text(15,-72)[r]{\small$W$}
\Text(55,10)[l]{\small$e_L$}
\Text(55,-73)[l]{\small$e_L$}
\Text(55,-49)[l]{\small$\nu_{eL}$}
\Text(55,-10)[l]{\small$\nu_{eL}$}
\Text(45,35)[r]{\small$u_L$}
\Text(50,-93)[r]{\small$u_L$}
\Text(-10,-93)[l]{\small$d_L$}
\Text(-10,35)[l]{\small$d_L$}
\ArrowLine(20,25)(50,25)
\ArrowLine(20,-85)(50,-85)
\ArrowLine(220,25)(250,25)
\ArrowLine(220,-85)(250,-85)
\Photon(250,25)(250, 0)3 4
\Text(255,12)[l]{\small$W$}
\Text(252,-110)[c]{$2\beta 0\nu$}
\ArrowLine(250,0)(280,0)
\ArrowLine(250,-30)(250,0)
\Text(250.5,-30)[c]{$\times$}
\ArrowLine(250,-30)(250,-60)
\ArrowLine(250,-60)(280,-60)
\Photon(250,-60)(250,-85)3 4
\Text(255,-72)[l]{\small$W$}
\Text(285,0)[l]{\small$e_L$}
\Text(285,-60)[l]{\small$e_L$}
\Text(255,-30)[l]{\small$m_L$}
\Text(245,-15)[r]{\small$\nu_{eL}$}
\Text(245,-45)[r]{\small$\nu_{eL}$}
\Text(275,35)[r]{\small$u_L$}
\Text(280,-93)[r]{\small$u_L$}
\Text(220,-93)[l]{\small$d_L$}
\Text(220,35)[l]{\small$d_L$}
\ArrowLine(250,25)(280,25)
\ArrowLine(250,-85)(280,-85)
\end{picture}
\caption[]{$2\beta$ decay: normal (left) and neutrinoless (right)}
\label{fig:dbd}
\end{figure}

The best hope of observing a rare process of this type seems to be the
search for neutrinoless double--beta decay ($2\beta 0\nu$), the right
diagram of \Fref{fig:dbd}. The background to this process is the
standard double--beta decay depicted on the left of \Fref{fig:dbd},
which has been observed to take place with a lifetime of $T_{2\beta
2\nu} > 10^{19}$--$10^{21}$ years.

If the source of $L$ violation is just the Majorana $\nu$ mass, the
inverse lifetime for this process is given by
\begin{eqnarray}
T^{-1}_{2\beta 0\nu} \simeq 
  \underbrace{G^{0\nu}}_\text{Phase}
\underbrace{\left|M^{0\nu}\right|^2}_{\rm Nuclear M.E.}
\underbrace{\left|\sum_i \left(V^{ei}_\text{MNS}\right)^2 m_i
\right|^2}_{|m_{ee}|^2} ,
\end{eqnarray}
where $m_{ee}$ is the 11 entry in the neutrino mass matrix in the
flavour basis. In spite of the suppression in the neutrino mass (over
the energy of this proccess), the neutrinoless mode has a larger phase
factor than the $2\nu$ mode, and as a result the lifetime is expected
to be of the order
\begin{equation}
T^{-1}_{2\beta 0\nu} \sim  \left(\frac{m_\nu}{E}\right)^2 10^{9}\; T^{-1}_{2\beta 2 \nu},
\end{equation}
which could be observable for neutrino masses in the $\UeVZ$
range. Several experiments have set stringent upper bounds on
$|m_{ee}|$ and there is even a controversial positive signal, as shown
in \Tref{tab:2b0n}.

\begin{table}
\centering
\begin{tabular}{@{}lll@{}}                         \hline\hline
\Rule[-0.5em]{1.5em}
Experiment 
 & Nucleus 
  & $|m_{ee}|$                                     \\\hline
\Rule{1.2em}
Heidelberg-Moscow I 
 & \Isotope[76]{Ge} 
  & $< 0.34$--$1.1\UeV  (90\% \text{ CL} )$ \cite{heidelmoscow} \\
\Rule{1.2em}
Heidelberg-Moscow II 
 & \Isotope[76]{Ge} 
  & $0.2$--$0.6\UeV$ \cite{klapdor}                     \\
\Rule{1.2em}
CUORICINO 
 & \Isotope[120]{Te} 
  &  $<0.2$--$1.1\UeV (90\% \text{ CL})$ \cite{cuoricino}\\
\Rule[-0.5em]{1.5em}
NEMO-3 
 & \Isotope[100]{Mo} 
 & $<0.6$--$2\UeV (90\% \text{ CL})$ \cite{nemo3}        \\\hline\hline
\end{tabular}
\caption[]{Present bounds from various neutrinoless double-beta-decay
           experiments}
\label{tab:2b0n} 
\end{table}

\subsection{Neutrino mixing}
\label{sec:mixing}

\begin{figure}
\centering
\includegraphics[angle=270,width=.4\linewidth]{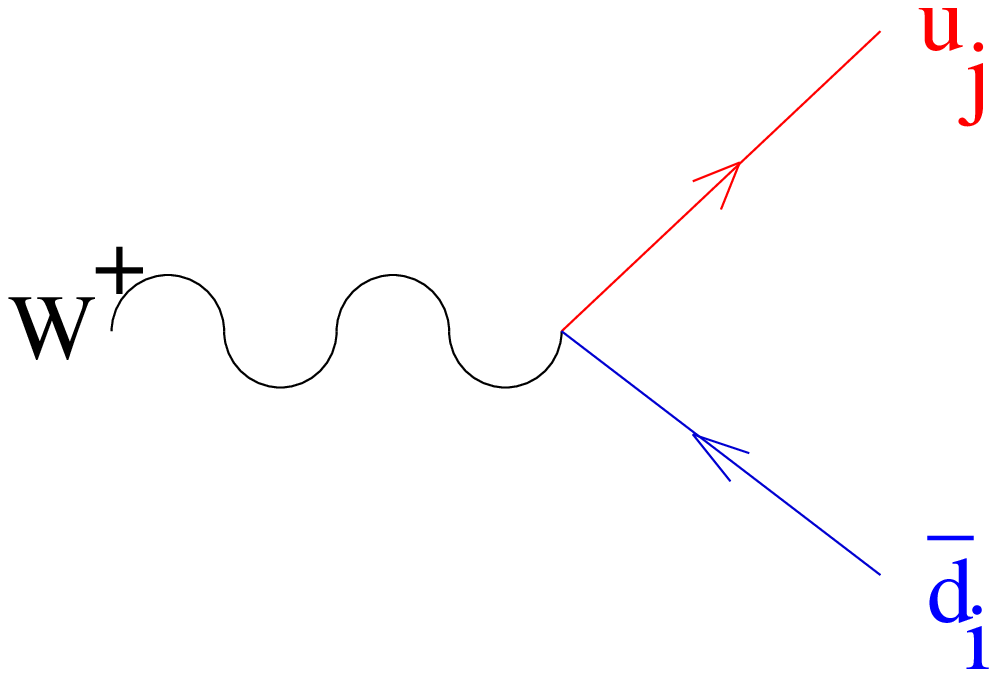}\qquad
\includegraphics[angle=270,width=.4\linewidth]{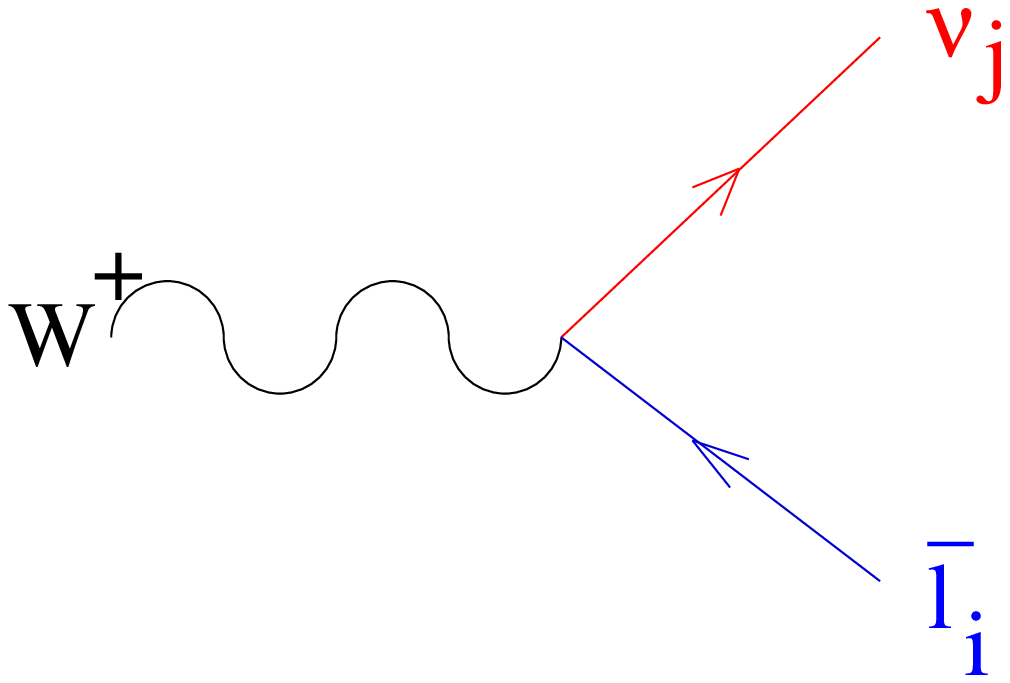}
\caption[]{Quark and lepton mixing}
\label{fig:mix}
\end{figure}

Generically, neutrino masses imply neutrino mixing
\cite{pontecorvo,mns}, because the Yukawa couplings need not be
flavour diagonal:
\begin{eqnarray}
\mathcal{L}_m^{\rm Dirac} = \overline{\nu^i_L} \left(\lambda_\nu  v \right)_{ij} \nu^j_R  + \text{ h.c. } \\
\mathcal{L}_m^{\rm Majorana} = \frac{1}{2} \frac{v^2}{M} {\nu^i_L}^T C \left(\alpha_\nu\right)_{ij}  \nu^j_L + \text{ h.c. } 
\end{eqnarray}
Instead, in the mass eigenbasis for all the leptons, the charged weak couplings are not diagonal, in complete analogy with the quark flavour sector (see \Fref{fig:mix}): 
\begin{eqnarray}
\mathcal{L}^{\rm Dirac} = {\bar l^i_L} \gamma_\mu W_\mu^+ V^{ij}_\text{MNS} \nu^j_L + \frac{1}{2} \overline{\nu^i_L} \gamma_\mu Z_\mu \nu^i_L + \overline{\nu^i_L} m_i \nu^i_R + \text{ h.c. } \\
\mathcal{L}^{\rm Majorana} = {\bar l^i_L} \gamma_\mu W_\mu^+ {\tilde V}^{ij}_\text{MNS} \nu^j_L + \frac{1}{2} \overline{\nu^i_L} \gamma_\mu Z_\mu \nu^i_L + \frac{1}{2} {\nu^i_L}^T C m_i \nu^i_L + \text{ h.c. } 
\end{eqnarray}

The number of  parameters that are in principle observable in the
lepton mixing matrix ($V_\text{MNS}$ for Dirac and ${\tilde
V}_\text{MNS}$ for Majorana) can easily be computed by counting the
number of independent real and imaginary elements of the Yukawa
matrices and eliminating those that can be absorbed in field
redefinitions. The allowed field redefinitions are the unitary
rotations of the fields that leave the Lagrangian invariant in the
absence of lepton masses, but are not symmetries of the full
Lagrangian when lepton masses are included.

In the Dirac case, it is possible to rotate independently the
left-handed lepton doublet, together with the right-handed charged
leptons and neutrinos, that is $U(n)^3$, for a generic number of
families $n$. However, this includes total lepton number which remains
a symmetry of the massive theory and thus cannot be used to reduce the
number of physical parameters in the mass matrix. The
parameters that can be absorbed in field redefinitions are thus the
parameters of the group $U(n)^3/U(1)$ (that is $\frac{3 (n^2-n)}{2}$
real, $\frac{3(n^2+n)-1}{2}$ imaginary).

In the case of Majorana neutrinos, there is no independent
right-handed neutrino field, nor is lepton number a good
symmetry. Therefore the number of field redefinitions is the number of
parameters of the elements in $U(n)^2$ (that is $n^2-n$ real and
$n^2+n$ imaginary).

The resulting real physical parameters are the mass eigenstates and
the mixing angles, while the resulting imaginary parameters are
CP-violating phases. All this is summarized in
\Tref{table:mix}. Dirac and Majorana neutrinos differ only in the
number of observables phases. For three families ($n=3$), there is
just one Dirac phase and three in the Majorana case.

\begin{table}
\caption[]{Number of real and imaginary parameters in the Yukawa
           matrices, of those that can be absorbed in field
           redefinitions. The difference between the two is the number
           of observable parameters: the lepton masses ($m$), mixing
           angles ($\theta$), and phases ($\phi$).}
\label{table:mix}
\centering
\begin{tabular}{@{}l|l|l|l|l|l@{}} \hline\hline
 & Yukawas 
  & Field redefinitions 
   & $No.~Êm$ 
    & $No.~ \theta$ 
     & $No.~ \phi$                 \\ \hline 
 &&&&&                           \\ 
Dirac 
 & $\lambda_l,~\lambda_\nu$ 
  & $U(n)^3/U(1)_L$   
   &&&                           \\  
 & $4 n^2$   
  &  $\dfrac{3(n^2-n)}{2},~\dfrac{3(n^2+n)-1}{2}$ 
   & $2 n$ 
    & $\dfrac{n^2-n}{2}$ 
     & $\dfrac{(n-2)(n-1)}{2}$   \\
 &&&&&                           \\ \hline
 &&&&&                           \\
Majorana 
 & $\lambda_l,~\alpha_\nu^T=\alpha_\nu$ 
  & $U(n)^2$ 
   &&&                           \\  
 & $3 n^2 + n$ 
  & $n^2-n,~n^2+n$ 
   & $2 n$ 
    & $\dfrac{n^2-n}{2}$ 
     &  $\dfrac{n^2-n}{2}$       \\ 
 &&&&&                           \\\hline\hline
\end{tabular}
\end{table}

A standard parametrization of the mixing matrices is given by
\begin{eqnarray}
V_\text{MNS} &=& 
\begin{pmatrix}
1 & 0 & 0           \\
0 & c_{23} & s_{23} \\
0 & -s_{23} & c_{23} 
\end{pmatrix}
\begin{pmatrix}
c_{13} & 0 & s_{13} \\
0 & 1 & 0           \\
-s_{13} & 0  & c_{13}
\end{pmatrix}
\begin{pmatrix}
c_{12} & s_{12} e^{i \delta}& 0  \\
-s_{12}e^{i\delta} & c_{12}  & 0 \\\
0 & 0 & 1 
\end{pmatrix}\\
{\tilde V}_\text{MNS} &=& 
V_\text{MNS}(\theta_{12},\theta_{13}, \theta_{23}, \delta) 
\begin{pmatrix}
  1 & 0 & 0             \\
  0 & e^{i \alpha_1} & 0\\
0 & 0 & e^{i \alpha_2}
\end{pmatrix}\SPp.
\label{mns}
\end{eqnarray}

\section{Neutrino oscillations}

The fact that neutrinos are such weakly interacting particles allows
them to have coherence over very long distances.  For example, a
neutrino with an energy of $\mathcal{O}(1\UMeV$) moving in lead, which
has a density of $\rho = 7.9\Ug/\UcmZ^3$, has a mean free path $l \sim
\frac{1}{\sigma \rho} \sim 4 \times 10^{16}$ metres $\sim 4$
light-years.

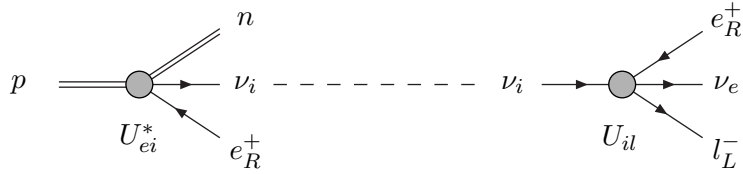
\begin{figure}
\centering
\begin{picture}(300,60)(0,20)
  \Line(30,51)(60,51) \Line(30,49)(60,49) \Text(15,50)[]{$p$}
  \Line(60,51)(90,71) \Line(60,49)(90,69) \Text(100,75)[]{$n$}
  \ArrowLine(90,30)(60,50) \Text(100,25)[]{$e^+_R$}
  \ArrowLine(65,50)(90,50) \Text(100,50)[]{$\nu_i$}
  \GCirc(60,50){5}{.7} \Text(60,30)[]{$U_{ei}^*$}
  \DashLine(110,50)(185,50)5
  \ArrowLine(210,50)(240,50) \Text(200,50)[]{$\nu_i$}
  \ArrowLine(240,50)(270,30) \Text(280,25)[]{$l^-_L$}
  \ArrowLine(270,70)(240,50) \Text(280,75)[]{$e^+_R$}
  \ArrowLine(245,50)(270,50) \Text(280,50)[]{$\nu_e$}
  \GCirc(240,50){5}{.7} \Text(240,30)[]{$U_{il}$}
\end{picture}
\caption[]{Neutrino oscillations}
\label{fig:nuosc}
\end{figure}

Neutrinos are necessarily produced in a flavour eigenstate, that is, in
a precise combination of the mass eigenstates, which are the true
eigenstates of the free Hamiltonian. After some distance $L$, where
neutrinos have evolved freely, the mass eigenstate components in the
original flavour state get different phases and, as a result, there is a
non-zero probability that the flavour measured at $L$ is a different
one \cite{pontecorvo}, as shown in \Fref{fig:nuosc}.

There has been a lot of discussion about what is the rigorous way to
define such a transition probability. This is not
straightforward because, in quantum field theory (which is required
since neutrinos are relativistic), we are used to considering processes in
which there is no knowledge of the position in space or time where the
interaction took place, and it is then a good approximation to consider
asymptotic states that are simply plane waves, with well-defined
energy--momentum. In this case this is not possible, because we must
distinguish the macroscopic distance that separates the source of
neutrinos and the detector. This implies that it cannot be a good
approximation to consider asymptotic states of well-defined momentum
at least in the direction between source and detector. This fact has
often confused the derivation and even led to incorrect results.
 
Let us consider that neutrinos are produced as wave packets localized
around the source position $x_0 = (t_0, \vec{x_0})$ in a flavour state
$\alpha$:
\begin{equation}
|\nu_\alpha(x)\rangle = \sum_j  V_{\alpha j} 
  \int \frac{d^3 k}{(2 \pi)^3} f_j(\vec{k})  
  e^{-i k_0^j (t-t_0)} 
  e^{i \vec{k} (\vec{x}-\vec{x_0})}  |\nu_j\rangle\SPp,
\label{is}
\end{equation}
where ${k^j_0}^2 = \vec{k}^2 + m_j^2$, since the state being
asymptotic must be on-shell and $V_{\alpha j}$ is the mixing
matrix. The wave packets $f_j(\vec{k})$ depend on the production
process (uncertainty in momentum of the initial states, kinematics),
but we do not need to know the exact form. For example we can consider
a Gaussian:
\begin{eqnarray}
f_i(\vec{k}) \sim e^{-(\vec{k}- \vec{\bar{q}}^i)^2/(2 \sigma^2_i)}\SPp.
\label{eq:gauss}
\end{eqnarray}
We expect that, neglecting neutrino masses, the wave packets are the
same for all the mass eigenstates:
\begin{eqnarray}
f_i(\vec{k}) \sim 
f(\vec{k}) + O\left(m_i/|\vec{k}|\right) \sim e^{-(\vec{k}- \vec{q})^2/(2 \sigma^2)}\SPp.
\end{eqnarray}
Let us forget about the proper normalization of the state for the time
being.  Let us consider that the neutrino produced is moving in the
direction of a detector located at some distance down the beam line
$L$ in the $\hat z$ direction (therefore $\vec{q}=(0,0,q_z)$), where
we want to measure the flavour of the state in \Eref{is}. The
probability that we measure a state with flavour $\beta$ at any point
$x$ is $\sim |\langle \nu_\beta |\nu_\alpha(x)\rangle|^2$, where
\begin{eqnarray}
|\nu_\beta \rangle = \sum_j  V_{\beta j} |\nu_j \rangle .
\end{eqnarray} 
The amplitude is then
\begin{eqnarray}
\langle \nu_\beta |\nu_\alpha(x)\rangle = \sum_i~V^*_{\beta i}
V_{\alpha i} \int d^3 k~f_i(\vec{k}) e^{-i k^i_0 (t-t_0) } e^{i
\vec{k} (\vec{x}-\vec{x_0}) }.
\end{eqnarray} 
Note that we  measure neither the time of the measurement nor
the spatial $\hat x$ and $\hat y$ components, so we can integrate over
them:
\begin{eqnarray}
P(\nu_\alpha\rightarrow\nu_\beta) \sim \int dt~d{\hat x}~d{\hat y}
|\langle \nu_\beta |\nu_\alpha(x)\rangle|^2 = \sum_{i,j} V^*_{\beta i}
V_{\alpha i} V_{\beta j} V^*_{\alpha j} \times \nonumber\\
\int_{\vec{k}} \int d k'_z~f_i(\vec{k}) f^*_j(\vec{k'})
\delta\left(\sqrt{m_i^2+ k_z^2+k^2_{x}+k^2_y} -\sqrt{m_j^2+
  {k'_z}^2+k^2_x+k^2_y}\right)~e^{i (k_z-k_z') L}.
\end{eqnarray} 
Up to exponentially small terms and neglecting effects of
$O(m_i/|\vec{k}|)$ everywhere else than in the phase factor (where
they are enhanced by $L$), we obtain
\begin{eqnarray}
P(\nu_\alpha\rightarrow\nu_\beta) \sim \sum_{i,j} V^*_{\beta i}
V_{\alpha i} V_{\beta j} V^*_{\alpha j} \int_{\vec{k}} |f(\vec{k})|^2
\frac{|\vec{k}|}{|k_z|} e^{-i \frac{\Delta m^2_{ji} L}{2 |k_z|}},
\end{eqnarray} 
where $\Delta m^2_{ji} = m_i^2-m_j^2$. 

Now, we have to care about the normalization. The simplest way to
compute it is by requiring that the probability be one if
$\alpha=\beta$ in the case of zero or equal neutrino masses
(\ie $\Delta m_{ji}^2=0$). Doing this we finally obtain
\begin{eqnarray}
P(\nu_\alpha\rightarrow\nu_\beta) = 
\sum_{i,j} V^*_{\beta j} V_{\alpha j} V_{\beta i} V^*_{\alpha i} 
\int_{\vec{k}}~e^{-i \frac{\Delta m^2_{ij} L}{2 |k_z|}} 
                     \frac{|\vec{k}|}{|k_z|} |f(\vec{k})|^2 /
\int_{\vec{k}}~\frac{|\vec{k}| }{|k_z|} |f(\vec{k})|^2 \nonumber\\
\simeq \sum_{i,j} V^*_{\beta j} V_{\alpha j} V_{\beta i} 
                  V^*_{\alpha i} e^{- i \frac{\Delta m^2_{ij} L}{2 |{q}_z|}}\SPp,
\label{eq:approxosc}
\end{eqnarray}
where in the last equality we have assumed that the phase factor does
not change very much in the range of momenta of the wave packet, so
that it can be taken out of the integral. The probability for the
flavour transition is thus a periodic function of the distance between
source and detector, hence the name \emph{neutrino oscillations}
first described by Pontecorvo \cite{pontecorvo}.

Defining $W_{\alpha\beta}^{jk}\equiv \,[V_{\alpha j}V_{\beta j}^*
V_{\alpha k}^*V_{\beta k}]$ and using the unitarity of the mixing
matrix, we can rewrite the probability in the more familiar way:
\begin{eqnarray}
P(\nu_\alpha& \rightarrow & \nu_\beta) \, = \delta_{\alpha\beta}\, 
-4\; \sum_{k>j}\,\rm{Re}[W_{\alpha\beta}^{jk}]\, 
\sin^2\!\left(\frac{\Delta m^2_{jk}\,L}{4 E_\nu}\right)
\,\\ 
&\pm &\, 2  \,
\sum_{k>j}\, \rm{Im}[W_{\alpha\beta}^{jk}]\, 
\sin\!\left(\frac{\Delta m^2_{jk}\,L}{2 E_\nu}\right),
\label{eq:prob}
\end{eqnarray}
where the $\pm$ refers to neutrinos/antineutrinos and $|\vec{q}|=|q_z| \simeq E_\nu$. 

We refer to an \emph{appearance} or \emph{disappearance} oscillation
probability when the initial and final flavours are different
($\alpha\neq\beta$) or the same ($\alpha = \beta$), respectively. Note
that oscillation probabilities show the expected GIM suppression of
any flavour changing process: they vanish if the neutrinos are
degenerate.

\begin{figure}
\centering
\includegraphics[angle=270,width=.47\linewidth]{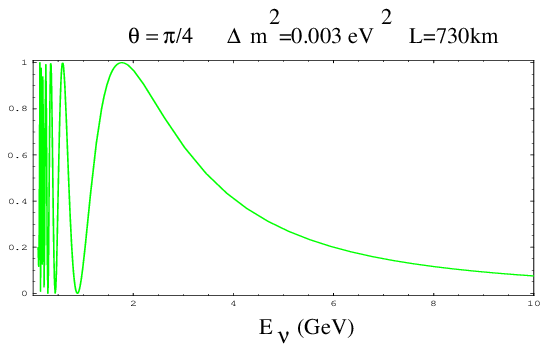}\quad
\includegraphics[angle=270,width=.47\linewidth]{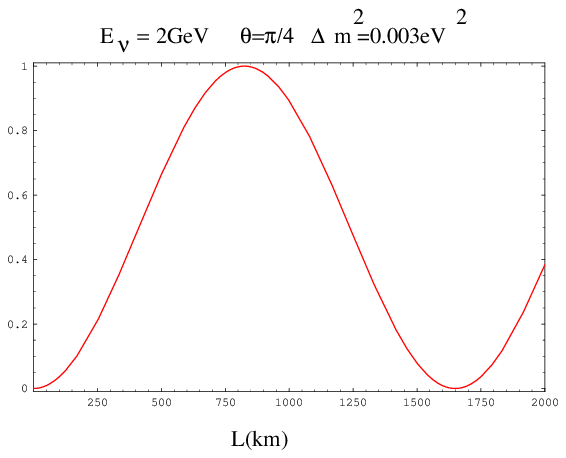}
\caption[]{Two-family oscillation probability as a function of the
           neutrino energy at fixed baseline of $L=730\Ukm$ (left) and
           as a function of the baseline at fixed neutrino energy
           $E_\nu=2\UGeV$ (right)}
\label{fig:osc}
\end{figure}

In the simplest case of two-family mixing, the mixing matrix depends on just one
mixing angle:
\begin{eqnarray}
V_\text{MNS} = \begin{pmatrix}
\cos \theta & \sin \theta \cr
-\sin \theta & \cos \theta 
          \end{pmatrix}\SPp,
\end{eqnarray}
and there is only one mass square difference $\Delta m^2$.  The
oscillation probability of \Eref{eq:prob} simplifies to the well-known
expression
\begin{eqnarray}
P(\nu_\alpha\rightarrow \nu_\beta) =  
\sin^2 2\theta \; \sin^2\!\left(\frac{\Delta m^2 \,L}{4 E_\nu}\right), \quad
\alpha\neq\beta\SPp.
\label{eq:wk}
\end{eqnarray} 
The probability is the same for neutrinos and antineutrinos because
there are no imaginary entries in the mixing matrix.  It is a
sinusoidal function of the distance between source and detector, with
a period determined by the oscillation length:
\begin{eqnarray} 
L_\text{osc}~(\UkmZ) = 2
\pi \frac{E_\nu(\UGeVZ)}{1.27 \Delta m^2 (\UeVZ^2)}\SPp,
\end{eqnarray}
which is proportional to the neutrino energy and inversely
proportional to the neutrino mass square difference.  The amplitude of
the oscillation is determined by the mixing angle. It is maximal for
$\sin^2 2 \theta =1$ or $\theta=\pi/4$. This oscillation probability
as a function of the neutrino energy and the baseline is shown in
\Fref{fig:osc}

It is important to stress that there is an intrinsic limit to
coherence, since the size of the wave packet is non-zero. Indeed the
last equality of \Eref{eq:approxosc} requires that the phase factor
varies slowly in the range of momenta of the wave packet. This
condition is not satisfied when $L$ becomes too large. The decoherence
length, $L_D$, can be estimated as
\begin{eqnarray}
\left|\frac{\Delta m^2_{ij} L_D}{2} 
      \left(\frac{1}{|q_z|} - \frac{1}{|q_z| +\sigma}\right)
\right| \sim 2 \pi \Rightarrow  L_D \sim  L_\text{osc} \frac{|q_z|}{\sigma}\SPp.
\end{eqnarray}
that is the phase factor changes by $2\pi$ when the momentum in the
$\hat z$ direction varies within one $\sigma$ from the central value,
where $\sigma$ is the width of the wave packet in momentum space [see
\Eref{eq:gauss}] When the baseline satisfies $L \gg L_D$, neutrinos do
not oscillate because the phase factor averages to zero all the terms
with $i\neq j$ in \Eref{eq:approxosc}. The flavour transition
probability then becomes independent of $L$:
\begin{eqnarray}
P\left(\nu_\alpha \rightarrow \nu_\beta\right) = \sum_i |V_{\alpha i} V_{\beta i}|^2 = 2 \cos^2\theta \sin^2\theta  = \frac{1}{2} \sin^2 2\theta .
\end{eqnarray}
In practice, the smearing in $L$ and $E_\nu$ produces the same
effect. When $L \gg L_\text{osc}$, the oscillations are so fast that any
real experiment will measure the average:
\begin{eqnarray}
\langle P\left(\nu_\alpha \rightarrow \nu_\beta\right) \rangle = 
  \frac{1}{2} \sin^2 2 \theta, 
\end{eqnarray}
which is exactly the same result as in the case of no coherence.

Note that the 'smoking gun' for neutrino oscillations is not the flavour
transition, which can occur in the presence of neutrino mixing without
oscillations, but the peculiar $L/E_\nu$ dependence.  An idealized
experiment looking for neutrino oscillations should then be able to
tell flavour on one hand and should be performed at a baseline such
that $L \sim L_{\rm osc}(E_\nu)$ in order to observe the oscillatory
pattern, which measures the neutrino mass square difference. Note that
neutrino oscillations are not sensitive to the absolute mass scale
though.

\subsection{Matter effects}

When neutrinos propagate in matter (Earth, Sun, etc.), the amplitude for their
propagation is modified owing to coherent forward scattering on electrons
and nucleons \cite{wolf}:

\begin{center}
\begin{picture}(300,100)(35,-50)
\ArrowLine(40,50)(70,25)
\ArrowLine(40,-50)(70,-25)
\Photon(70,25)(70, -25)3 4
\Text(105,0)[r]{$W^\pm$}
\Text(95,35)[r]{$e$}
\Text(110,-35)[r]{$\nu_e$}
\Text(40,-35)[l]{$e$}
\Text(40,35)[l]{$\nu_e$}
\ArrowLine(70,25)(100,50)
\ArrowLine(70,-25)(100,-50)
\ArrowLine(240,50)(270,25)
\ArrowLine(240,-50)(270,-25)
\Photon(270,25)(270, -25)3 4
\Text(295,0)[r]{$Z^0$}
\Text(335,35)[r]{$\nu_{e, \mu,\tau}$}
\Text(335,-35)[r]{$p, n, e$}
\Text(210,-35)[l]{$p,n,e$}
\Text(210,35)[l]{$\nu_{e,\mu,\tau}$}
\ArrowLine(270,25)(300,50)
\ArrowLine(270,-25)(300,-50)
\end{picture}
\end{center}

The effective Hamiltonian density 
resulting from the charged current interaction is
\begin{eqnarray}
\mathcal{H}_{\text{CC}}= \sqrt{2} G_F \,[\bar{e}\gamma_\mu P_L \nu_e]
[\bar{\nu}_e\gamma^\mu P_L e]
= \sqrt{2} G_F\,[\bar{e}\gamma_\mu P_L e]
[\bar{\nu}_e\gamma^\mu P_L \nu_e]. 
\end{eqnarray}
Since the medium is not polarized, the expectation value of the
electron current is simply the number density of electrons:
\begin{eqnarray}
\langle \bar{e}\gamma_\mu P_L e \rangle_{\rm unpol. medium} & = & \delta_{\mu 0} N_e.
\end{eqnarray}
Including also the neutral current interactions in the same way, the 
effective Hamiltonian for neutrinos in the presence of matter is
\begin{eqnarray}
\mathcal{H}^{eff}= \mathcal{H}^\text{vac} + \bar{\nu} V_m \gamma^0 (1-\gamma_5)\nu \\\nonumber\\
V_m = \begin{pmatrix}
       (\frac{G_F}{\sqrt{2}} \left(N_e  -\frac{N_n}{2}\right) 
         & 0 
           & 0\\
       0 
         &   \frac{G_F}{\sqrt{2}} \left(-\frac{N_n}{2}\right)  
           & 0 \\
       0 
         & 0 
           & \frac{G_F}{\sqrt{2}} \left(-\frac{N_n}{2}\right)
\end{pmatrix}\SPp,
\end{eqnarray}
where $N_n$ is the number density of neutrons.  The matter potential
in the center of the Sun is $V_e \sim 10^{-11}\UeV$ and in the Earth
$V_e \sim 10^{-13}\UeV$. In spite of these tiny values, these effects
are non-negligible in neutrino oscillations.

The plane wave solutions to the modified Dirac equation satisfy a 
different dispersion relation and as a result,  
the phases of neutrino oscillation phenomena change. The new dispersion relation becomes
\begin{equation}
E - V_m - M_\nu = 
  \left(\pm |\vec{p}| - V_m \right) 
  \frac{1}{E + M_\nu - V_m} \left( \pm |\vec{p}| - V_m \right) \;\;\; h=\pm, 
\end{equation}
where $h=\pm$ indicate the two helicity states and we have neglected
effects of $\mathcal{O}(V M_\nu)$. This is a reasonable approximation
since $m_\nu \gg V_m$. For the positive energy states we then have
\begin{equation}
E > 0 \;\;\;\; E^2 = |\vec{p}|^2 + M_\nu^2 + 4 E V_m \;\;\ h=-\;\;\;  E^2 = |\vec{p}|^2 + M_\nu^2,\;  h=+,   
\end{equation}
while for the negative energy ones $V_m \rightarrow - V_m$ and $h\rightarrow -h$. 

The effect of matter can be simply accommodated in an effective mass matrix:
\begin{equation}
{\tilde M}_\nu^2 = M_\nu^2 \pm 4 E V_m.
\end{equation}
The effective mixing matrix ${\tilde V}_\text{MNS}$ is the one that takes us from the original flavour basis
to that which diagonalizes this effective mass matrix:
\begin{eqnarray}
\begin{pmatrix} {\tilde m}^2_1 & 0 & 0 \\
                0 & {\tilde m}^2_2 & 0 \\
                0 & 0 & {\tilde m}^2_3 
\end{pmatrix} = {\tilde V}^\dagger_\text{MNS} 
\left( M_\nu^2 \pm  4 E 
     \begin{pmatrix} V_e & 0   & 0 \\
                     0 & V_\mu & 0 \\
                     0 & 0 & V_\tau 
     \end{pmatrix} 
\right) {\tilde V}_\text{MNS}.
\end{eqnarray}
Note that the number of physical parameters
is the same but the effective mixing angles and masses depend on the energy. 

\subsection{Neutrino oscillations in constant matter}

In the case of two flavours, the effective mass and mixing angle have relatively 
simple expressions:
\begin{eqnarray}
\sin^2 2\tilde\theta =\frac{\left(\Delta m^2 \sin 2\theta\right)^2}
{\left(\Delta m^2\cos 2\theta \mp 2 \sqrt{2}\,G_F E \,N_e\right)^2
+\left(\Delta m^2 \sin 2\theta\right)^2}\, \\
\Delta {\tilde m}^2 = \sqrt{\left(\Delta m^2\cos 2\theta\mp 2 \sqrt{2} E \,G_F\,N_e 
\right)^2+\left(\Delta m^2 \sin 2\theta\right)^2},\; 
\end{eqnarray}
where the sign $\mp$ corresponds to neutrinos/antineutrinos.  The
corresponding oscillation amplitude has a resonance \cite{wolf,ms},
when the neutrino energy satisfies
\begin{eqnarray}
\sqrt{2}\,G_F\,N_e \mp \frac{\Delta m^2}{2 E}\cos 2\theta = 0 \;\;\; \Rightarrow \;\;\;\sin^2 2 \tilde \theta = 1\;\;\; \Delta {\tilde m}^2 = \Delta m^2 \sin 2 \theta .
\label{res1}
\end{eqnarray} 
The oscillation amplitude is therefore maximal independently of the value of the vacuum 
mixing angle. 

We also note that
\begin{itemize}
\item oscillations vanish at $\theta=0$, because the oscillation length 
becomes infinite for $\theta=0$;
\item the resonance is only there for $\nu$ or $\bar{\nu}$  but not 
both; 
\item the resonance condition depends on the sign$(\Delta m^2 \cos 2 \theta)$:
\begin{center}
resonance observed in $\nu$ $\rightarrow$ sign($\Delta m^2 \cos 2 \theta$) $> 0$,\\
resonance observed in $\bar{\nu}$ $\rightarrow$ sign($\Delta m^2 \cos 2 \theta$) $< 0$.
\end{center} 
\end{itemize}

\subsection{Neutrino oscillations in variable matter}

In the Sun the density of electrons is not constant. However, if the 
variation is sufficiently slow, the eigenstates
of $H_{eff}$ change slowly with the density and we can 
assume that the neutrino produced in a local eigenstate remains in the same 
eigenstate along the trajectory. This is the so-called \emph{adiabatic approximation}.

Let us suppose that neutrinos are crossing the Sun. We consider here 
two-family mixing for simplicity. At any point in 
the trajectory, it is possible to diagonalize the Hamiltonian 
fixing the matter density to that at the given point. The
resulting eigenstates can be written as
\begin{eqnarray}
& |{\tilde\nu}_1\rangle = |\nu_e\rangle \,\cos\tilde\theta - |\nu_\mu\rangle \, \sin\tilde \theta , 
\\
& |{\tilde\nu}_2\rangle = |\nu_e\rangle \,\sin\tilde\theta  + |\nu_\mu\rangle \,\cos\tilde\theta .
\label{eq:eigenmat}  
\end{eqnarray}
Neutrinos are produced close to the centre $x=0$ where the electron density, 
$N_e(0)$,
is very large. Let us suppose that it satisfies
\begin{eqnarray}
2 \sqrt{2} G_F N_e(0) \gg \Delta m^2 \cos 2 \theta.
\end{eqnarray}
Then the diagonalization of the mass matrix at this point gives
\begin{eqnarray}
\tilde \theta \simeq \frac{\pi}{2} \Rightarrow  |\nu_e\rangle \simeq |{\tilde\nu}_2\rangle 
\end{eqnarray}
in such a way that an electron neutrino is mostly the second mass
eigenstate.  When neutrinos exit the Sun, at $x=R_\odot$, the matter
density falls to zero, $N_e(R_\odot) = 0$, and the local effective
mixing angle is the one in vacuum, ${\tilde \theta} = \theta$.  If
$\theta$ is small, the eigenstate ${\tilde\nu}_2$ is mostly $\nu_\mu$
according to \Eref{eq:eigenmat}.

\begin{figure}
\centering
\includegraphics[width=.5\linewidth]{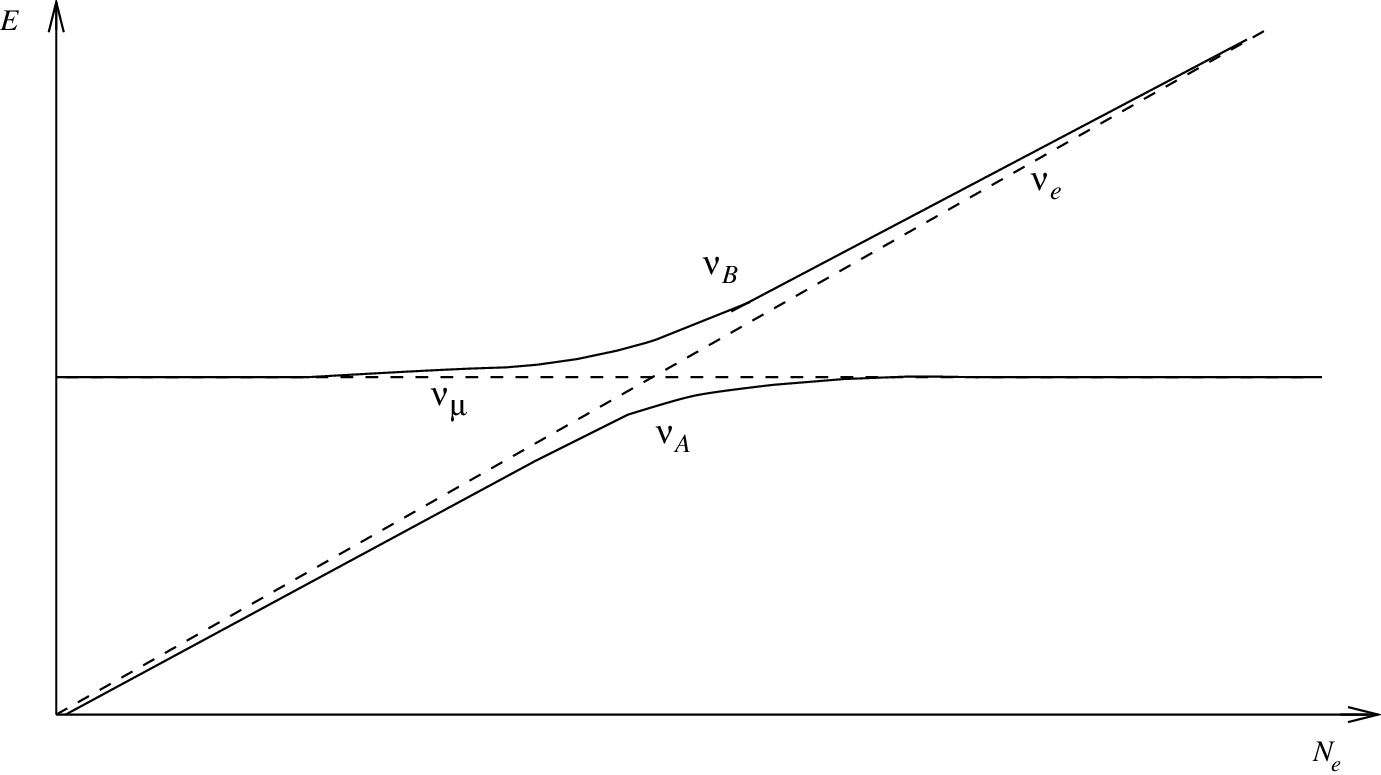}
\caption[]{Evolution of the eigenstates as a function of the distance to the 
           centre of the Sun}
\label{fig:levcross}
\end{figure}

\begin{figure}
\centering 
\includegraphics[width=.6\linewidth]{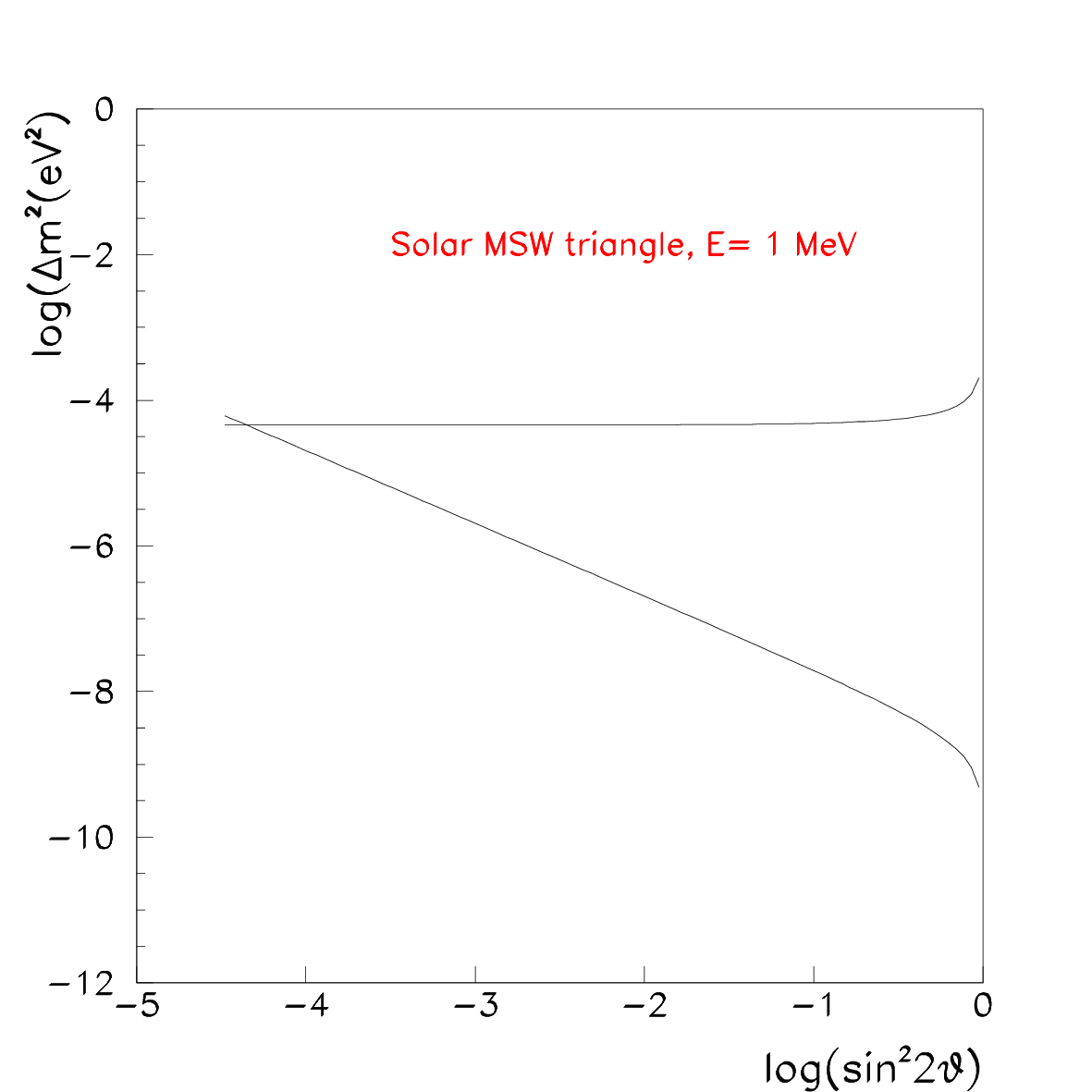}
\caption[]{MSW triangle: in the region between the two lines the
           resonance and adiabaticity conditions are both satisfied
           for neutrinos of energy 1\UMeV}
           \label{fig:mswtrian} 
\end{figure}

Therefore an electron neutrino produced at $x=0$ is mostly the eigenstate
$\tilde \nu_2$, but 
this eigenstate outside the Sun is mostly $\nu_\mu$. There is maximum $\nu_e \rightarrow \nu_\mu$ conversion if the adiabatic approximation is a good one. 
This is the famous MSW effect \cite{wolf,ms}. The evolution of the eigenstates 
is shown in \Fref{fig:levcross}: the MSW effect would occur when there is a level crossing in the absence of mixing. 
 The conditions for this to happen are:
\begin{itemize}
\item \emph{Resonant condition}: the density at the production is above the critical one 
\begin{equation}
N_e(0) > \frac{\Delta m^2 \cos 2\theta}{2 \sqrt{2} E G_F}. 
\end{equation}
\item \emph{Adiabaticity}: the splitting of the levels is large compared 
to energy injected in the system by the variation of $N_e(r)$. A measurement of this is given by $\gamma$ which should be much larger than one:
\begin{equation}
\gamma = \frac{\sin^2 2 \theta}{\cos 2\theta}
         \frac{\Delta m^2}{2 E} \frac{1}{|\nabla \log N_e(r)|} > \gamma_\text{min} > 1, 
\end{equation}
where $\nabla = \partial/\partial r$.
\end{itemize}
At fixed energy both conditions give the famous MSW triangles, if
plotted on the plane $(\log(\sin^2 2 \theta), \log(\Delta m^2))$:
\begin{eqnarray}
\log\!\left(\Delta m^2\right) < 
\log\!\left( \frac{2 \sqrt{2} G_F N_e(0) E}{\cos 2\theta}\right)  \\
\log\!\left(\Delta m^2\right) > 
\log\!\left( \gamma_{\rm min} 2 E 
  \nabla \log N_e \frac{\cos 2\theta}{\sin^2 2\theta}\right). 
\end{eqnarray}
For example, taking $N_e(r) = N_c \exp(-r/R_0),  R_0=
R_\odot/10.54, N_c = 1.6 \times 10^{26}\Ucm^{-3}, E =1\UMeV$, these
curves are shown in \Fref{fig:mswtrian}.

\begin{figure}[htb] 
\centering
\includegraphics[width=.5\linewidth]{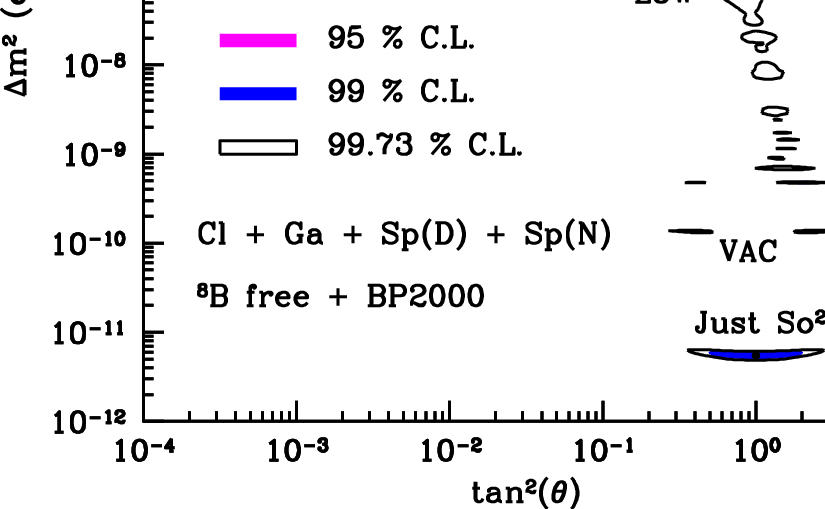}
\caption[]{Neutrino oscillation solutions to the solar neutrino
           deficit in year 2000 (taken from Ref.~\cite{bks})}
\label{fig:sol2000}
\end{figure}

As we shall see, the deficit of electron neutrinos coming from the Sun
has been interpreted in terms of an MSW effect in neutrino propagation
in the Sun. Before the recent experiments SNO and KamLAND that we shall
discuss in Section~\ref{sec:solar}, there were several solutions
possible inside the expected MSW triangle: SMA, LMA and LOW as shown
in \Fref{fig:sol2000}. The famous SMA and LOW solutions are now
history.

\section{Evidence for neutrino oscillations} 

Nature has been kind enough to provide us with two natural sources of
neutrinos (the Sun and the atmosphere) where neutrino flavour
transitions have been observed in a series of ingenious experiments,
that started back in the 1960s with the pioneering experiment of
R.~Davies.  This effort has already been rewarded once with the Nobel
prize of 2002.

\subsection{ The solar puzzle}
\label{sec:solar}

\begin{figure}
\centering
\includegraphics[angle=-90,width=.7\linewidth]{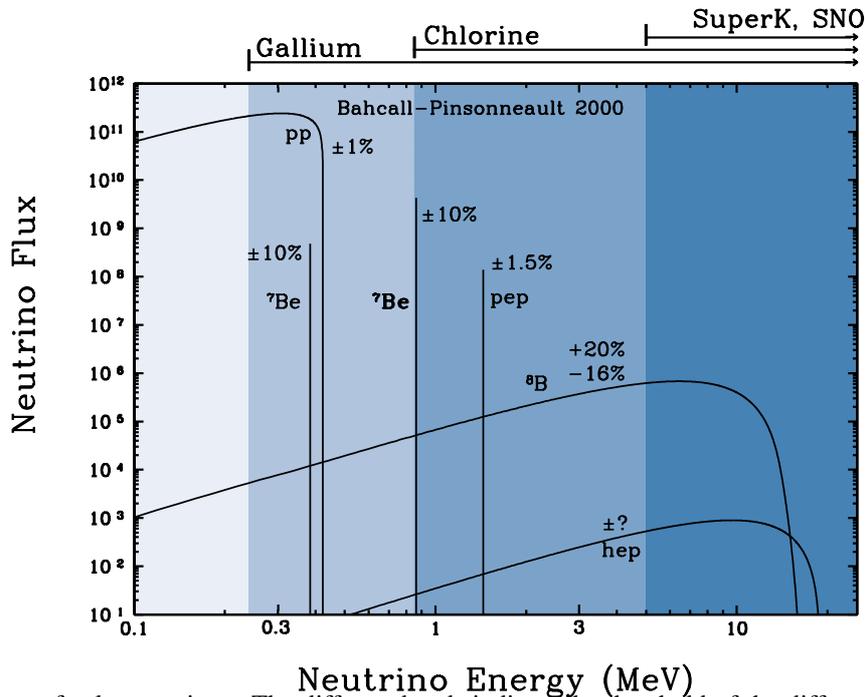}
\caption[]{Spectrum of solar neutrinos. The different bands indicate
           the threshold of the different detection
           techniques.}
\label{fig:bp2000}
\end{figure}

The Sun is an intense source of neutrinos produced in the chain of 
 nuclear reactions that burn hydrogen into helium:
\begin{eqnarray}
4 p \longrightarrow \Isotope[4]{He} + 2 e^+ + 2 \nu_e. 
\end{eqnarray}
The expected spectral flux of $\nu_e$ in the absence of oscillations 
is shown in \Fref{fig:bp2000}. The prediction of this flux obtained 
by J.~Bahcall and collaborators \cite{bp} is the result of a detailed 
simulation of the solar interior and has been improved over many years. 
It is the so-called standard solar model (SSM).

Neutrinos coming from the Sun have been detected with several
experimental techniques that have a different neutrino energy
threshold as indicated in \Fref{fig:bp2000}. On the one hand, the
radiochemical techniques, used in the experiments Homestake
(chlorine, $^{37}$Cl)\cite{homestake}, Gallex/GNO 
\cite{gallex-gno} and Sage \cite{sage} (using gallium, $^{71}$Ga, and
germanium, $^{71}$Ge, respectively), can count the total number of
neutrinos with a rather low threshold ($E_\nu > 0.81\UMeV$ in Homestake
and $E_\nu > 0.23\UMeV$ in Gallex and Sage), they cannot get any
information on the directionality, the energy of the neutrinos, nor the
time of the event.  On the other hand, Kamiokande \cite{kamio}
pioneered a new technique to observe solar neutrinos using water
Cherenkov detectors. The signal comes from elastic neutrino scattering
on electrons (ES), $\nu_e \; + \; e^- \;\rightarrow \; \nu_e \; +
e^-$, that can be observed from the Cherenkov radiation emitted by the
recoiling electrons. These are real-time experiments that provide
information on the directionality and the energy of the neutrinos by
measuring the recoiling electron. Unfortunately, the threshold
for these types of experiments is much higher, $\geq 5\UMeV$. All these
experiments have consistently observed a number of solar neutrinos
between 1/3 and 1/2 of the number expected in the SSM and for a long time
this was referred to as the \emph{solar neutrino problem or deficit}.

The progress in this field over the past ten years has been enormous
culminating in a solution to this puzzle that no longer relies on
the predictions of the standard solar model.

There have  been three milestones.  

\begin{figure}
\centering
\includegraphics[width=.47\linewidth]{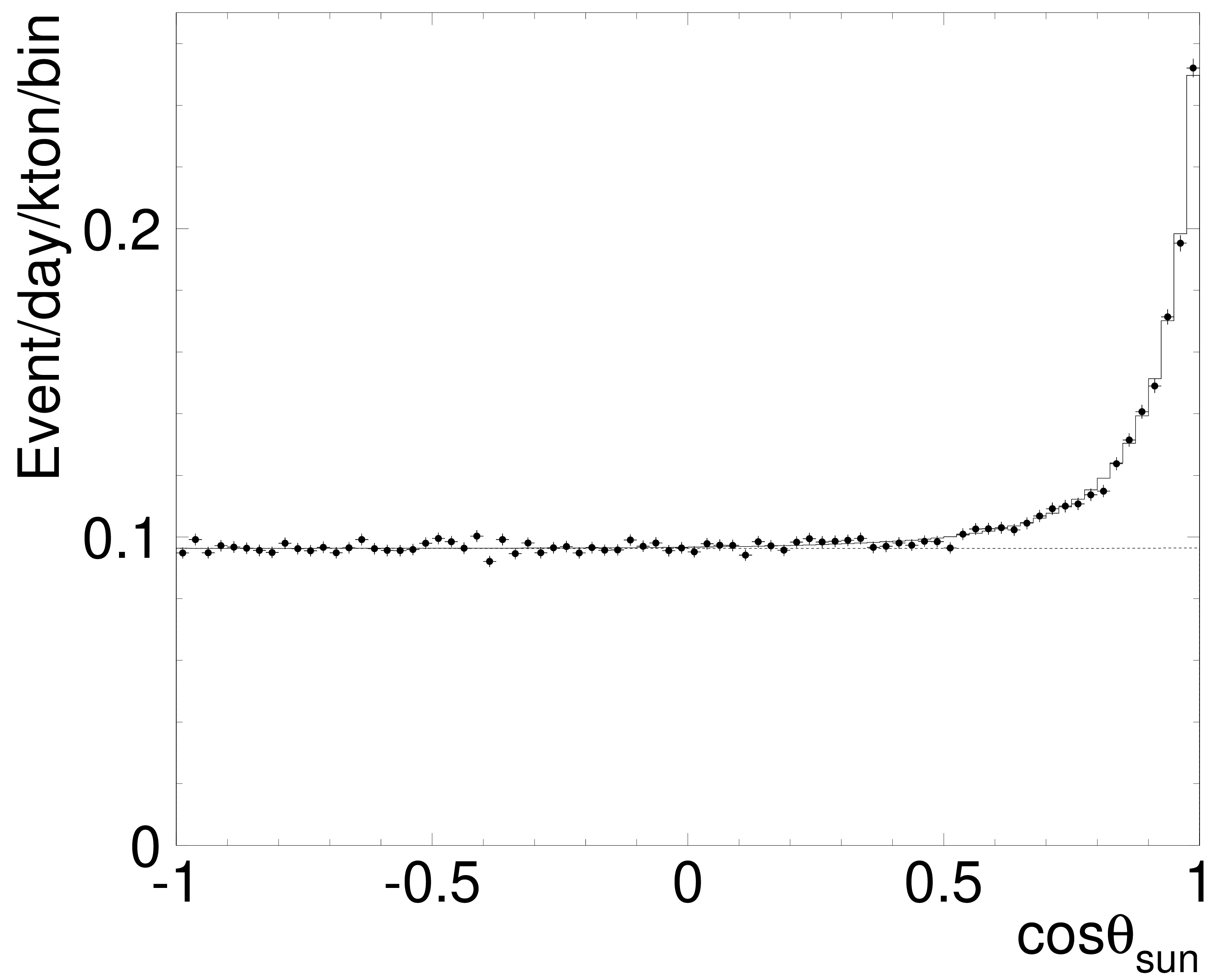}\hfill
\includegraphics[width=.47\linewidth]{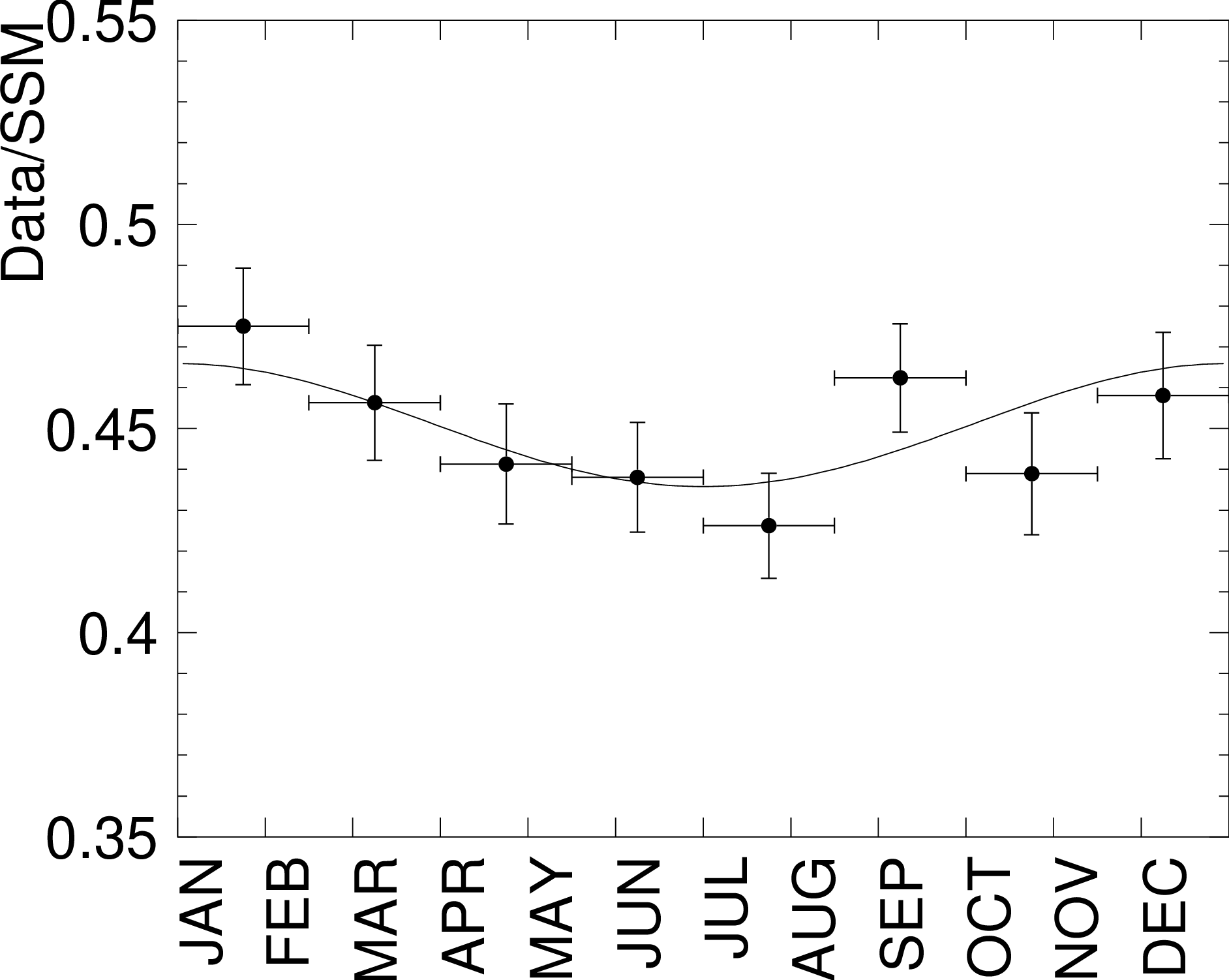}
\caption[]{Left: distribution of solar neutrino events as a function
           of the zenith angle of the Sun. Right: seasonal variation of the solar
           neutrino flux in SuperKamiokande.}
\label{fig:mile1}
\end{figure}

\begin{figure}
\centering
\includegraphics[width=.47\linewidth]{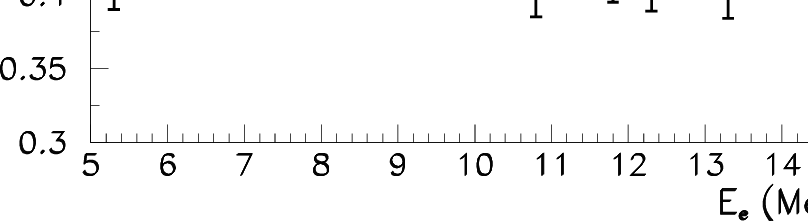}\hfill
\includegraphics[width=.47\linewidth]{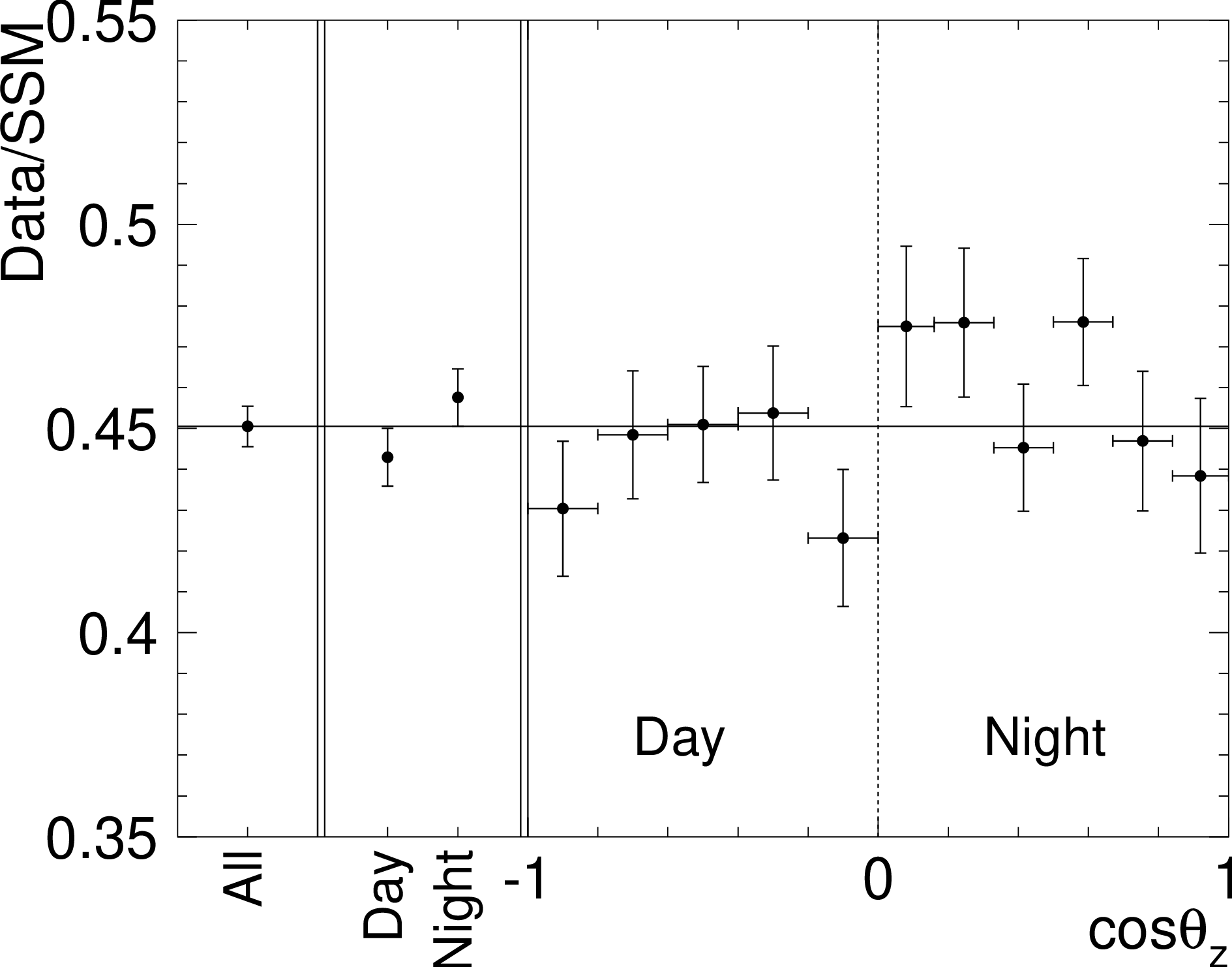}
\caption[]{Left: Distribution of the solar neutrino events as a
           function of the electron energy. Right: Day--night
           distribution of the solar neutrino events in
           SuperKamiokande.}
\label{fig:mile2}
\end{figure}

{\bf 1998}: SuperKamiokande \cite{sk} measured the solar neutrino
 deficit with unprecedented precision. Furthermore the measurement of
 the direction of the events demonstrated that the neutrinos measured
 definitely come from the Sun: the left plot of \Fref{fig:mile1} shows
 the distribution of the events as a function of the zenith angle of
 the Sun.  A seasonal variation of the flux is expected since the
 distance between the Earth and the Sun varies seasonally. The right
 plot of \Fref{fig:mile1} shows that the measured variation is in
 perfect agreement with that expectation.
If the deficit of $\nu_e$ in the Sun is interpreted in terms of
neutrino oscillations, two very important observables to discriminate
between different solutions are the spectral distribution of the
events shown in the left plot of \Fref{fig:mile2}, which shows a
rather flat spectrum, and the day/night asymmetry. The latter is
important because neutrinos arriving from the Sun at night have to
cross the Earth and some of the possible solutions are such that
matter effects in neutrino propagation in the Earth are relevant.
The analysis of solar data in year 2000 in terms of neutrino oscillations
of the $\nu_e$ into some other type indicated a number of possible solutions
as shown in \Fref{fig:sol2000}.

{\bf  2001}: The SNO experiment \cite{sno}  measured the flux
of solar neutrinos using the three reactions:   
\begin{eqnarray}
(\text{CC})&
 \quad\nu_{e}  + d \rightarrow p + p +  e^{-}  \;\;\;\;\;
  & \quad E_\text{thres} > 5\UMeV \\
(\text{NC})&
 \quad\nu_{x}  + d \rightarrow p + n +  \nu_x \;\;\;x=e,\mu,\tau 
  & \quad E_\text{thres} > 2.2 \UMeV  \\
(\text{ES})&
 \quad\nu_e \; + \; e^- \;\rightarrow \; \nu_e \; + e^- 
  & \quad E_\text{thres} > 5\UMeV 
\end{eqnarray}
Since the CC reaction is only sensitive to electron neutrinos, while
the NC one is sensitive to all the types that couple to the $Z^0$
boson, the comparison of the fluxes measured with both reactions can
establish if there are $\nu_\mu$ and $\nu_\tau$ in the solar flux
independently of the normalization given by the SSM. The neutrino
fluxes measured by the three reactions by SNO are:
\begin{equation}
\phi^{\text{CC}} = 1.67(9)   \times 10^6\Ucm^{-2}\UsZ^{-1}, \;\;\;
\phi^{\text{NC}} = 5.54(48)  \times 10^6\Ucm^{-2}\UsZ^{-1}, \;\;\; 
\phi^{\text{ES}} = 1.77(26)  \times 10^6\Ucm^{-2}\UsZ^{-1} .
\end{equation}

\begin{figure}
\centering
\includegraphics[width=.7\linewidth]{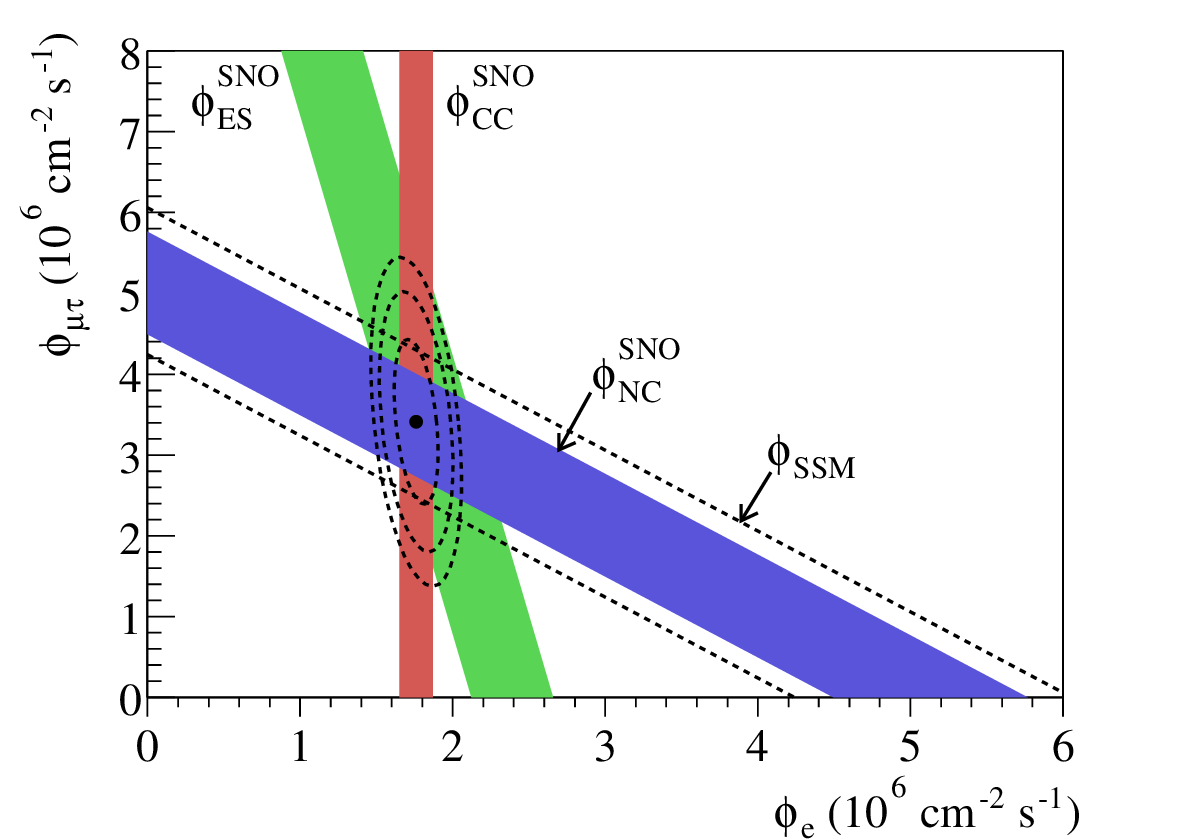}
\caption[]{Flux of $\nu_\mu$ and $\nu_\tau$ versus the flux of $\nu_e$
           in the solar neutrino flux as measured from the three
           reactions observable in the SNO experiment. The dashed band
           shows the prediction of the SSM, which agrees perfectly
           with the flux measured with the NC reaction (from
           Ref.\cite{sno_new}).}
\label{fig:compa}
\end{figure}

These measurements demonstrate that the Sun shines $(\nu_\mu,
\nu_\tau)$ about two times more than it shines $\nu_e$, which
constitutes the first direct demonstration of flavour transitions in
the solar flux! Furthermore the NC flux that measures all active
species in the solar flux, is compatible with the total $\nu_e$ flux
expected according to the SSM as shown in \Fref{fig:compa}.

\begin{figure}
\centering
\includegraphics[width=.7\linewidth]{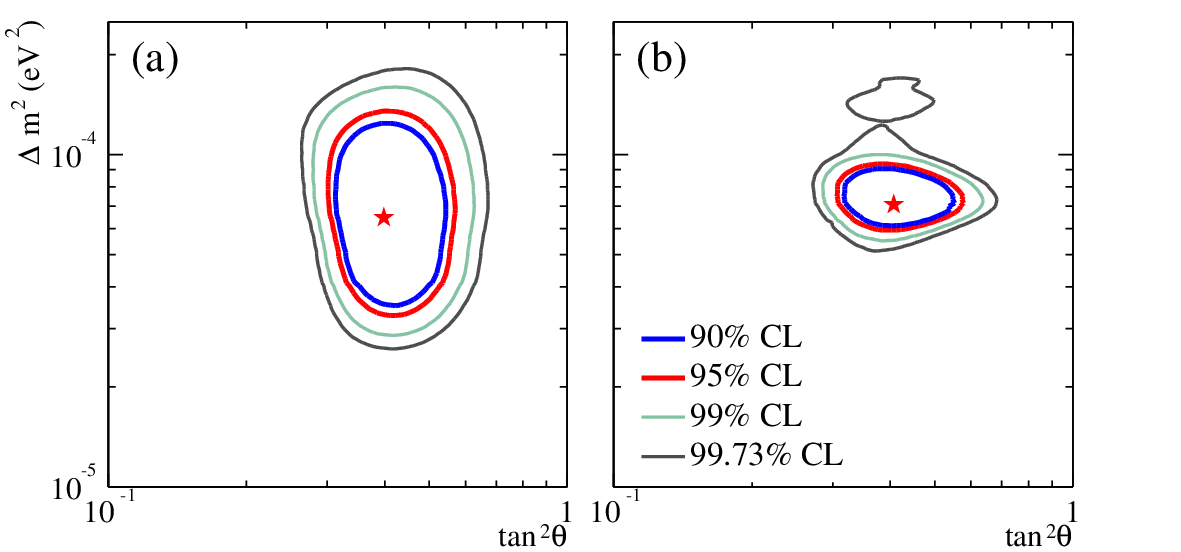}
\caption[]{Left: Analysis of all solar data in terms of neutrino
           oscillations.  Right: Analysis including also KamLAND
           data (from Ref.~\cite{sno}).}
\label{fig:sno}
\end{figure}

The post-SNO global fits of all solar data shown in
\Fref{fig:sno} (left) in terms of neutrino oscillations are quite
different from those in \Fref{fig:sol2000}. Of all the possible
solutions, only the one at the largest mixing angle and mass square
difference survives, the famous LMA solution.

{\bf 2002}: The solar oscillation is confirmed with reactor neutrinos
in the KamLAND experiment \cite{kamland}. This is 1kton of liquid
scintillator which measures the flux of reactor neutrinos produced in
a cluster of nuclear plants around Kamioka. The average distance is
$\langle L \rangle = 175\Ukm$.  Neutrinos are detected via inverse
$\beta$-decay which has a threshold energy of about $2.6\UMeV$:
\begin{eqnarray}
\bar{\nu}_e + p \rightarrow e^+ + n \qquad E_\text{th} > 2.6\UMeV \SPp.
\end{eqnarray}
The fortunate circumstance that  
\begin{eqnarray}
\langle E_\nu(1\UMeV)\rangle/L(100\Ukm) \sim 10^{-5}\UeV^2
\end{eqnarray}
is in the range indicated by solar data, and that the expected mixing
angle is large, implies that a large depletion of the expected
antineutrino flux (which is known to a few per cent accuracy) should
be observed together with a significant energy dependence.

\begin{figure}
\centering
\includegraphics[angle=270,width=.7\linewidth]{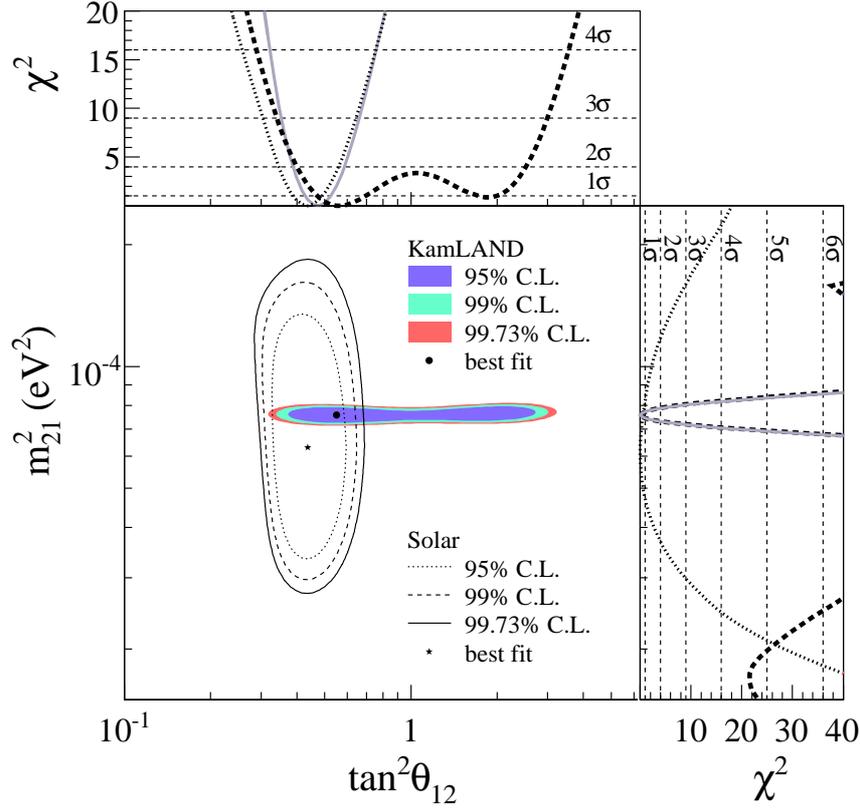}
\caption[]{Analysis of all solar and KamLAND data in terms of
           oscillations (from Ref.~\cite{kamland08})}
\label{fig:kamosc}
\end{figure}

\Fref[b]{fig:kamland2008} shows the latest KamLAND results
\cite{kamland08} for the spectral distribution of events as well as as
a function of the ratio $E_\nu/L$.
\begin{figure}
\centering
\includegraphics[angle=270,width=.47\linewidth]{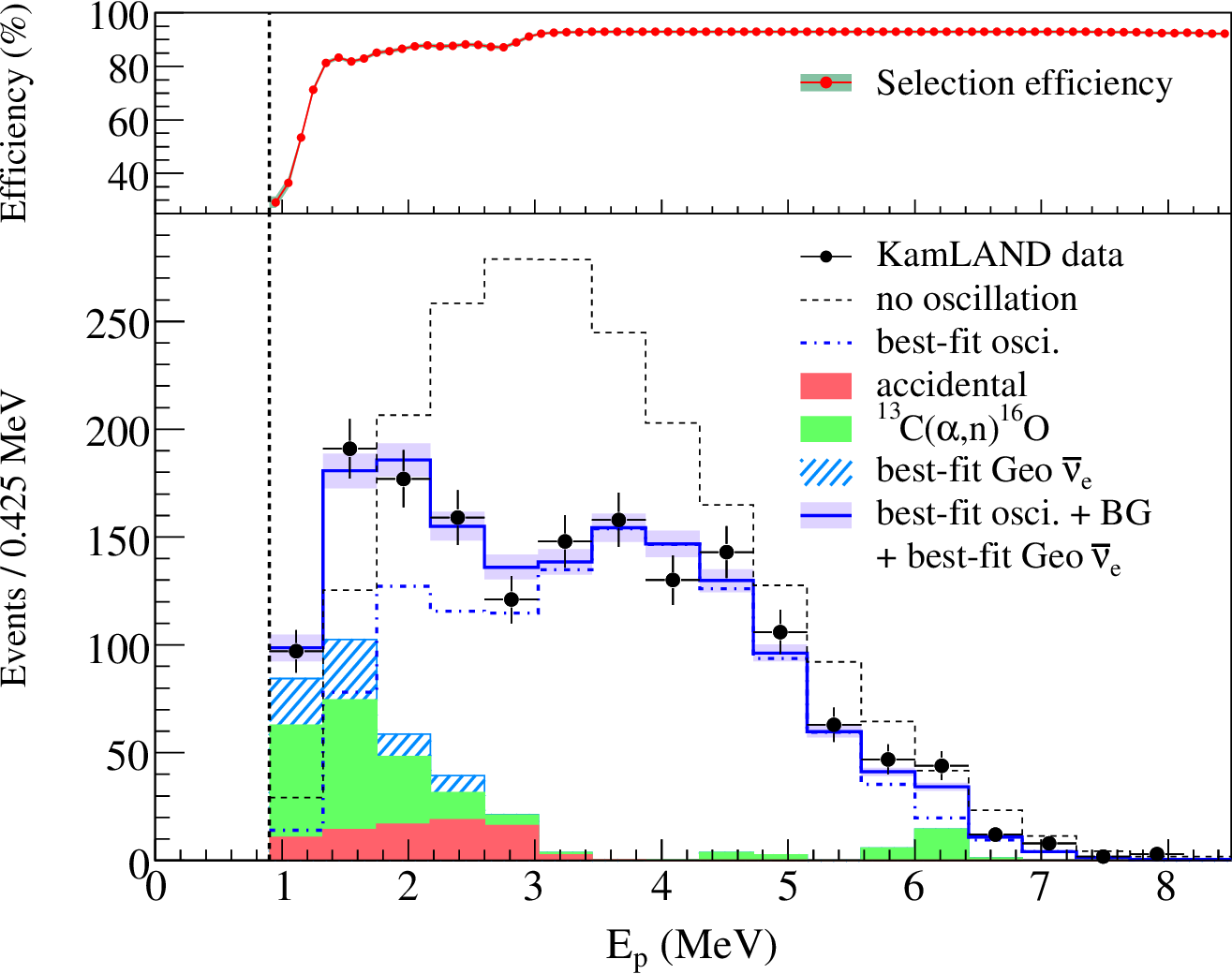}
\includegraphics[angle=270,width=.47\linewidth]{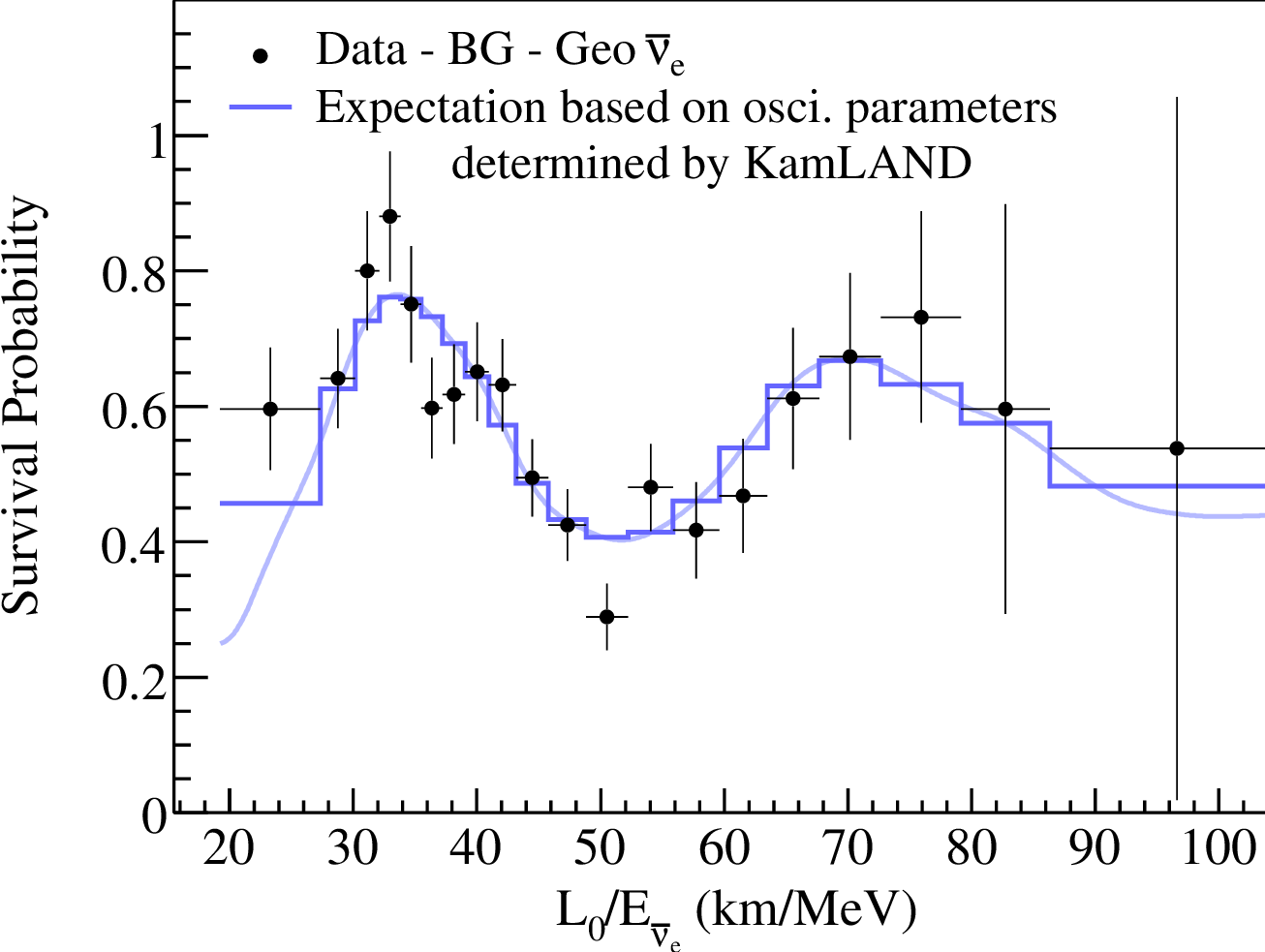}
\caption[]{Spectral distribution of the $\bar{\nu}_e$ events in
           KamLAND (left) and $E_\nu/L$ dependence (right). The data
           are compared to the expectation in the absence of
           oscillations and to the best fit oscillation
           hypothesis (from Ref.~\cite{kamland08}).}
\label{fig:kamland2008}
\end{figure}
They have recently lowered the energy threshold and have sensitivity to
geoneutrinos. The measurements of geoneutrinos could have important
implications in geophysics.  Concerning the sensitivity to the
oscillation parameters, \Fref{fig:kamosc} shows the present
determination of the solar oscillation parameters from KamLAND and
other solar experiments. The precision in the determination of $\Delta
m^2_\text{solar}$ is spectacular and shows that neutrino experiments are
entering the era of precision physics.

\begin{figure}
\centering
\includegraphics[width=.47\linewidth]{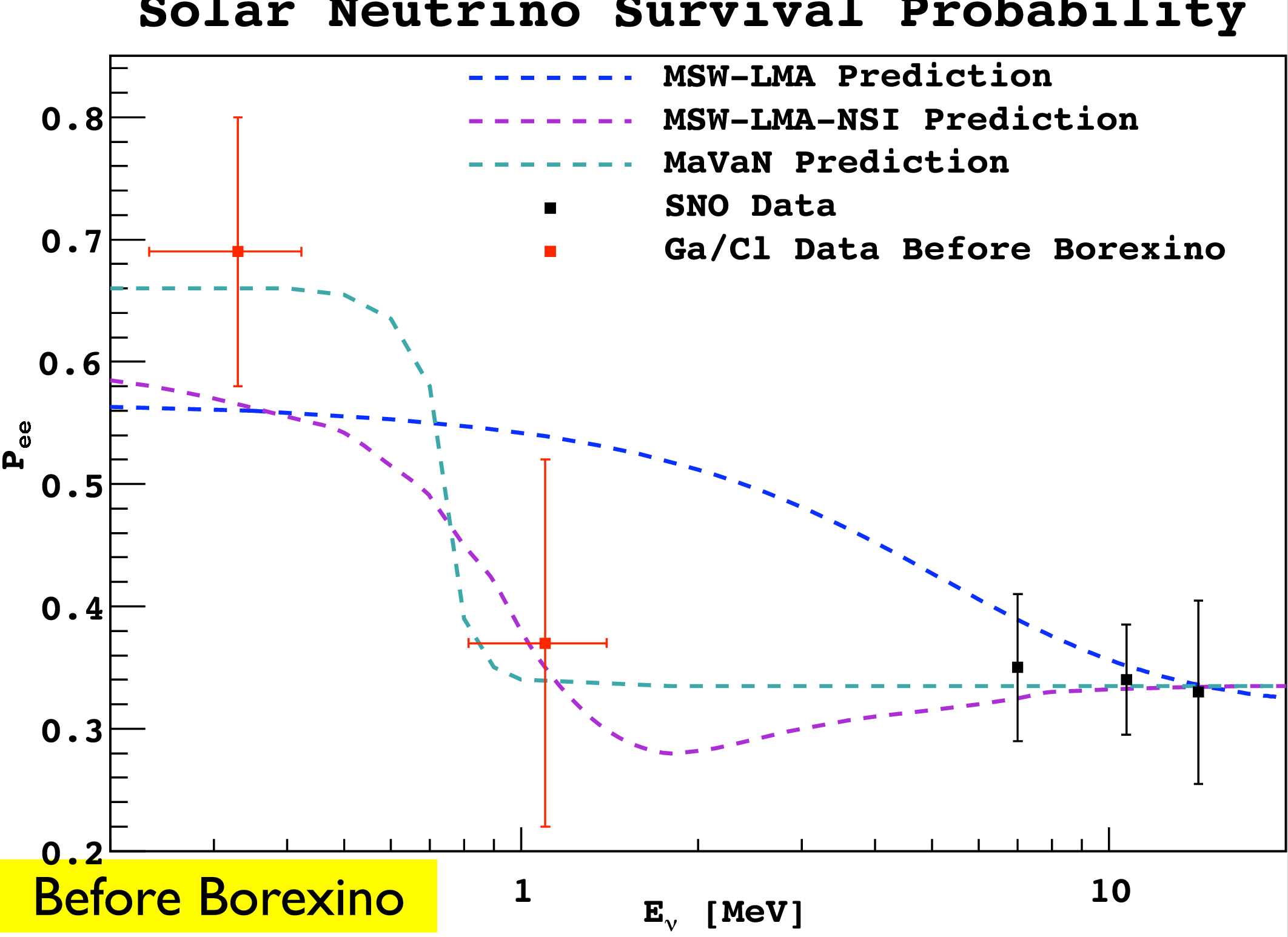}\hfill
\includegraphics[width=.47\linewidth]{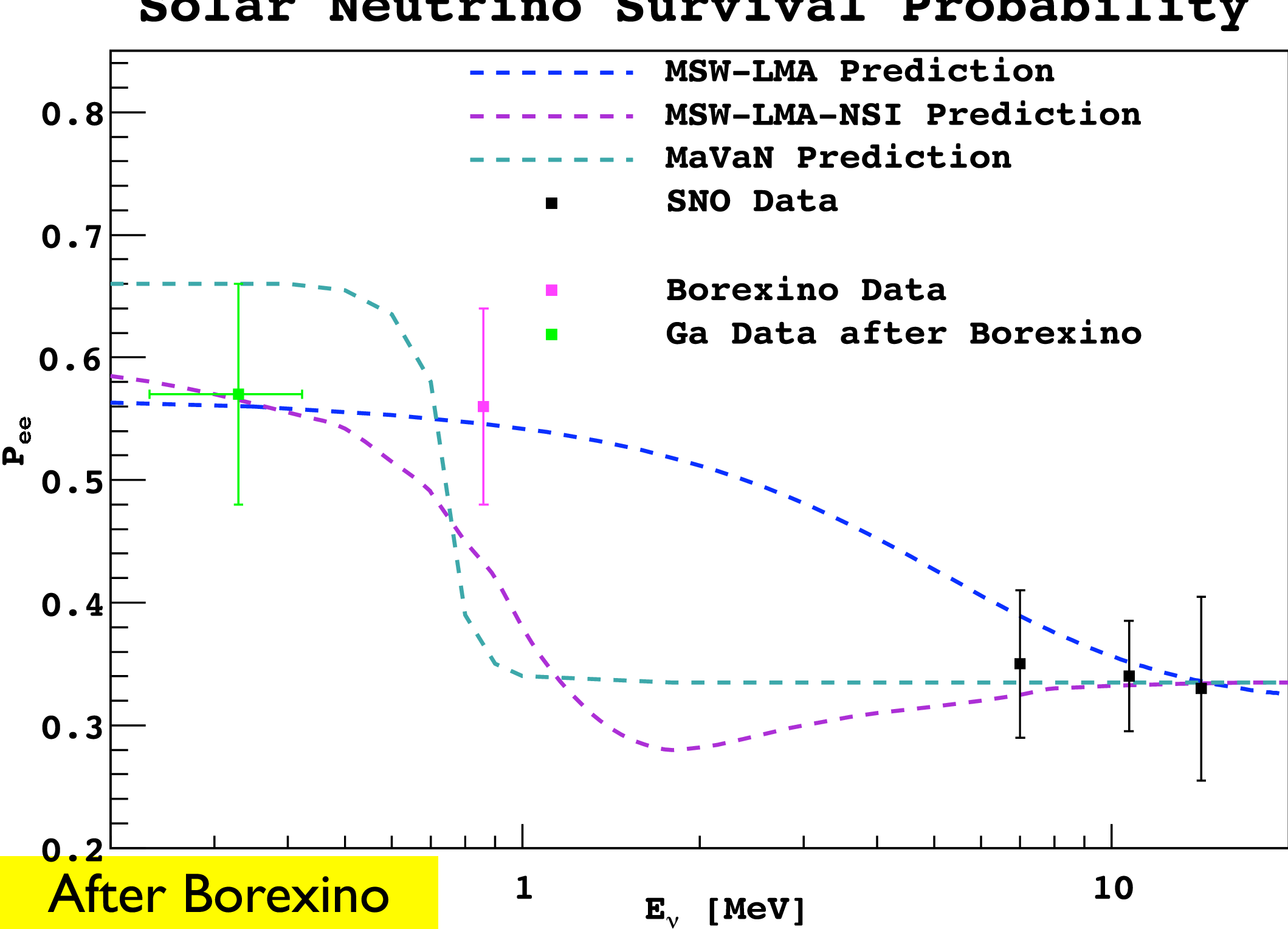}
\caption[]{Comparison of solar neutrino fluxes measured by the
           different experiments before Borexino (left) and after
           (right). Presented by the Borexino Collaboration at
           Neutrino 2008.}
\label{fig:borexino}
\end{figure}

Last year new data was presented by a new solar neutrino experiment
Borexino \cite{borexino}. It is the lowest-threshold real-time solar
neutrino experiment and the only one that could measure the flux of
the monocromatic $^7$Be neutrinos:
\[
\Phi(\Isotope[7]{Be}) = 5.08(25) \times 10^{9}\Ucm^{-2}\UsZ^{-1}\SPp.
\]
 The relevance of Borexino is illustrated in \Fref{fig:borexino}. The
result is in agreement with the oscillation interpretation of other
solar and reactor experiments and it adds further information to
disfavour alternative exotic interpretations of the data.

In summary, solar neutrinos experiments have made fundamental
discoveries in particle physics and are now becoming useful for other
applications, such as a precise understanding of the Sun and the
Earth.

\subsection{Atmospheric neutrino anomaly}
 
\begin{figure}
\includegraphics[width=\linewidth]{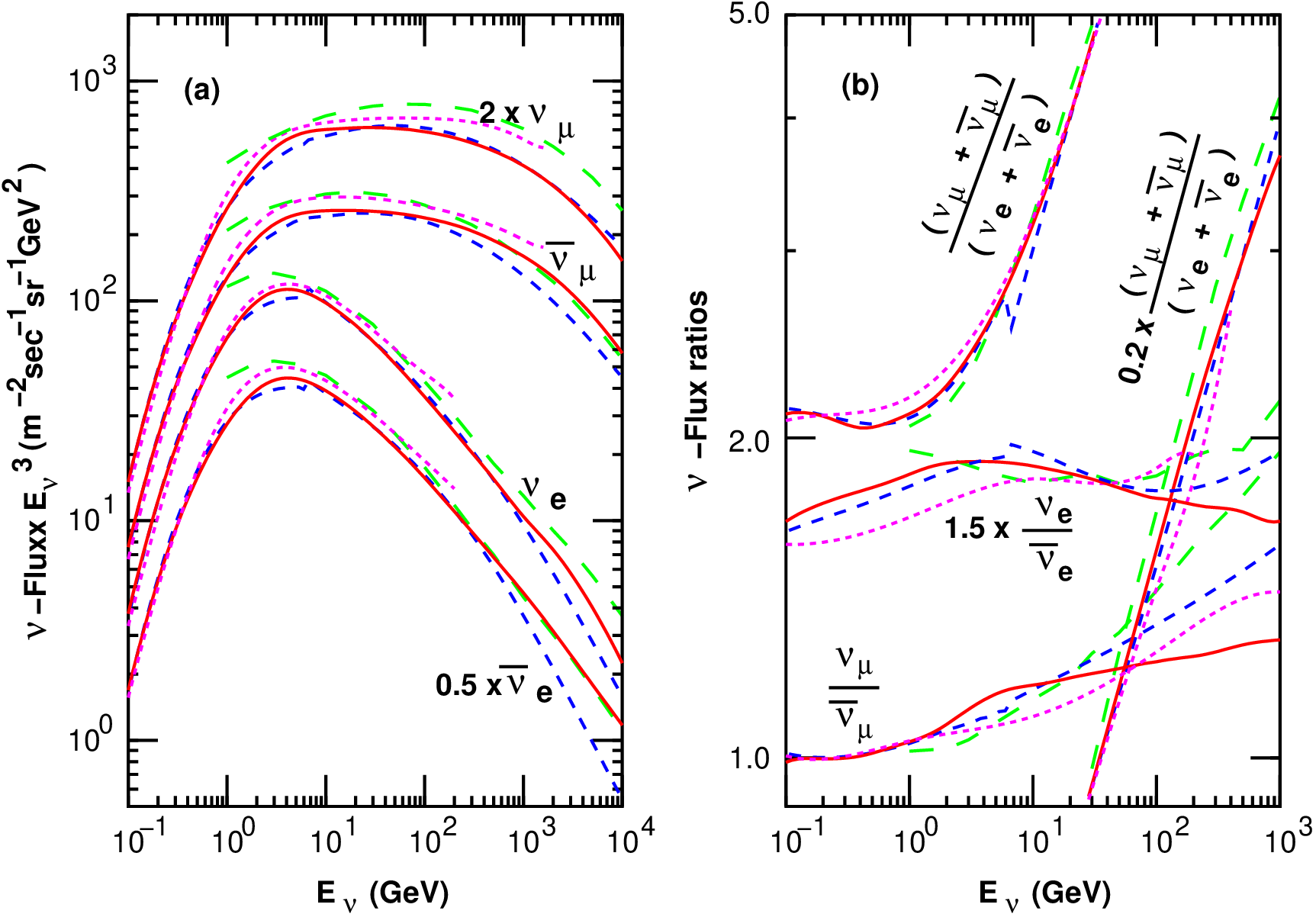}
\caption[]{Comparison of the predictions of different Monte Carlo simulations of the
           atmospheric neutrino fluxes averaged over all directions
           (left) and of the flux ratios $(\nu_\mu +
           \bar\nu_\mu)/(\nu_e + \bar\nu_e)$, $\nu_\mu / \bar\nu_\mu$, 
           and $\nu_e/\bar\nu_e$ (right). The solid line corresponds
           to a recent full 3D simulation.  Taken from the last
           reference in Ref.~\cite{atmflux}.}
\label{fig:atmflux}
\end{figure}

Neutrinos are also produced in the atmosphere when primary cosmic rays
impinge on it producing $K,\pi$ that subsequently decay. The fluxes of
such neutrinos can be predicted within a 10--20$\%$ accuracy to be
those in the left plot of \Fref{fig:atmflux}.

\begin{figure}
\centering
\includegraphics[width=.5\linewidth]{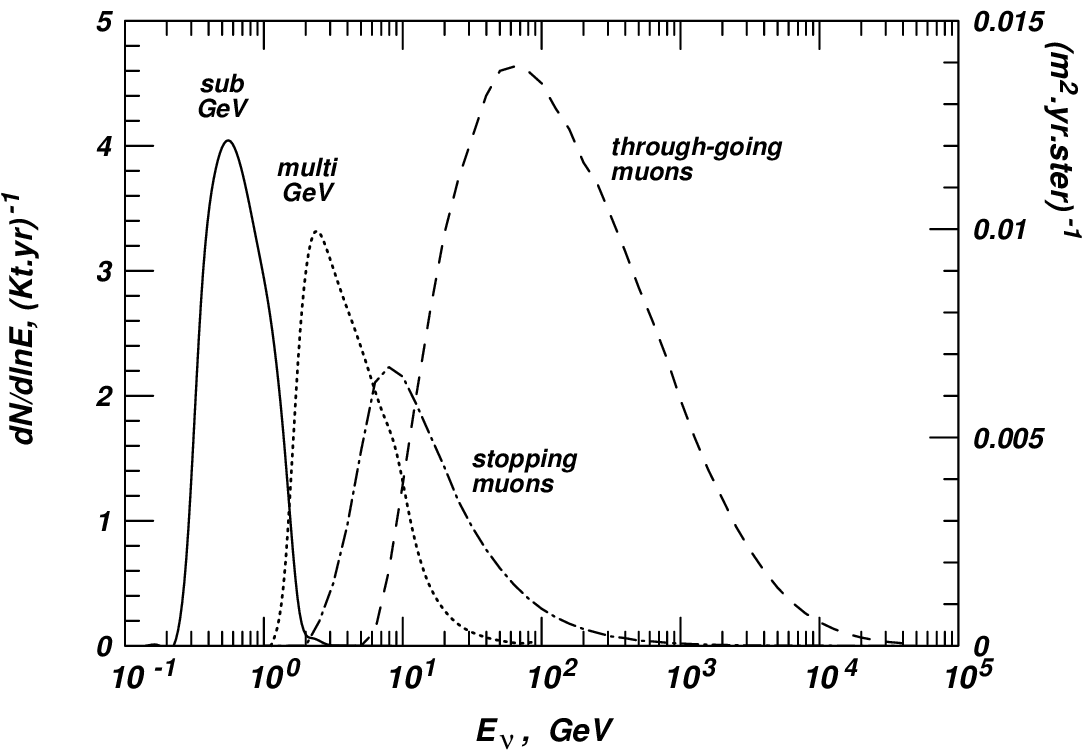}\quad
\includegraphics[width=.4\linewidth]{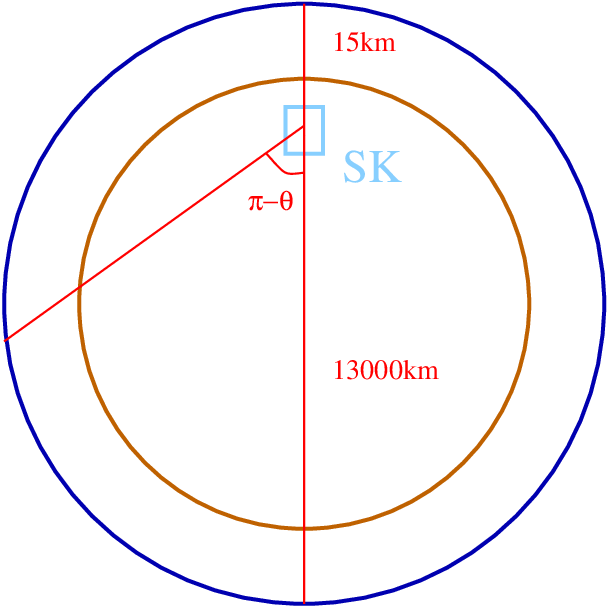}
\caption[]{Left: Parent neutrino energies of the different samples
           considered in Superkamiokande: sub-GeV, multi-GeV, stopping
           and through-going muons. Right: Distances travelled by
           atmospheric neutrinos as a function of the zenith
           angle.}
\label{fig:sksample}
\end{figure}

Clearly, atmospheric neutrinos are an ideal place to look for neutrino
oscillation since the $E_\nu/L$ span several orders of magnitude, with
neutrino energies varying from a few hundred MeV to $10^3\UGeV$ and
distances between production and detection varying from
$10$--$10^4\Ukm$, as shown in \Fref{fig:sksample} (right).

Many of the uncertainties in the predicted fluxes cancel when the
ratio of muon to electron events is considered. The first indication
of a problem was found when a deficit was observed precisely in this
ratio by several experiments: Kamiokande \cite{kamio-atmos}, IMB
\cite{imb}, Soudan2 \cite{soudan2}, Macro \cite{macro}.

\begin{figure}
\begin{center}
\includegraphics[width=.32\linewidth]{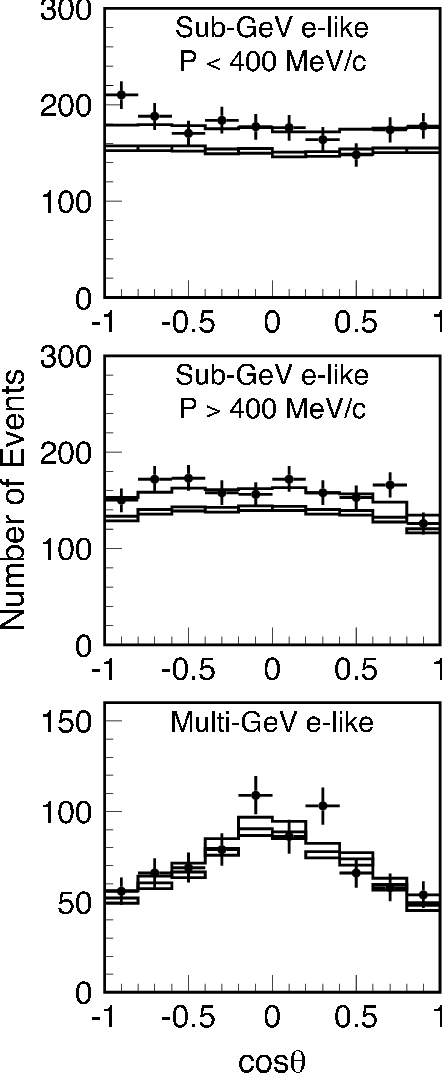}\hfill
\includegraphics[width=.64\linewidth]{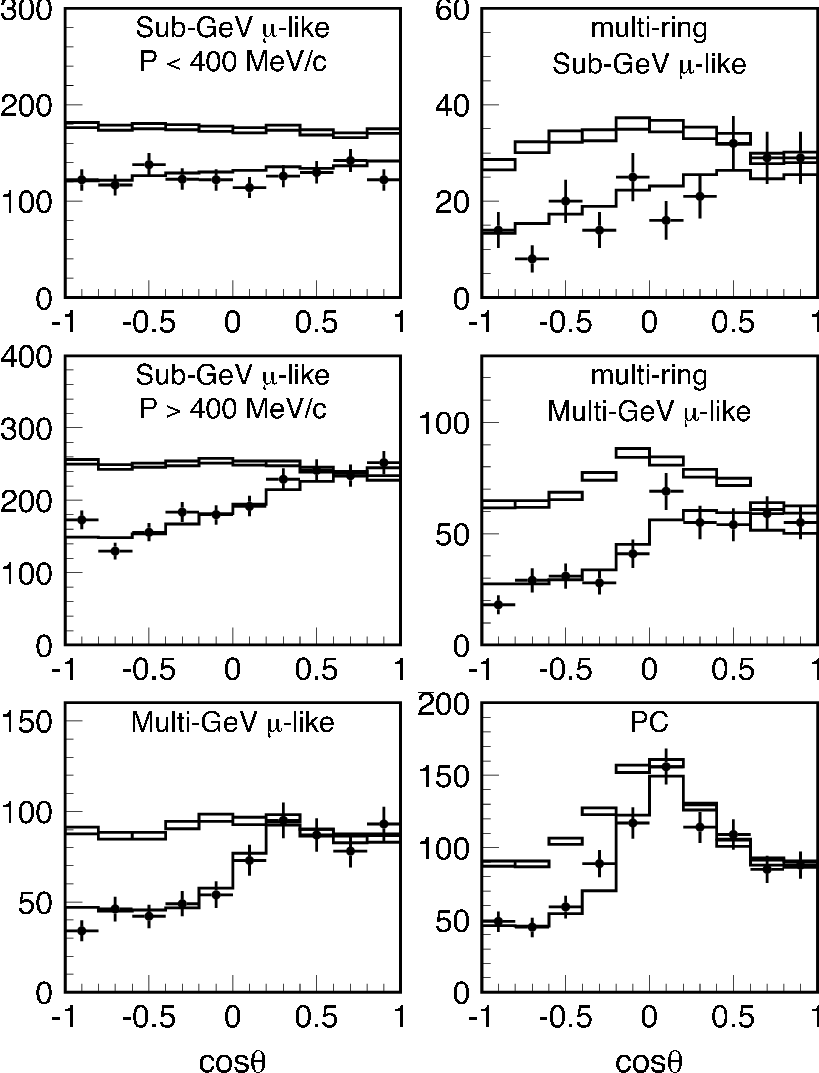}\\
\includegraphics[width=.64\linewidth]{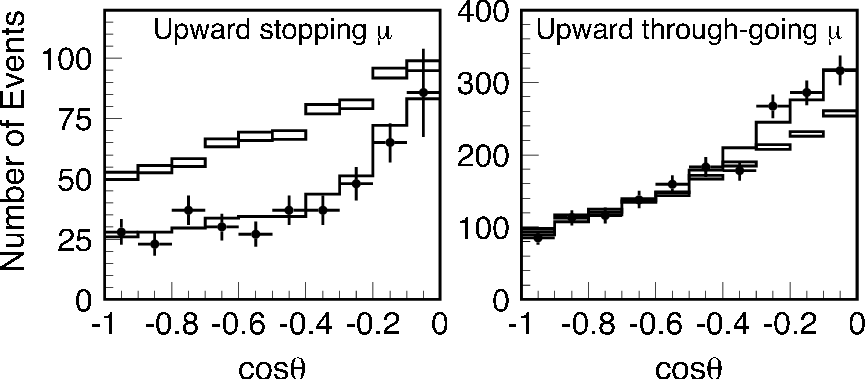}
\caption[]{Zenith angle distribution for fully-contained single-ring
           $e$-like and $\mu$-like events, multi-ring $\mu$-like
           events, partially contained events, and upward-going
           muons. The points show the data and the solid lines show
           the Monte Carlo events without neutrino oscillation.  The
           dashed lines show the best-fit expectations for $\nu_\mu
           \leftrightarrow \nu_\tau$ oscillations (from Ref.~
           \cite{moriond_sk}).}
\label{fig:skzen}
 \label{fig:zenith_angle}
\end{center}
\end{figure}

In 1998, SuperKamiokande clarified to a large extent the origin of
this anomaly \cite{sk-atmos}.  This experiment can distinguish muon
and electron events, measure the direction of the outgoing lepton (the
zenith angle with respect to the Earth's axis) which is correlated to
that of the neutrino ( the higher the energy the higher the
correlation), in such a way that they could measure the variation of
the flux as a function of the distance travelled by the
neutrinos. Furthermore, they considered different samples of events:
sub-GeV (lepton with energy below $1\UGeV$) ), multi-GeV (lepton with
energy above $1\UGeV$), together with stopping and through-going muons
that are produced on the rock surrounding Superkamiokande. The
different samples correspond to different parent neutrino energies as
can be seen in \Fref{fig:sksample} (left).
The number of events for the different samples as a function of the zenith 
angle of the lepton are shown in \Fref{fig:skzen}. 

\begin{figure}
\centering
\includegraphics[width=.6\linewidth]{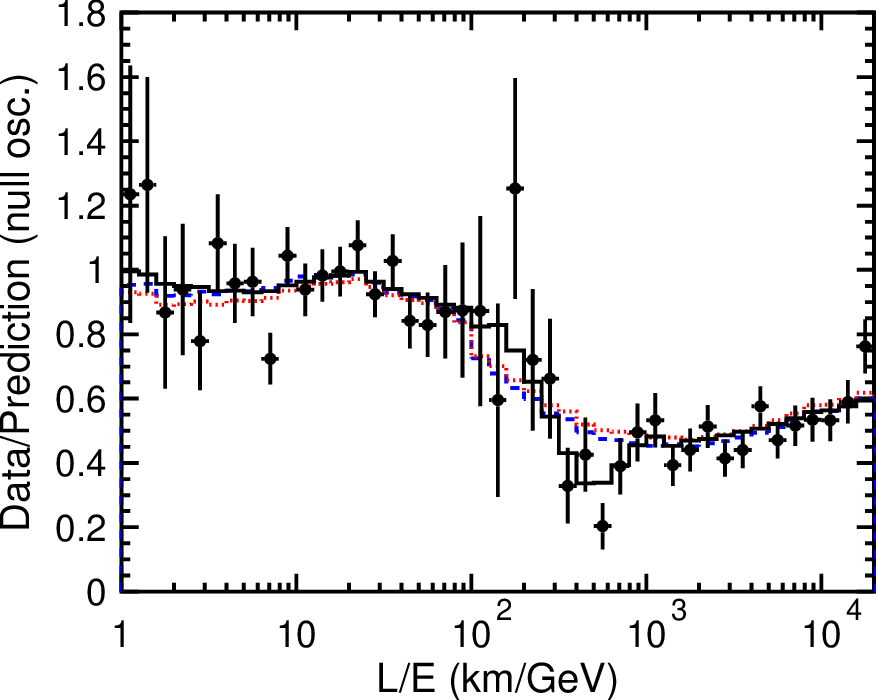}
\caption[]{Ratio of the data to the non-oscillated Monte Carlo events
           (points) with the best-fit expectation for 2-flavour
           $\nu_\mu \leftrightarrow \nu_\tau$ oscillations (solid
           line) as a function of $E_\nu/L$ (from Ref.~\cite{sk_atmos_el}).}
\label{fig:sk_el}
\end{figure}

While the electron events observed are in agreement with predictions,
a large deficit of muon events was found with a strong dependence on
the zenith angle: the deficit was almost 50$\%$ for those events
corresponding to neutrinos coming from below $\cos \theta = -1$, while
there is no deficit for those coming from above. The quality of the
fit to the neutrino oscillation hypothesis $\nu_\mu \rightarrow
\nu_\tau$ is shown in the plot. The perfect fit to the oscillation
hypothesis is rather non-trivial given the sensitivity of this
measurement to the $E_\nu$ (different samples) and $L$ (zenith angle)
dependence. The significance of the $E_\nu/L$ dependence has been
presented recently by the SuperKamiokande Collaboration
\cite{sk_atmos_el}, as shown in \Fref{fig:sk_el}.

Appropriate neutrino beams to search for the atmospheric oscillation
can easily be produced at accelerators if the detector is located at a
long baseline of a few hundred kilometres, since
\begin{equation}
|\Delta m^2_\text{atmos}| \sim \frac{E_\nu(1-10\UGeV)}{L(10^2-10^3\Ukm)}.
\end{equation}
A \emph{conventional} neutrino beam is produced from protons hitting a
target and producing $\pi$ and $K$:
\begin{eqnarray}
p \;\;\rightarrow\;\; \text{Target} \rightarrow \pi^+, K^+ \rightarrow 
 & \nu_\mu  (\% \nu_e, \bar{\nu}_{\mu}, \bar{\nu}_e) \\
   &  \nu_\mu   \rightarrow \nu_x .
\end{eqnarray}
Those of a selected charge are focused and are left to decay in a long
decay tunnel producing a neutrino beam of mostly muon neutrinos (or
antineutrinos) with a contamination of electron neutrinos of a few per
cent. The atmospheric oscillation can be established by studying, as a
function of the energy, either the disappearance of muon neutrinos or, 
if the energy of the beam is large enough, the appearance of $\tau$
neutrinos.

\begin{figure}
\centering
\includegraphics[width=.6\linewidth]{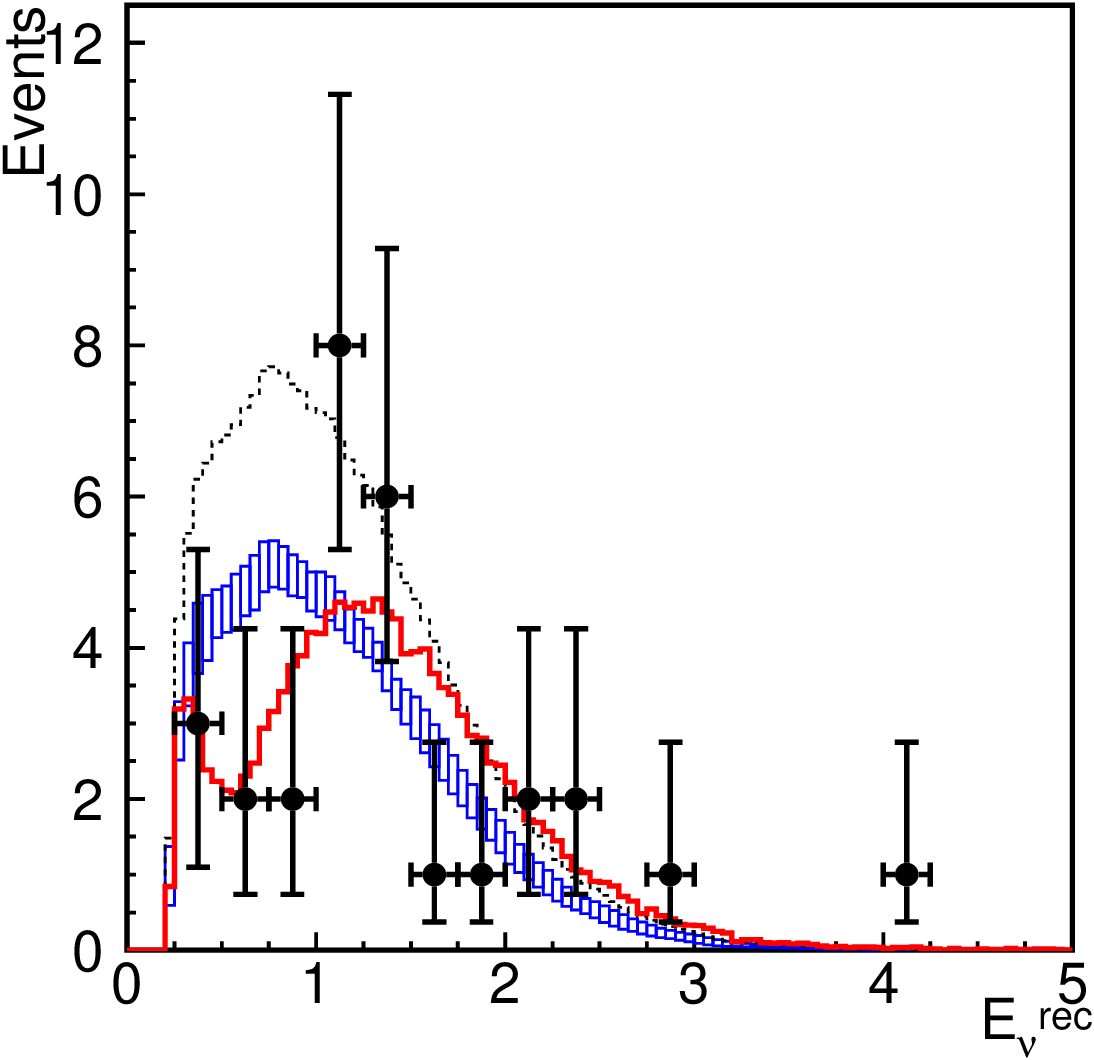}
\caption[]{Distribution of $\nu_\mu$ events in K2K as a function of
           the \emph{reconstructed} neutrino energy (from Ref.~
           \cite{k2k})}
\label{fig:k2k}
\end{figure}

\begin{figure}
\centering 
\includegraphics[width=.47\linewidth]{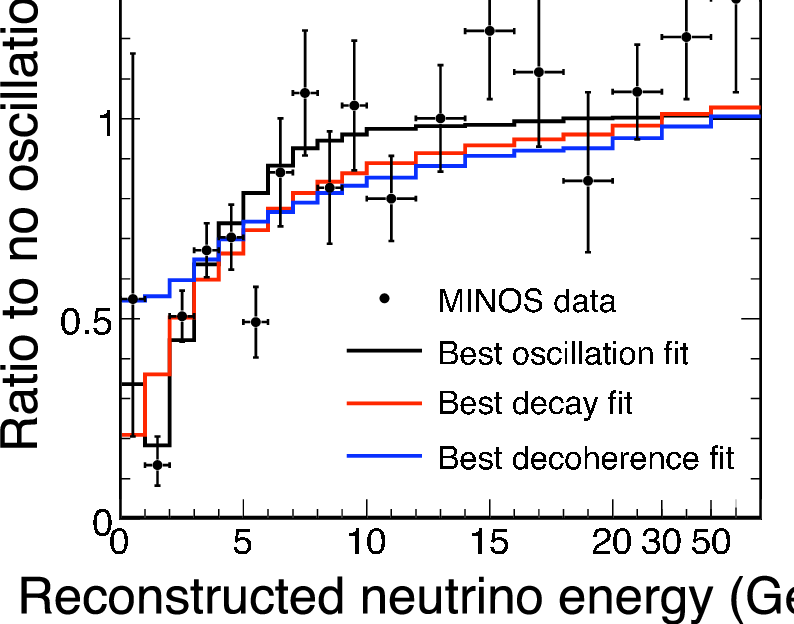}\hfill
\includegraphics[width=.47\linewidth]{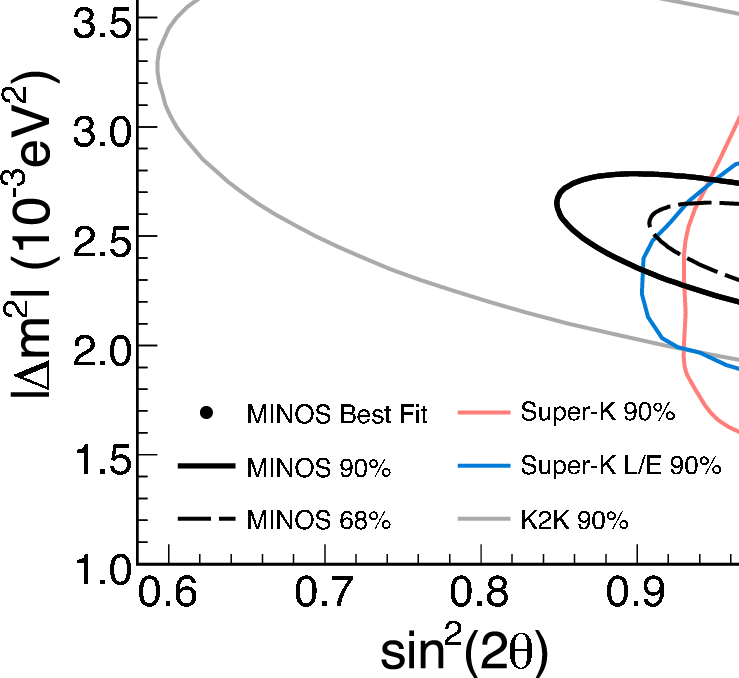}
\caption[]{Left: Ratio of measured to expected (in absence of
           oscillations) neutrino events in MINOS as a functions of
           neutrino energy. Right: Determination of oscillation
           parameters from MINOS data compared to K2K and Super-K.}
\label{fig:minos}
\end{figure}

There are three such conventional beams: KEK--Kamioka (L = 235\Ukm),
Fermilab--Soudan (L = 730\Ukm), CERN-Gran Sasso (L = 730\Ukm). The latter being
the only one sensitive to $\nu_\tau$ appearance.  The K2K experiment
at Kamioka has already presented a positive signal for $\nu_\mu$
disappearance \cite{k2k}, confirming the atmospheric oscillation. Their
result is shown in \Fref{fig:k2k}.
More recently also the MINOS experiment has presented a positive
result as shown in \Fref{fig:minos}.

\subsection{Reactor experiments in the atmospheric range}

\begin{figure}
\centering
\includegraphics[width=.47\linewidth]{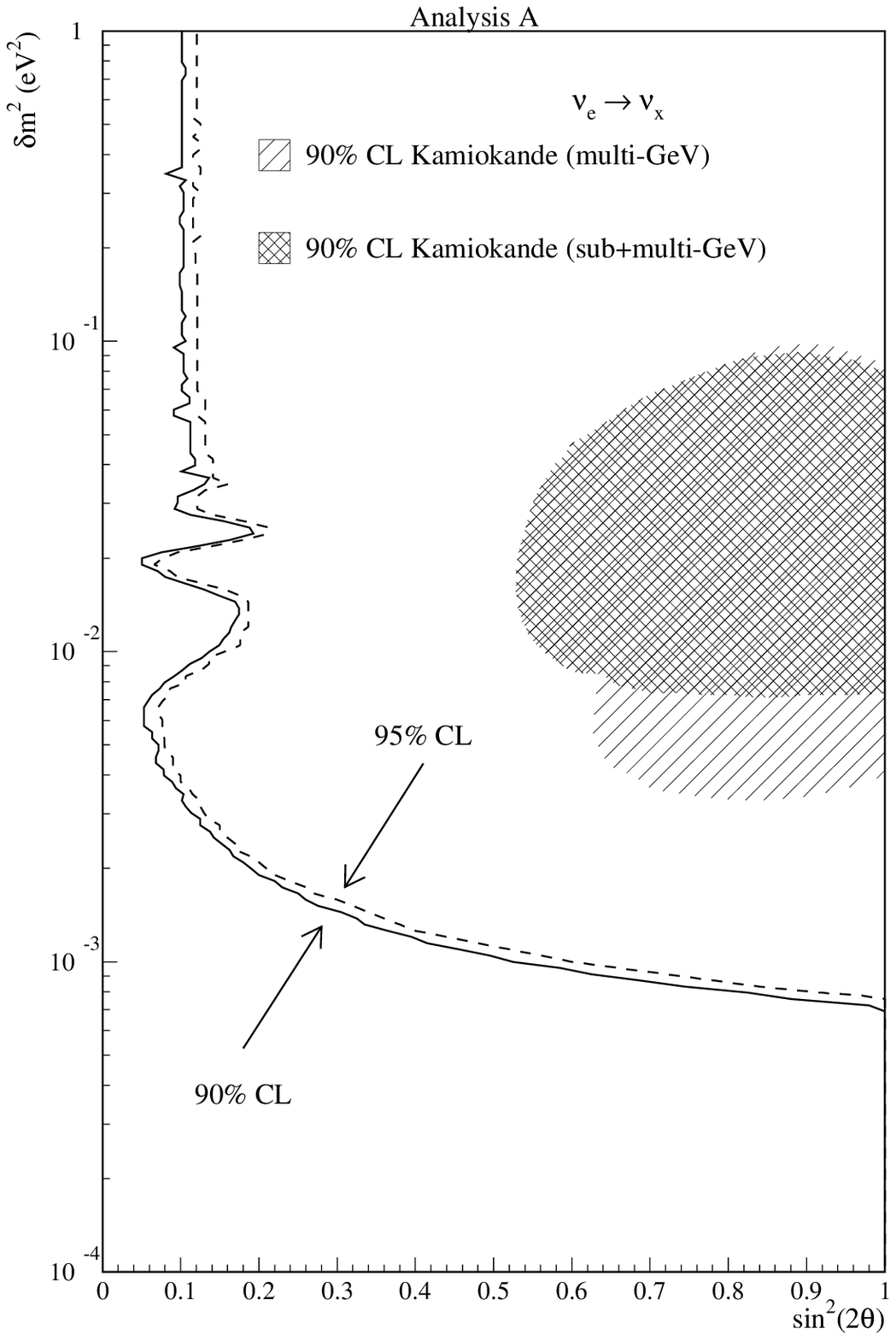}
\caption[]{Range of oscillation parameters for the oscillation
           $\bar{\nu}_e \rightarrow \bar{\nu}_x$ excluded by the Chooz
           data (from Ref.~\cite{chooz})}
\label{fig:chooz}
\end{figure}

Experiments that look for the disappearance of reactor $\bar{\nu}_e$
at an $E_\nu/L \sim \Delta m^2_\text{atmos}$ have also been performed
\cite{bugey, paloverde,chooz}. The most sensitive of these has been
Chooz \cite{chooz}. No disappearance of $\bar{\nu}_e$ was observed,
which excludes the parameter range shown in \Fref{fig:chooz}.
Although SuperKamiokande had already established that atmospheric
$\nu_e/\bar{\nu}_e$ do not seem to oscillate in the atmospheric range,
the sensitivity of SuperKamiokande to this oscillation turns out to be
much worse than that of Chooz because of the presence of electron and
muon neutrinos in the atmospheric flux. It is in the context of three-neutrino mixing that the negative signal of Chooz has been most
relevant, as we shall see.

\subsection{LSND}

\begin{figure}
\centering
\includegraphics[width=.5\linewidth]{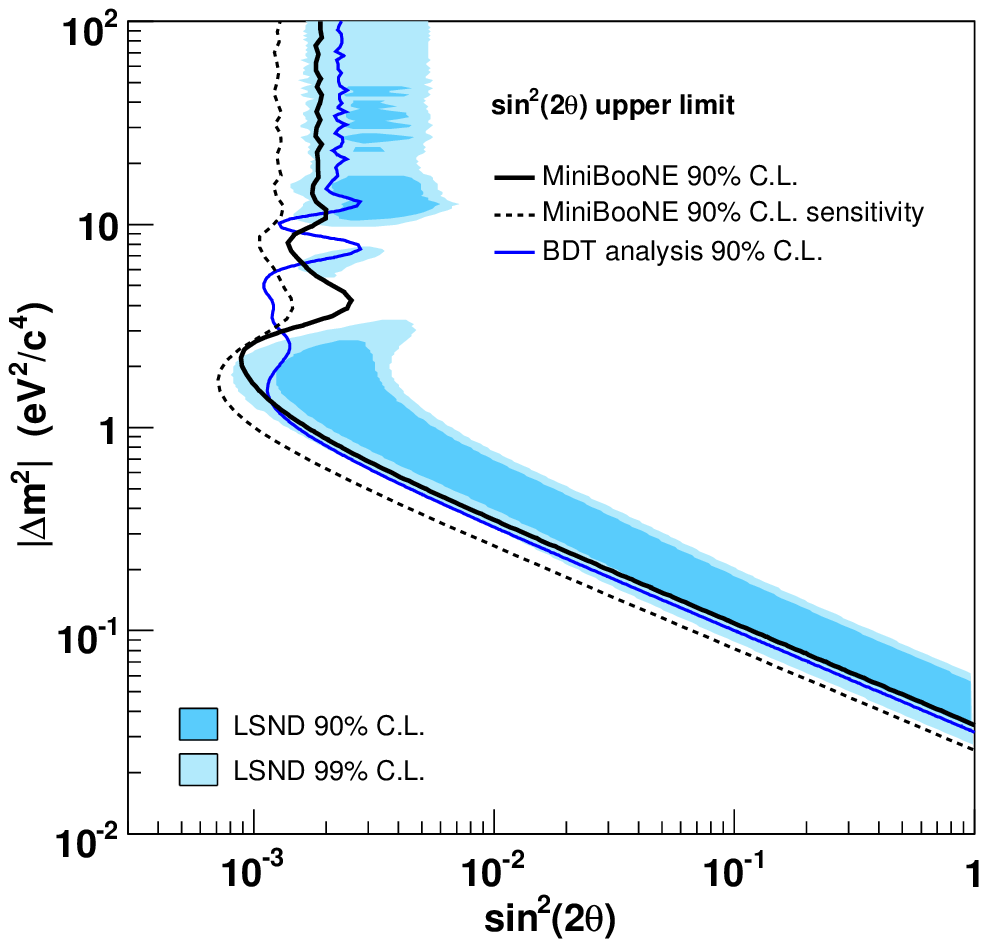}
\caption[]{Range of parameters for the oscillation $\nu_\mu
           \rightarrow \nu_e$ that could explain LSND data and those
           excluded by MiniBOONE (from Ref.~\cite{miniboone})}
\label{fig:miniboone}
\end{figure}

Finally, an accelerator experiment, LSND, has found an appearance
signal that could be interpreted in terms of neutrino flavour
transitions \cite{lsnd}.  They observed a surplus of electron events
in a muon neutrino beam from $\pi^+$ decaying in flight (DIF) and a
surplus of positron events in a neutrino beam from $\mu^+$ decaying at
rest (DAR). The interpretation of this data in terms of neutrino
oscillations gives the range shown by a coloured band in
\Fref{fig:miniboone}:

\[
\begin{array}{lll} 
\pi^+ \rightarrow
 & \mu^+  
  & \nu_\mu \\ 
 &&\nu_\mu \rightarrow \nu_e \quad\text{DIF}~(28\pm 6/10\pm2)\\
 & \mu^+  
  & \rightarrow e^+ \nu_e  \bar{\nu}_\mu\\
 && \hspace{10mm} \bar{\nu}_\mu \rightarrow\bar{\nu}_e \quad\text{DAR}~(64\pm18/12\pm3)
\end{array}
\]
Part of this region was already excluded by the experiment KARMEN
\cite{karmen} that has unsuccessfully searched for $\bar{\nu}_\mu
\rightarrow \bar{\nu}_e$ in a similar range.

In 2006 the first results from MiniBOONE were presented. This
experiment was designed to search for $\nu_\mu \rightarrow \nu_e$
transitions in the region of the LSND signal. They did not find
confirmation of LSND as shown in \Fref{fig:miniboone}

\subsection{Three-neutrino mixing}

As we have seen there is experimental evidence for neutrino
oscillation pointing to three distinct neutrino mass square
differences:
\begin{equation}
\underbrace{ |\Delta m^2_\text{Sun}| \;\;\;}_{\sim 8 \cdot 10^{-5}\UeV^2}     \ll\;\;\;\;
\underbrace{|\Delta m^2_\text{atmos}|}_{\sim 2.5\cdot 10^{-3}\UeV^2} \;\;\;\; \ll\;\;\;
\underbrace{|\Delta m^2_\text{LSND}|}_{> 0.1\UeV^2} 
\end{equation}
Clearly the mixing of the three standard neutrinos $\nu_e, \nu_\mu,
\nu_\tau$ can only explain two of the anomalies, so the explanation of
the three sets of data would require the existence of a sterile $\nu$
species, since only three light neutrinos can couple to the $Z^0$
boson.

The existence of extra light sterile neutrinos could 
accomodate a third splitting, but all such scenarios give a very poor
fit to all data.

It is now the standard scenario to consider three-neutrino mixing
dropping the LSND result.  The two independent neutrino mass square
differences are assigned to the solar and atmospheric ones:
\begin{equation}
\Delta m^2_{13} = m_3^2 - m_1^2 = \Delta m^2_\text{atmos}, \qquad  
\Delta m^2_{12} = m_2^2 - m_1^2 = \Delta m^2_\text{Sun}\SPp.
\end{equation}
With this convention, the mixing angles $\theta_{23}$ and
$\theta_{12}$ in the parametrization of \Eref{mns} correspond
approximately to the ones measured in atmospheric and solar
oscillations, respectively. This is because solar and atmospheric
anomalies approximately decouple as independent 2-by-2 mixing
phenomena thanks to the hierarchy between the two mass splittings,
$|\Delta m^2_\text{atmos}| \gg |\Delta m^2_\text{Sun}|$ , on the one
hand and the fact that the angle $\theta_{13}$, which measures the
electron component of the third mass eigenstate element $\sin
\theta_{13} = \left(V_\text{MNS}\right)_{e 3}$, is small.

To see this, let us first consider the situation in which $E_\nu/L
\sim \Delta m^2_{13}$. We can thus neglect the solar mass square
difference in front of the atmospheric one and $E_\nu/L$. The
oscillation probabilities obtained in this limit are given by
\begin{eqnarray}
P(\nu_e\to\nu_\mu)&\simeq&s_{23}^2\,\sin^2 2\theta_{13}\,
  \sin^2\!\left(\frac{\Delta m_{13}^2 L}{4 E_\nu}\right), \\
P(\nu_e\to\nu_\tau)&\simeq&c_{23}^2\,\sin^2 2\theta_{13}\,
  \sin^2\!\left(\frac{\Delta m_{13}^2 L}{4 E_\nu}\right),\,\\
P(\nu_\mu\to\nu_\tau)&\simeq&c_{13}^4\,\sin^2 2\theta_{23}\,
  \sin^2\!\left(\frac{\Delta m_{13}^2 L}{4 E_\nu}\right). 
\label{approx_atmos}
\end{eqnarray}
Only two angles enter these formulae: $\theta_{23}$ and
$\theta_{13}$. The latter is the only one that enters the
disappearance probability for $\nu_e$ in this regime:
\begin{equation}
P(\nu_e\to\nu_e)= 1 - P(\nu_e\to\nu_\mu) - P(\nu_e\to\nu_\tau) 
\simeq \sin^2 2\theta_{13}\,\sin^2\left(\frac{\Delta m_{13}^2 L}{4 E_\nu}\right)\SPp. 
\end{equation}
This is precisely the measurement of the Chooz experiment. Therefore the
result of Chooz constrains the angle $\theta_{13}$ to be unobservably
small.

If $\theta_{13}$ is set to zero in \Eref{approx_atmos}, the only
probability that survives is the $\nu_\mu \rightarrow \nu_\tau$ one,
which has the same form as a 2-family mixing formula \Eref{eq:wk} if
we identify
\begin{equation}
(\Delta m^2_\text{atmos},\theta_\text{atmos}) \rightarrow 
(\Delta m^2_{13}, \theta_{23})\SPp.
\end{equation}   

Instead if $E_\nu/L \sim \Delta m^2_{12}$, the atmospheric oscillation
its too rapid and gets averaged out. The survival probability for
electrons in this limit is given by:
\begin{equation}
P(\nu_e\to \nu_e) 
  \simeq c_{13}^4 
    \left(1-\sin^2 2\theta_{12}\,\sin^2\!\left(\frac{\Delta m_{12}^2 L}{4 E_\nu}\right)\, 
    \right) 
  + s_{13}^4.  
\end{equation}
Again it depends only on two angles, $\theta_{12}$ and $\theta_{13}$, 
and in the limit in which the latter is zero, the survival probability
measured in solar experiments has the form of two-family mixing if we
identify
\begin{eqnarray}
(\Delta m^2_\text{Sun},\theta_\text{Sun}) \rightarrow (\Delta m^2_{12}, \theta_{12})\SPp.
\end{eqnarray}
The results that we have shown of solar and atmospheric experiments
have been analysed in terms of 2-family mixing. The previous argument
indicates that when fits are done in the context of 3-family mixing
nothing changes very much, thanks to the strong constrain set by Chooz
on $\theta_{13}$.

\Fref[b]{fig:ggm} shows the result of a recent global analysis of all
data for the different parameters.
\begin{figure}
\centering
\includegraphics[width=.9\linewidth]{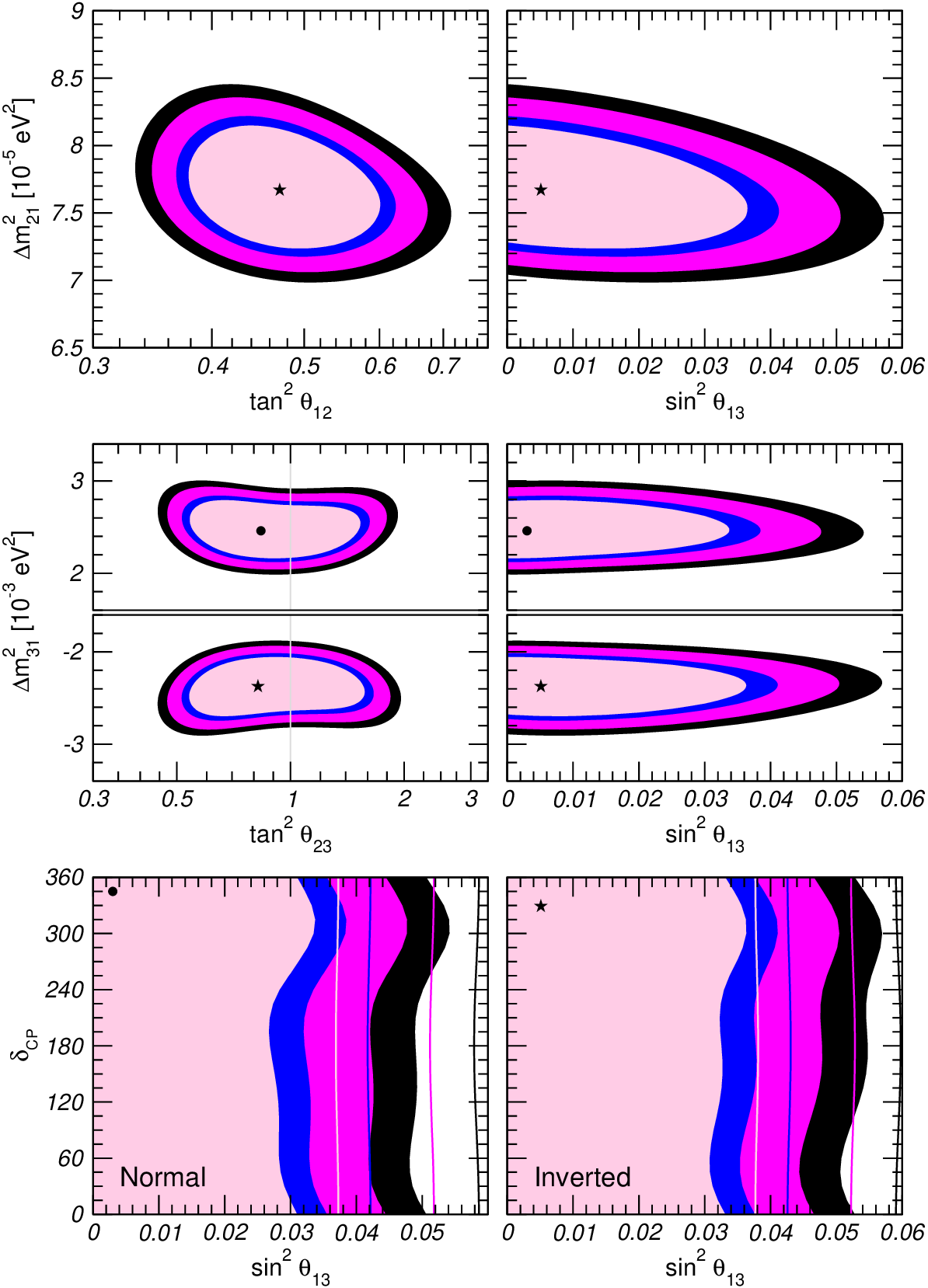}
\caption[]{Fits to the standard $3\nu$-mixing scenario including all
           available neutrino oscillation data (from Ref.~\cite{ggm})}
\label{fig:ggm}
\end{figure}
The $2\sigma$ limits are
\begin{eqnarray}
\theta_{23} = 36.9^\circ - 51.3^\circ \qquad
\theta_{12} = 32.3^\circ - 37.8^\circ \qquad
\theta_{13} < 10.3^\circ\nonumber\\
\Delta m^2_{12} = 7.66(35) \times 10^{-5}\UeV^2 \qquad
\Delta m^2_{23} = 2.38(27) \times 10^{-3}\UeV^2 \SPp.
\end{eqnarray}

\begin{figure}
\centering
\includegraphics[width=.7\linewidth]{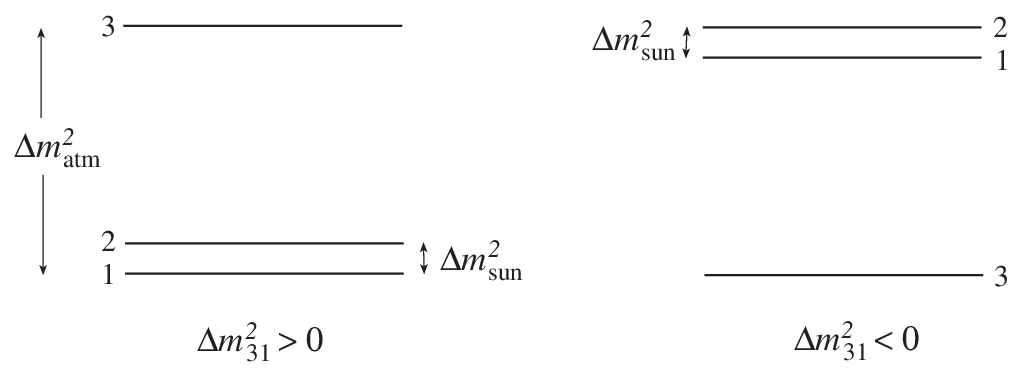}
\caption[]{Possible neutrino spectra consistent with solar and
           atmospheric data}
\label{spectrum}
\end{figure}

In summary, all the data, except LSND, can be explained if the neutrino 
spectrum has a structure as shown in \Fref{spectrum}. 
The neutrino mixing matrix is approximately given by 
\begin{equation}
|V_\text{MNS}| \simeq 
\begin{pmatrix}
        0.77--0.86 & 0.5--0.63  & 0--0.22    \\
        0.22--0.56 & 0.44--0.73 & 0.57--0.80 \\
        0.21--0.55 & 0.40--0.71 & 0.59--0.82 
\end{pmatrix}\SPp,
\end{equation}
and we do not know anything about the phases $(\delta,\alpha_1,\alpha_2)$. 
Note the striking difference between this mixing matrix and the CKM matrix which 
is approximately diagonal:
\begin{equation}
 V_\text{CKM}  \simeq 
\begin{pmatrix}
  1 & O(\lambda) & O(\lambda^3)   \\
  O(\lambda) & 1 & O(\lambda^2)   \\
  O(\lambda^3) & O(\lambda^2) & 1 
\end{pmatrix}
\quad \lambda \sim 0.2.
\end{equation}
The main features are
\begin{itemize}
\item Large mixing angles, in particular one is close to maximal.
\item There is an intriguing near tri-bimaximal mixing pattern
\[
V_\text{tri-bi} \simeq 
\begin{pmatrix}
 \sqrt{\frac{2}{3}} &  \sqrt{\frac{1}{3}} & 0\\
-\sqrt{\frac{1}{6}} &  \sqrt{\frac{1}{3}} & \sqrt{\frac{1}{2}} \\
 \sqrt{\frac{1}{6}} & -\sqrt{\frac{1}{3}} & \sqrt{\frac{1}{2}} 
\end{pmatrix} .
\]
\end{itemize}

\section{Prospects in neutrino physics}

After the next generation of neutrino experiments that are under
construction, we shall probably still be far from having complete
knowledge of the neutrino mass matrix.  There remain several
fundamental questions to be answered:
\begin{itemize}
\item[1.] Are neutrinos Dirac or Majorana particles?
\item[2.] Is total lepton number conserved or violated?
\item[3.] What is the absolute neutrino mass scale? Is it a new physics scale?
\item[4.] What is the neutrino mass spectrum: \ie $\Delta
          m^2_\text{atmos} > {\rm or} < 0$ ?
\item[5.] Is there CP violation in the lepton sector? 
\item[6.] What is the value of $\theta_{13}$?
\end{itemize}

The best hope addressing the first three questions lies in more
precise experiments searching for neutrinoless double-$\beta$ decay,
measuring the end-point of $\beta$ decay as well as cosmological
measurements.  \Fref[b]{fig:fogli} shows the present constraints on
the combination of parameters that is directly measured in
$2\beta0\nu$ experiments:
\begin{equation}
m_{\beta\beta} \equiv |m_{ee}| = 
|c_{13}^2 ( m_1 c_{12}^2 + m2 e^{i \alpha_1} s_{12}^2 ) 
 + m_3 e^{i \alpha_2} s_{13}^2|\SPp,
\end{equation}
and in cosmology:
\begin{equation}
\Sigma \equiv m_1+ m_2+ m_3 \,. 
\end{equation}
The cosmological data included in this fit is only that from the
cosmic microwave background (CMB).

Note that a lot of information on $m_{\beta\beta}$ is already provided
by neutrino oscillation experiments. If the hierarchy is inverse ($m_3
\ll m_1, m_2 \sim \sqrt{|\Delta m^2_\text{atmos}|}$), there is a lower
bound on $m_{\beta\beta} \geq 10^{-2}\UeV$, as shown by the red (I.H.)
band. Instead, if the hierarchy is normal $m_3 \sim \sqrt{|\Delta
m^2_\text{atmos}|} \gg m_1,m_2$, there is no lower bound because
neither $\theta_{13}$ nor $m_1$ is bounded from below, as shown by the
blue (N.H.) band.  The horizontal band shows the controversial claim
of a positive signal \cite{klapdor}.
\begin{figure}[ht]
\centering
\includegraphics[width=.5\linewidth]{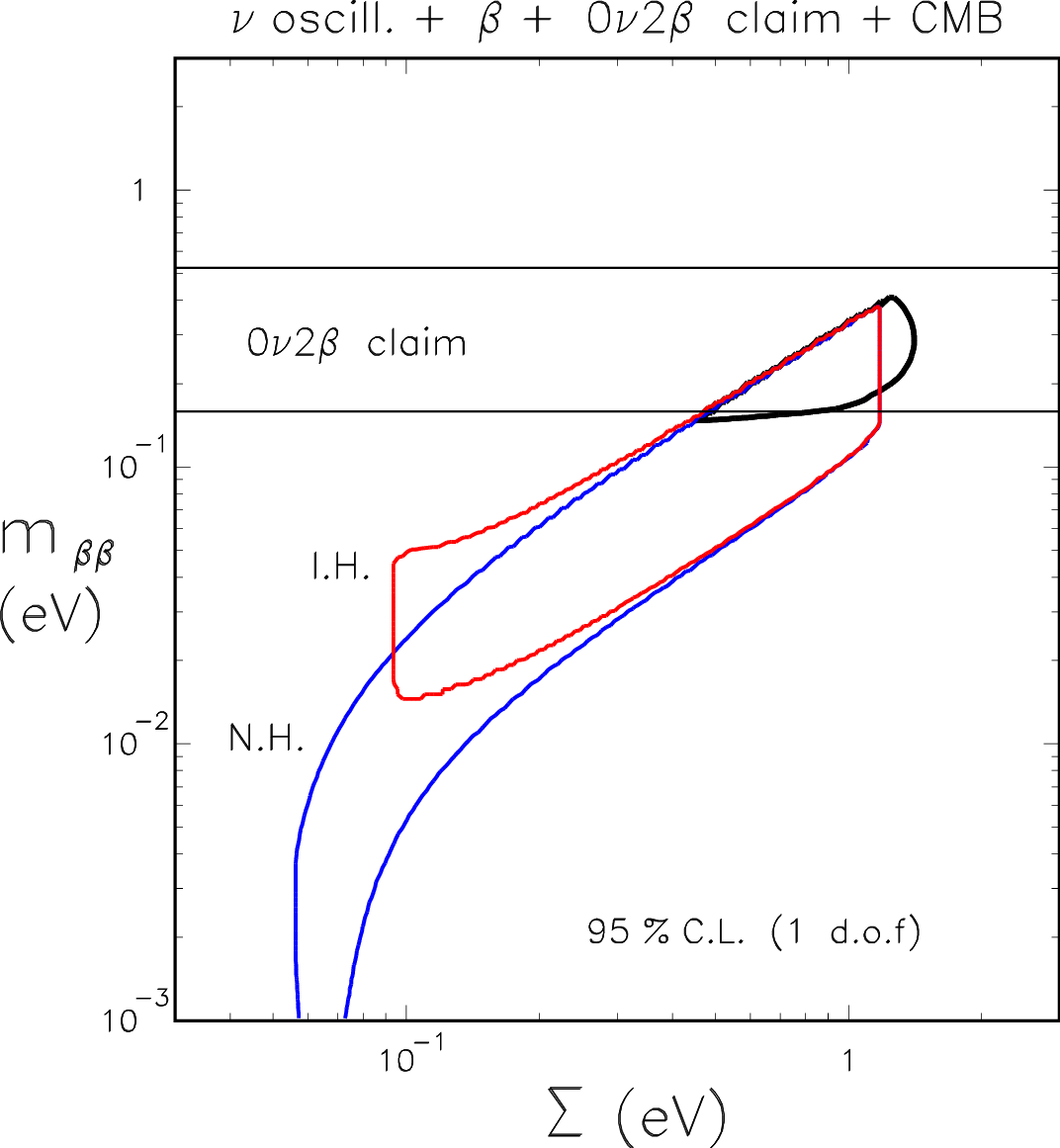}
\caption[]{Present constraints on $m_{\beta\beta}$ and $\Sigma$ from
           neutrino experiments and CMB data (from Ref.~\cite{fogli})}
\label{fig:fogli} 
\end{figure}

A plethora of forthcoming experiments that will improve these
constraints are under construction.
  
KATRIN \cite{katrin} is an experiment to measure the spectrum of 
tritium $\beta$ decay that is expected to improve the sensitivity to the
element:
\begin{equation}
m_e \equiv \sqrt{m^2_1 c_{12}^2 c_{13}^2 + m^2_2 s_{12}^2 c_{13}^2 + m^2_3 s_{13}^2}   
\end{equation}
to about $0.2\UeV$, which is an improvement of one order of magnitude
with respect to the present limit in \Eref{mnue}. Concerning
$0\nu\beta\beta$ \cite{vogel} the next step of several experiments
using different detector techniques (CUORE, EXO, GENIUS, Majorana, etc.)
is to reach the level of precision of $ m_{\beta\beta} \sim 0.1\UeV$,
which would allow testing the positive claim in a definite way.
Further in the future there are also proposals to improve this
precision by another order of magnitude reaching the $10^{-2}\UeV$
level, which could be sufficient to explore the full parameter space
in the case of the inverse hierarchy. The measurement of a non-zero
$m_{\beta\beta}$ would not only prove that neutrinos are Majorana and
that lepton number is violated, but might give the best determination
of the lightest neutrino mass, and even help in establishing the
neutrino mass hierarchy.

Concerning cosmology, it is quite impressive that the sensitivity to
the neutrino matter component of the Universe has already reached the
\UeVZ{} range. Further significant improvements are expected in the near future (\eg by
PLANCK) that can push present limits by at least
one order of magnitude.

Concerning the last three fundamental questions above, they can be
studied in more precise neutrino oscillation experiments in the
atmospheric range (\ie $\langle E_\nu \rangle/L \sim \Delta
m^2_\text{atmos}$) optimized to measure the subleading transitions
involving $\nu_e$. In particular, $\nu_e \leftrightarrow \nu_{\mu}$
and $\bar{\nu}_e \leftrightarrow \bar{\nu}_{\mu}$ are the so-called
\emph{golden} measurements \cite{golden}, while the $\nu_e
\leftrightarrow \nu_{\tau}$ and $\bar{\nu}_e \leftrightarrow
\bar{\nu}_{\tau}$, being experimentally more challenging, are the
\emph{silver} ones \cite{silver}.

\subsection{CP violation in neutrino oscillations } 

As in the quark sector, the mixing matrix of three neutrinos has CP
violating phases. The so-called Dirac phase, $\delta$, induces CP
violation in neutrino oscillations, that is a difference between
$P(\nu_\alpha \rightarrow \nu_\beta)$ and $P(\bar{\nu}_\alpha
\rightarrow \bar{\nu}_\beta)$, for $\alpha \neq\beta$. As we saw in
the general expression of \Eref{eq:prob}, CP violation is possible
if there are imaginary entries in the mixing matrix that make
Im$[W_{\alpha\beta}^{jk}\equiv \,[U_{\alpha j}U_{\beta j}^* U_{\alpha
k}^*U_{\beta k}] \neq 0$. By CPT, disappearance probabilities cannot
violate CP however, because under CPT
\begin{equation}
P(\nu_\alpha \rightarrow \nu_\beta) = 
P(\bar{\nu}_\beta \rightarrow \bar{\nu}_\alpha)\SPp,
\end{equation}
so in order to observe a CP or T-odd asymmetry the initial and final
flavour must be different, $\alpha\neq\beta$:
\begin{equation}
A_{\alpha \beta}^{CP}\equiv\frac{P(\nu_\alpha\rightarrow \nu_\beta)-
P(\bar\nu_\alpha\rightarrow \bar\nu_\beta)}{P(\nu_\alpha\rightarrow
 \nu_\beta)+
P(\bar\nu_\alpha\rightarrow \bar\nu_\beta)} , \quad
 A_{\alpha \beta}^{T}\equiv\frac{P(\nu_\alpha\rightarrow \nu_\beta)-
P(\nu_\beta\rightarrow \nu_\alpha)}{P(\nu_\alpha\rightarrow \nu_\beta)+
P(\nu_\beta\rightarrow \nu_\alpha)}\SPp. 
\end{equation}
In the case of 3-family mixing it is easy to see that the CP(T)-odd
terms in the numerator are the same for all transitions $\alpha\neq
\beta$:
\begin{equation}
A^\text{CP(T)-odd}_{\nu_\alpha \nu_\beta} = 
\frac{\sin\delta c_{13}\sin
      2\theta_{13} \overbrace{\sin 2\theta_{12} \frac{\Delta m^2_{12} L}{4
      E_\nu}}^\text{solar} \overbrace{\;\sin 2\theta_{23}\sin^2\frac{\Delta
      m^2_{13} L}{4 E_\nu}}^\text{atmos}}
     {P^\text{CP-even}_{\nu_\alpha\nu_\beta}} .\;\;\;
\end{equation}
As expected, the numerator is GIM suppressed in all the $\Delta
m_{ij}^2$ and all the angles, because if any of them is zero, the
CP-odd phase becomes unphysical.

In order to maximize this asymmetry, it is necessary to perform
experiments in the atmospheric range $\langle E_\nu\rangle/L \sim
\Delta m^2_\text{atmos}$, so that the GIM suppression is minimized. In
this case, only two small parameters remain in the CP-odd terms: the
solar splitting, $\Delta m^2_\text{Sun}$ (\ie small compared to the other
scales, $\Delta m^2_\text{atmos}$ and $\langle E_\nu \rangle/L$), and
the angle $\theta_{13}$. The asymmetry is then larger in the
subleading transitions: $\nu_e\rightarrow \nu_\mu(\nu_\tau)$, because
the CP-even terms in the denominator are also suppressed by the same
small parameters.  Indeed a convenient approximation for the
$\nu_e\leftrightarrow\nu_\mu$ transitions is obtained expanding to
second order in both small parameters \cite{golden}:
\begin{eqnarray}
P_{\nu_ e\nu_\mu ( \bar \nu_e \bar \nu_\mu ) } & = & 
  s_{23}^2 \,\sin^2 2 \theta_{13} \, 
    \sin^2\!\left (\frac{\Delta m^2_{13} \, L}{4E_\nu} \right ) \quad 
\equiv P^\text{atmos} \nonumber\\ 
&+ &
  c_{23}^2 \, \sin^2 2 \theta_{12} \, 
    \sin^2\!\left(\frac{\Delta m^2_{12} \, L}{4 E_\nu} \right ) \quad
\equiv P^\text{solar} \nonumber \\ 
&+ & 
  \tilde J \cos \left ( \pm \delta - 
    \frac{\Delta m^2_{13} \, L}{4 E_\nu} \right ) \; 
  \frac{\Delta m^2_{12} \, L}{4 E_\nu} 
    \sin\left(\frac{\Delta m^2_{13} \, L}{4 E_\nu} \right )  \quad
\equiv P^\text{inter},
\end{eqnarray}
where ${\tilde J} \equiv c_{13}\; \sin 2 \theta_{13} \; \sin 2
\theta_{12}\; \sin 2 \theta_{23}$. This approximate formula is
obtained as an expansion to second order in the parameters
$\theta_{13}$ and $\Delta m^2_\text{Sun}$. The first term corresponds
to the atmospheric oscillation, the second one is the solar one and
there is an interference term which has the information on the phase
$\delta$. Depending on the value of $\theta_{13}$, it is possible that
the atmospheric term dominates over the other two, in such a way that
the CP-even terms are suppressed in $\theta_{13}^2$, or if it is the
solar term that dominates, the suppression is in $(\Delta
m^2_\text{Sun})^2$. The asymmetries in these two regimes show
therefore the following dependence on the small parameters:
\begin{eqnarray}
P^\text{atmos} \gg P^\text{solar} &\rightarrow & 
  A^{CP,T}_{{\nu}_e {\nu}_\mu({\nu}_\tau)}  
    \sim  \frac{\Delta m^2_{12} L/E_\nu}{\sin 2 \theta_{13}}, \nonumber\\
P^\text{solar} \gg P^\text{atmos}  &\rightarrow & 
  A^{CP,T}_{{\nu}_e {\nu}_\mu({\nu}_\tau)} 
    \sim  \frac{\sin 2 \theta_{13}}{\Delta m^2_{12} L/E_\nu}, \nonumber\\
P^\text{solar} \simeq  P^\text{atmos} &\rightarrow & 
  A^{CP,T}_{{\nu}_e {\nu}_\mu({\nu}_\tau)} = O(1)\SPp.
\end{eqnarray}
Therefore asymmetries in the subleading transitions are expected to be
rather large, specially when the solar and atmospheric terms are
comparable.

\begin{figure}
\centering
\includegraphics[width=.47\linewidth]{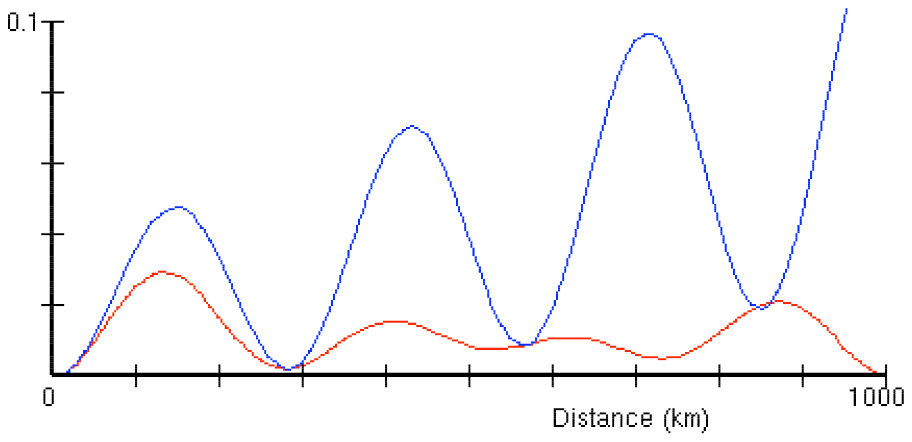}\hfill
\includegraphics[width=.47\linewidth]{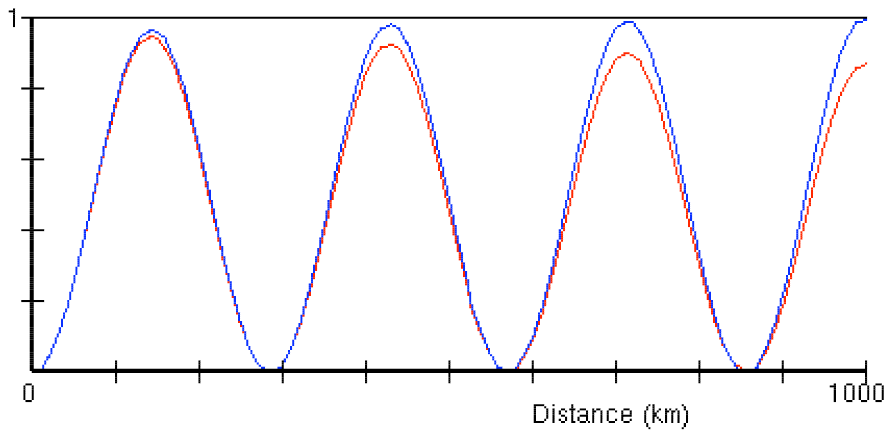}
\caption[]{Comparison of the
           $\nu_e\leftrightarrow\nu_\mu$/$\bar{\nu}_e\leftrightarrow\bar{\nu}_\mu$
           (left) and
           $\nu_\mu\leftrightarrow\nu_\tau$/$\bar{\nu}_\mu\leftrightarrow\bar{\nu}_\tau$
           (right) oscillation probabilities for $E_\nu = 500\UMeV$,
           $\theta_{13} = 8^\circ$ and $\delta=90^\circ$ as a function
           of the distance}
\label{fig:leadsub}
\end{figure}

In contrast, the asymmetries in the leading $\nu_\mu \rightarrow
\nu_\tau$ transition in the atmospheric range are much smaller,
because the CP-even terms are unsuppressed in each of the two small
parameters. The difference between the neutrino and antineutrino
oscillation probabilities for the leading and subleading channels are
shown in \Fref{fig:leadsub}.

\subsection{The neutrino spectrum}

\begin{figure}[t]
\centering
\includegraphics[width=.6\linewidth]{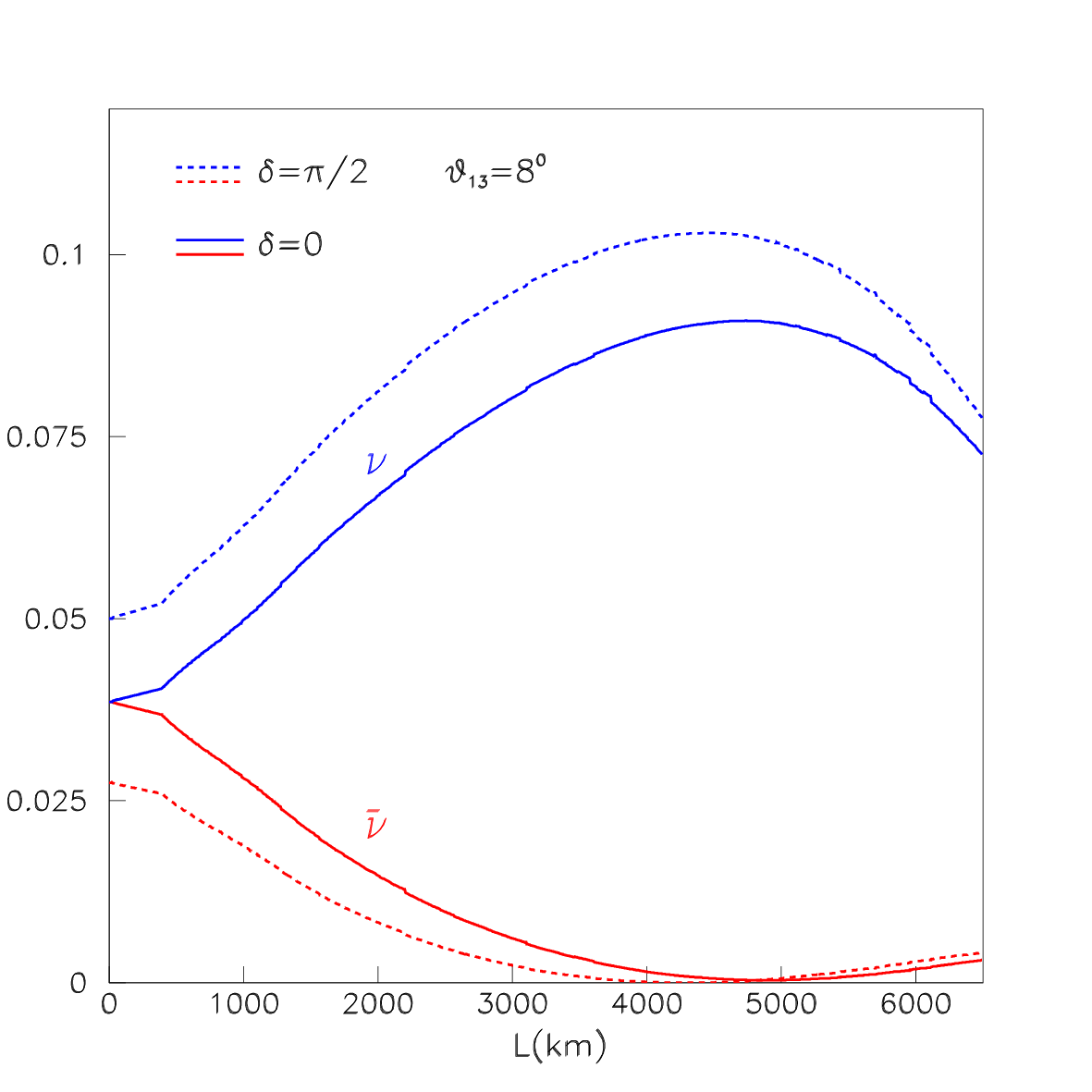}
\caption[]{$P(\nu_e\rightarrow\nu_\mu)$ and
           $P(\bar\nu_e\rightarrow\bar\nu_\mu)$ as a function of the
           baseline $L$ in kilometres, at a neutrino energy $E_\nu/L = |\Delta
           m^2_{13}|/2\pi$ and for $\theta_{13} =8^\circ$ and
           $\delta=0$ (solid) and $90^\circ$ (dashed)}
\label{fig:matter}
\end{figure}

The oscillation probabilities in matter can also be approximated by an
expansion to second order in the two small parameters: $\theta_{13}$
and $\Delta m^2_{12}$ \cite{golden}. The result has the same structure
as in vacuum:
\begin{eqnarray}
P_{\nu_e\nu_\mu({\bar \nu}_e {\bar \nu}_\mu)}  = 
  s^2_{23}\, 
    \sin^2 2 \theta_{13}\left(\frac{\Delta_{13}}{B_\pm}\right)^2 \; \; 
    \sin^2\!\left( \frac{B_{\pm} L}{2} \right) \nonumber \\
  +  c_{23}^2 \, 
    \sin^2 2 \theta_{12} \, \left(\frac{\Delta_{12}}{A}\right)^2 
    \sin^2\!\left(\frac{A \, L}{2} \right ) \nonumber \\
  + {\tilde J} \;\frac{\Delta_{12}}{A}\;  
    \sin(\frac{A L}{2}) \;\frac{\Delta_{13}}{B_\pm}\; 
    \sin\!\left(\frac{B_\pm L}{2}\right) \; 
    \cos\!\left( \pm \delta - \frac{\Delta_{13}  \, L}{2} \right )\SPp, 
\end{eqnarray}
where 
\begin{equation}
B_{\pm} = | A \pm \Delta_{13} | \qquad
\Delta_{ij} = \frac{\Delta m^2_{ij}}{2 E_\nu} \qquad
A= \sqrt{2} G_F N_e\SPp. 
\end{equation}
This formula shows a resonant enhancement of the atmospheric term in
the the neutrino or antineutrino oscillation probability (depending on
the sign of $\Delta m^2_{13}$) channel when
\begin{equation}
2 E_\nu A \sim |\Delta m^2_{13}|\SPp.
\end{equation}
Considering the electron number density in the Earth, the resonant
energy is $E_\nu \sim 10--20\UGeV$.  This resonance is illustrated in
\Fref{fig:matter}, which shows the $\nu_e\rightarrow \nu_\mu$
oscillation probability for neutrinos and antineutrinos, as a
function of the baseline, for neutrino energy constrained to the first
atmospheric peak, \ie  $E_\nu/L = |\Delta m^2_{13}|/2\pi$. The
difference between the neutrino and anti-neutrino oscillation
probabilities induced by matter effects becomes comparable to that due
to maximal CP-violation for $L=\mathcal{O}(1000)\Ukm$.  This is
approximately the baseline where matter effects and CP violation can
both be measured simultaneously. At much longer distances, matter
effects completely hide CP-violation effects and vice versa.

\subsection{ The measurement of $\theta_{13}$ and $\delta$ } 

\subsubsection{Theoretical challenge}

In the future, we shall face the challenge of extracting simultaneously
$\theta_{13}$, $\delta$ and also the hierarchy from the measurement of
the oscillation probabilities $\nu_\mu \leftrightarrow \nu_e$ and
$\bar{\nu}_\mu \leftrightarrow \bar{\nu}_e$. This turns out to be
non-trivial even in principle, because of the existence of
degeneracies \cite{burguet1}.  In fact, at fixed $E_\nu, L$ there are
generically two solutions for $(\theta_{13},\delta)$ that give the
same probabilities for neutrinos and antineutrinos.

\begin{figure}
\centering
\includegraphics[width=.6\linewidth]{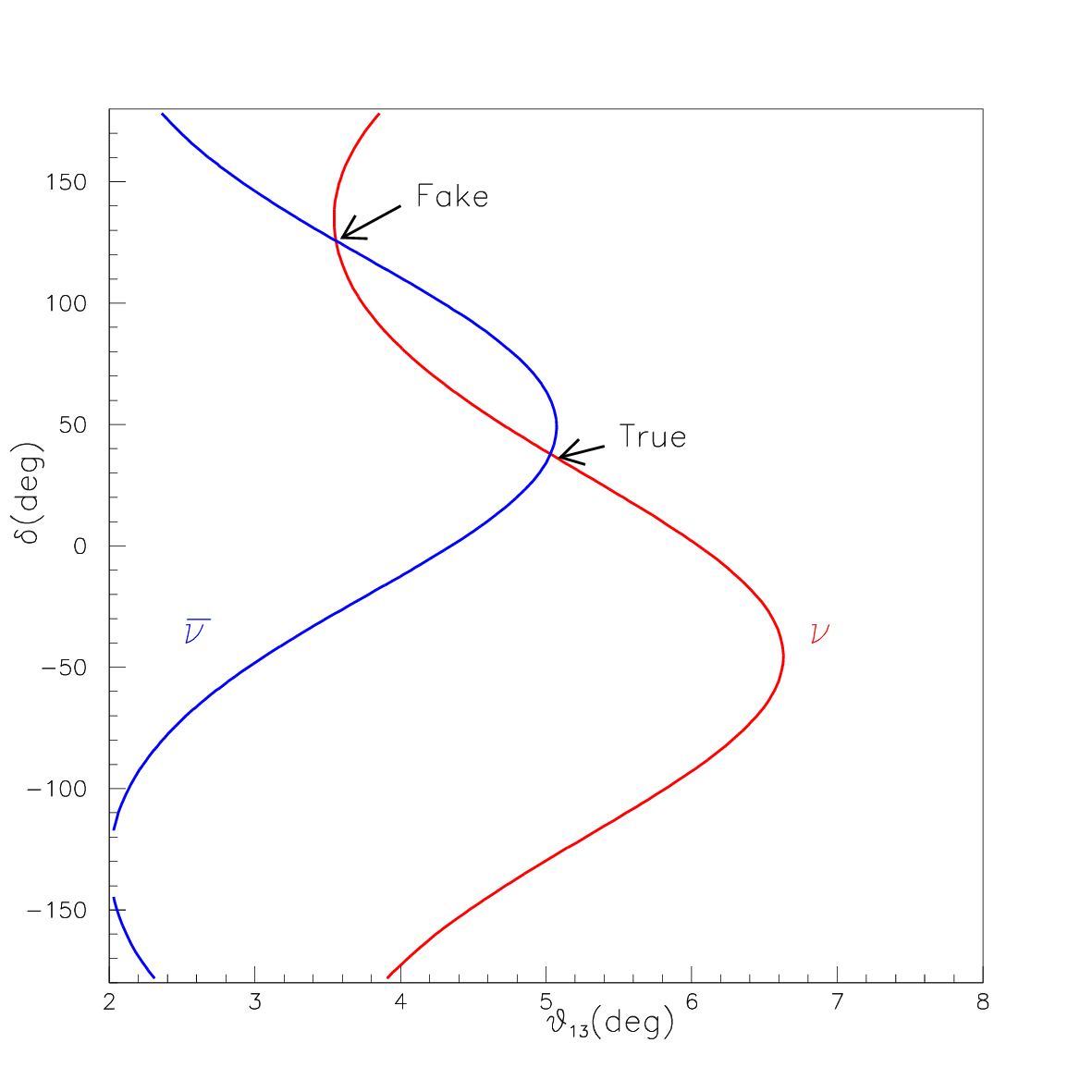}
\caption[]{Equiprobability curves $P_{\nu_e \nu_\mu} (E_\nu/L,
           \theta_{13}, \delta)$ = Meas$_1$ and $P_{\bar{\nu_e}
           \bar{\nu_\mu}} (E_\nu/L, \theta_{13}, \delta)$ = Meas$_2$
           on the plane $(\theta_{13},\delta)$. They generically cross
           at two points: the true solution $(\theta_{13}, \delta)$
           and a fake one.}
\label{fig:degen}
\end{figure}

This is due to the periodicity in $\delta$: if the equiprobability curves 
for neutrinos and antineutrinos on the plane $(\theta_{13}, \delta)$ cross 
at one point (at the true solution), they must cross at least once more
as shown in \Fref{fig:degen}. 

The fake solution has a strong dependence on the ratio $E_\nu/L$ in vacuum. 

Normally neutrino beams are not monochromatic, so $E_\nu/L$ is not
fixed.  If we consider as the measurement the integrated signals
(after integrating in energy the probability $\times$ flux $\times$
cross section), the same argument holds and a fake solution appears
generically although it has a more complicated dependence on $\langle
E_\nu \rangle$ and $L$.

Besides, the fact that other oscillation parameters will also not be
known at the time of this measurement, such as the sign($\Delta
m^2_{13}$) or sign($\cos \theta_{23}$), increases the difficulty
further: these unknowns will also bias the extraction of $\theta_{13}$
and $\delta$ leading to additional fake solutions, the so-called
eight-fold degeneracy \cite{8fold}.
 
Several strategies for resolving these degeneracies have been proposed.
Given the energy dependence of the fake solutions, it is very useful
to have a detector with good neutrino energy
resolution. \Fref[b]{fig:degen_spec} shows the oscillation probability as
a function of the neutrino energy for some values of $(\theta_{13},
\delta)$ with that corresponding to the fake solution
$(\theta_{13}^\text{fake}(\langle E_\nu \rangle/L), \delta^\text{fake}(\langle
E_\nu \rangle/L))$. The curves cross at $\langle E_\nu \rangle$ but
differ quite significantly at other energies.
\begin{figure}
\centering
\includegraphics[width=.6\linewidth]{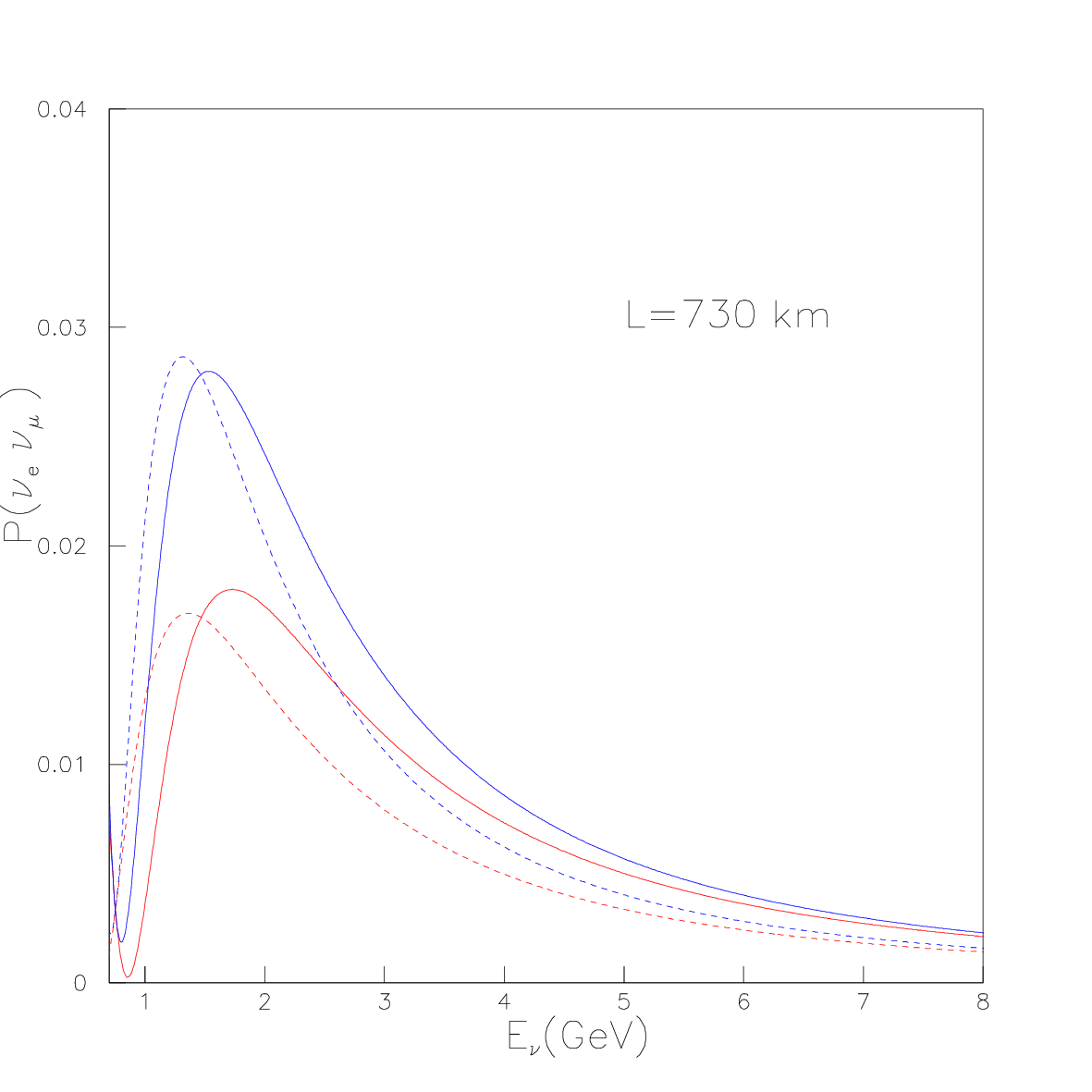}
\caption[]{Oscillation probability for neutrinos and antineutrinos as
           a function of the energy, for some true values of
           $\theta_{13}$ and $\delta$, and for the fake solutions
           (dashed curves)}
\label{fig:degen_spec}
\end{figure}

Another possibility is to consider performing several experiments with
differing $\langle E_\nu\rangle/L$ or with different matter effects.

Finally, the measurement of other oscillation probabilities beside the
golden one can help. For example, if a precise measurement of the
disappearance probability for $\nu_e$ is done in the atmospheric
range, with an improved Chooz-type experiment, this could provide a
measurement of $\theta_{13}$ that does not depend on $\delta$ at
all \cite{reactor_deg}.
  
Similarly, if we combine the golden measurement with the silver one:
$\nu_e \rightarrow \nu_\tau$ and $\bar{\nu}_e \rightarrow
\bar{\nu}_\tau$, the fake solutions can be excluded \cite{silver}.
 
\subsubsection{Experimental challenge}

\begin{figure}
\centering
\includegraphics[width=.6\linewidth]{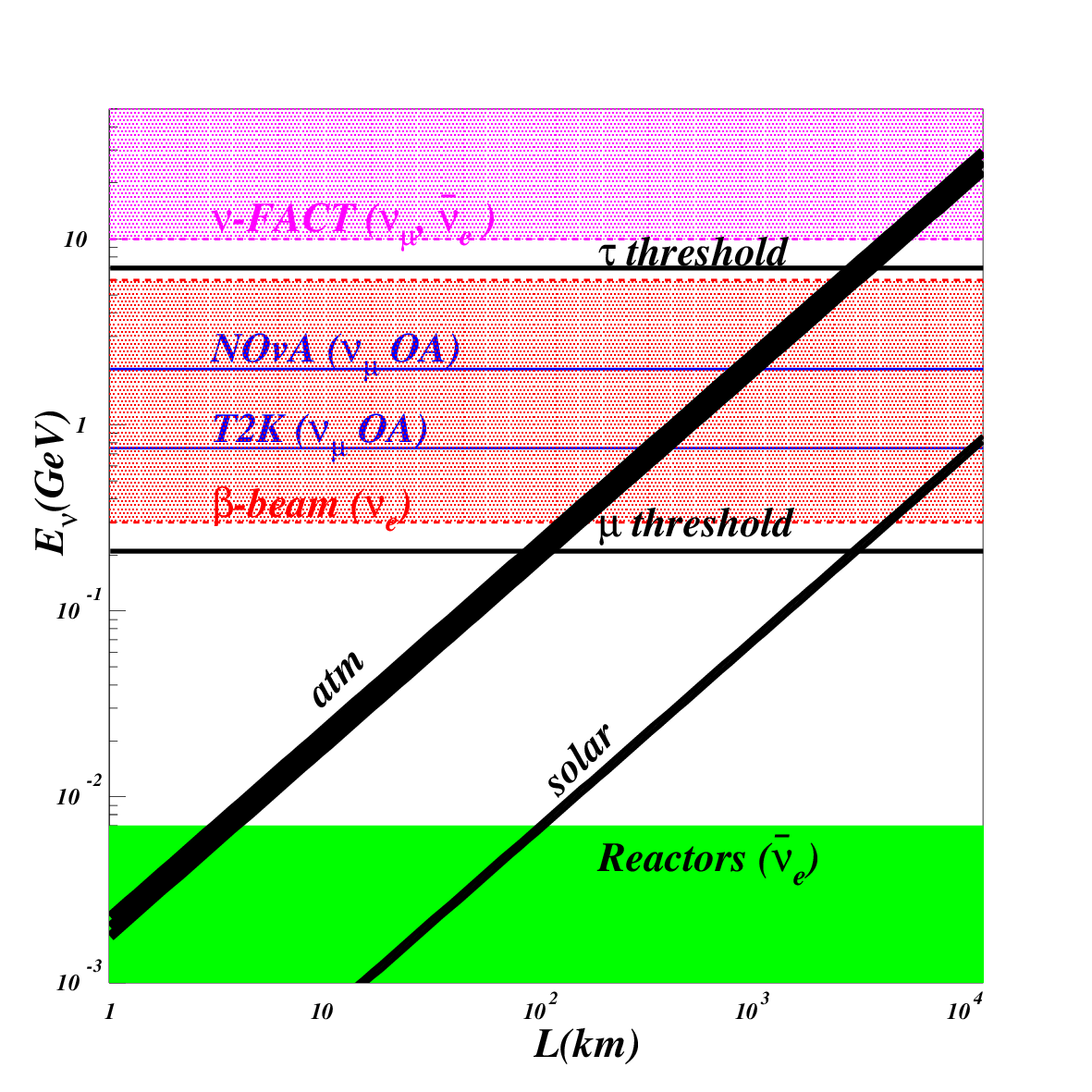}
\caption[]{Energy of the proposed future neutrino oscillation
           experiments: Nufact, $\beta$-beam, superbeams (T2K and
           NOvA) and reactors. The \emph{atm} and \emph{solar} black
           bands correspond to the first atmospheric and solar
           oscillation peaks, respectively.}
\label{fig:future} 
\end{figure}

The challenge is to measure for the first time the \emph{small}
subleading transitions $\nu_e\leftrightarrow\nu_\mu$ and
$\bar{\nu}_e\leftrightarrow \bar{\nu}_\mu$ with $\langle E_\nu
\rangle/L \sim |\Delta m^2_\text{atmos}|$. The need to be above the
muon threshold implies that rather long baselines are required as
shown in \Fref{fig:future}.
There are many ideas being pursued. Let us briefly describe the
different proposals.

\subsubsection{Future reactor experiments}

Reactor neutrinos have an energy in the range of MeV and therefore can
only look at the disappearance channel $\bar{\nu}_e \rightarrow
\bar{\nu}_e$.  It has been pointed out before that reactor neutrinos
have provided the most stringent limit on the angle $\theta_{13}$. A
future upgrade of this type of experiments is possible, by increasing
the detector size and reducing the systematics by intercepting the
beam with both a near and a far detector.  The experiment Double-Chooz
is under construction and expects to reach a sensitivity limit of
$\sin^2 2 \theta_{13} \geq 0.03$, with the advantage that being a
disappearance measurement, there is no ambiguity due to the CP phase
$\delta$ or any other parameter.

\subsubsection{Future superbeam experiments}
 
Neutrino beams produced at accelerators have already been constructed
to measure the disappearance of $\nu_\mu$ in the atmospheric range
(K2K and MINOS), as well as the apperance channel $\nu_\mu \rightarrow
\nu_\tau$ (OPERA).  As we have seen, these experiments have confirmed
the leading atmospheric oscillation, but they will improve the
sensitivity to the unknowns very little.

These \emph{conventional} beams result from the decay of pions and
kaons produced from an intense proton beam that hits a target. They
are thus mostly $\nu_\mu$ (or $\bar{\nu}_\mu$ depending on the
polarity) with a per cent contamination of $\nu_e$. Neutrino beams of
this type but with much higher intensity, the so-called
\emph{superbeams}, could be obtained with new megawatt proton sources,
however, the sensitivity to the subleading transition $\nu_\mu
\rightarrow \nu_e$ is limited by systematics. Not only can the flavour and
spectral composition of these beams not be determined with good
accuracy, but the irreducible background of $\nu_e$ is  the
limiting factor. One way to reduce this background is to use an
off-axis configuration. Pion decay kinematics implies that a detector
located off-axis intercepts a beam with a much better defined energy,
and this allows  the beam background to be reduced below the $1\%$ level.

Two projects using off-axis superbeams are being pursued. The first
one is T2K in Japan \cite{jhf}, that is expected to start taking data
in 2009. It will use the SuperKamiokande detector to intercept a beam
produced in J-PARC, which corresponds to a baseline of 295~km.  If
$\sin^2 2 \theta_{13} \geq 0.01--0.02$, an appearance of $\nu_e$ will be
observed, although the experiment will have no sensitivity to CP
violation nor to the mass hierarchy.  The second project is NOvA in the
USA \cite{numi}. The NUMI beam at Fermilab will be intercepted off-axis
by a new detector located $810\Ukm$ away. It is expected to reach a
similar sensitivity to $\theta_{13}$ as T2K, but if $\sin^2 2
\theta_{13} \geq 0.05$, the comparison of the $\nu$ and $\bar{\nu}$
appearance signals could provide the first determination of the
neutrino hierarchy.

\subsubsection{Neutrino factory and $\beta$ beams}

\begin{figure}
\centering
\includegraphics[width=.6\linewidth]{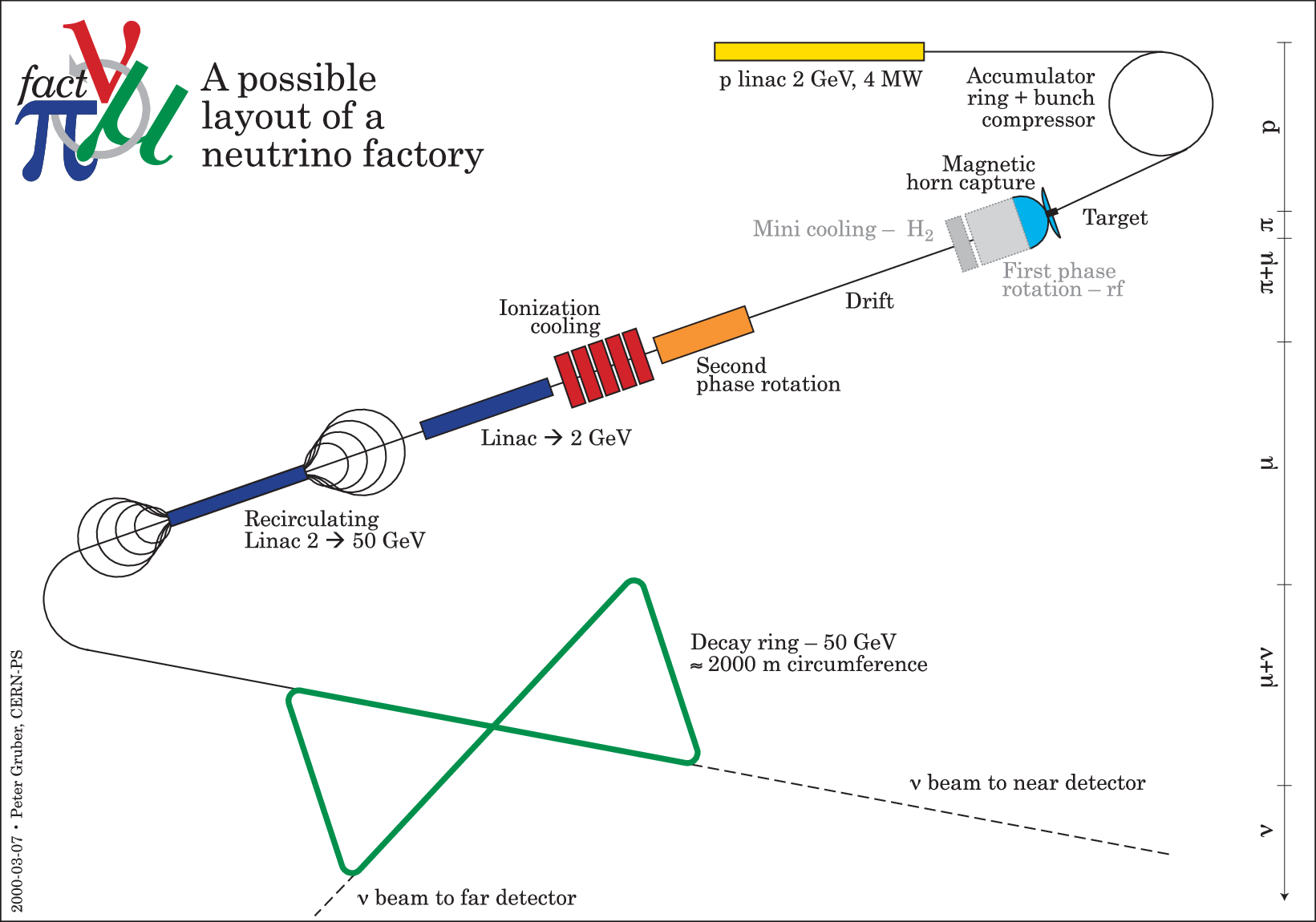}
\caption[]{Possible layout of a CERN-based Neutrino Factory complex}
\label{fig:nufact}
\end{figure}

The measurement of leptonic CP violation will probably require a
further step.  New ideas to obtain neutrino beams with reduced
systematics have been actively discussed in recent years. At the
\emph{Neutrino Factory} (NF) \cite{nufact} neutrinos are produced from
$\mu^+$ or $\mu^-$ which are accelerated to some reference energy and
are allowed to decay in a storage ring with long straight sections (see
\Fref{fig:nufact}).
 Subleading transitions can be searched for by looking
for wrong-sign muons in a massive magnetized detector:
\begin{eqnarray}
\mu^- \rightarrow e^-\,  
& \nu_\mu &  \, \bar{\nu}_e\, ;  \nonumber\\
& \;      & \bar{\nu}_e  \rightarrow \bar{\nu}_\mu \rightarrow \mu^+ \;\; \nonumber\\
\;\;
& \nu_\mu &  \rightarrow \nu_\mu\rightarrow \mu^- .
\end{eqnarray}

A similar situation is found in the case of the \emph{$\beta$ beam}
(BB)\cite{zucchelli}. This is a neutrino beam obtained from boosted
radioactive ions, such as $\Isotope[18][10]{Ne}$ or
$\Isotope[6]{He}^{++}$, which are accelerated and circulated in a
storage ring where they decay, producing a pure $\nu_e$ or
$\bar{\nu}_e$ beam, respectively (see \Fref{fig:bb}):
\begin{figure}
\centering
\includegraphics[width=.6\linewidth]{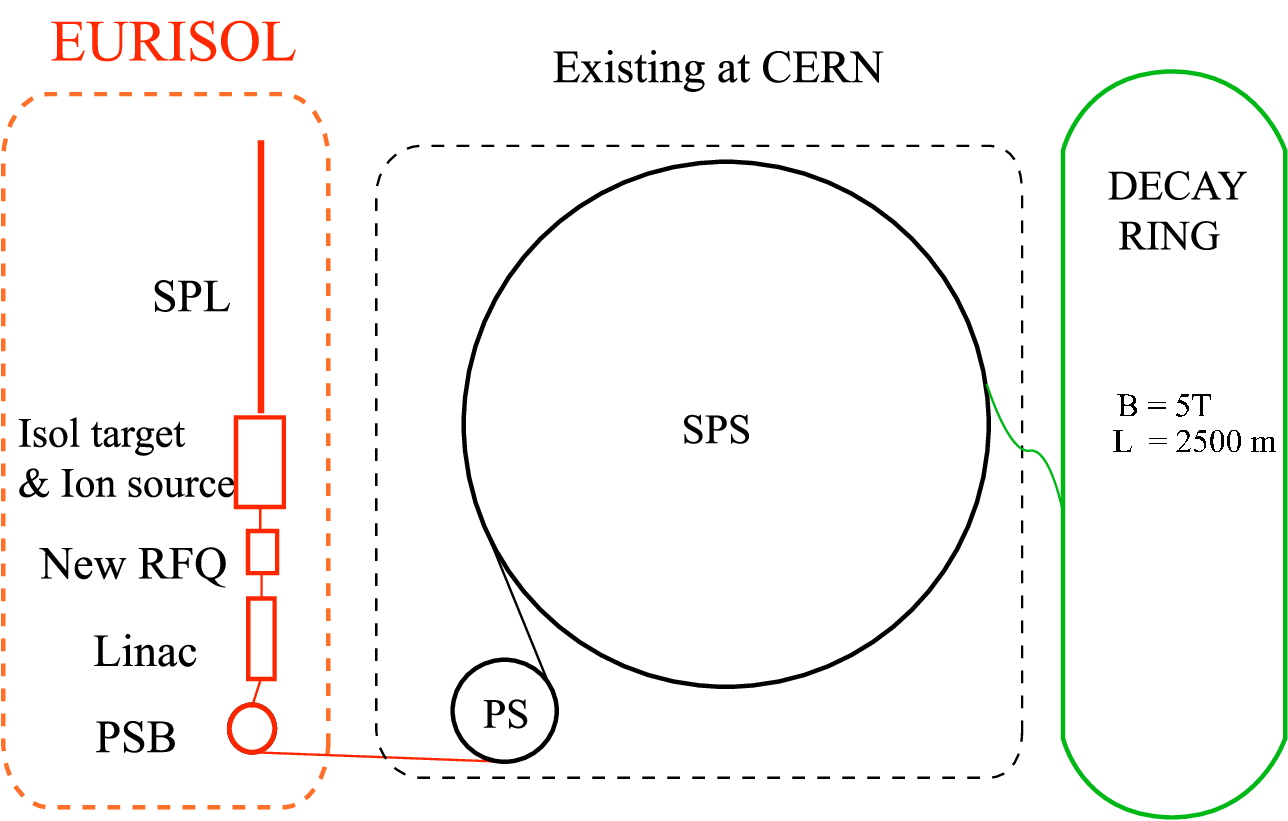}
\caption[]{Possible layout of a CERN-based $\beta$ beam}
\label{fig:bb}
\end{figure} 
\begin{eqnarray}
\Isotope[6]{He}^{++} \rightarrow \Isotope[6][3]{Li}^{+++} \; e^{-} 
 &\bar{\nu}_e \nonumber\\
 &\qquad\bar{\nu}_e \rightarrow \bar{\nu}_\mu \rightarrow \mu^+ \\
\Isotope[18][10]{Ne} \rightarrow \Isotope[18][9]{F}^{-} \; e^{+} 
 &\nu_e \nonumber \\
 &\qquad\nu_e \rightarrow \nu_\mu \rightarrow \mu^-. \nonumber
\end{eqnarray}
The golden transition can be searched for in this case by counting
muons. It is not necessary to measure their charge, so the detector
does not need to be magnetized.

The neutrino fluxes $\nu_e$ and $\bar{\nu}_e$ at the NF or BB can be
known with a very good accuracy, since they are easily obtained from
the number of muons or ions decaying in the storage ring and the
well-known muon or ion decay kinematics:
\begin{eqnarray}
\left.\frac{d\Phi^\text{NF}}{dS dy}\right|_{\theta\simeq 0}
\simeq \frac{N_\mu}{\pi L^2} {12 \gamma^2} y^2 (1-y), 
\end{eqnarray}
with $y=\frac{E_\nu}{E_\mu}$ and
\begin{eqnarray}
\left.\frac{d\Phi^\text{BB}}{dS dy}\right|_{\theta\simeq 0} \simeq 
\frac{N_\beta}{\pi L^2} \frac{\gamma^2}{g(y_e)} y^2 (1-y) \sqrt{(1-y)^2 - y_e^2} ,
\end{eqnarray}
and $y=\frac{E_\nu}{2 \gamma E_0}, y_e=m_e/E_0, g(y_e)\equiv
\frac{1}{60} \left\{ \sqrt{1-y_e^2} (2-9 y_e^2 - 8 y_e^4) + 15 y_e^4
\log\left[\frac{y_e}{1-\sqrt{1-y_e^2}}\right]\right\}$.  $N_\mu$ and
$N_\beta$ are the muons or ions decaying per year. Note that both
fluxes increase with the $\gamma$ factor of the parent particle as
$\gamma^2$.

\begin{figure}
\centering
\includegraphics[width=.47\linewidth]{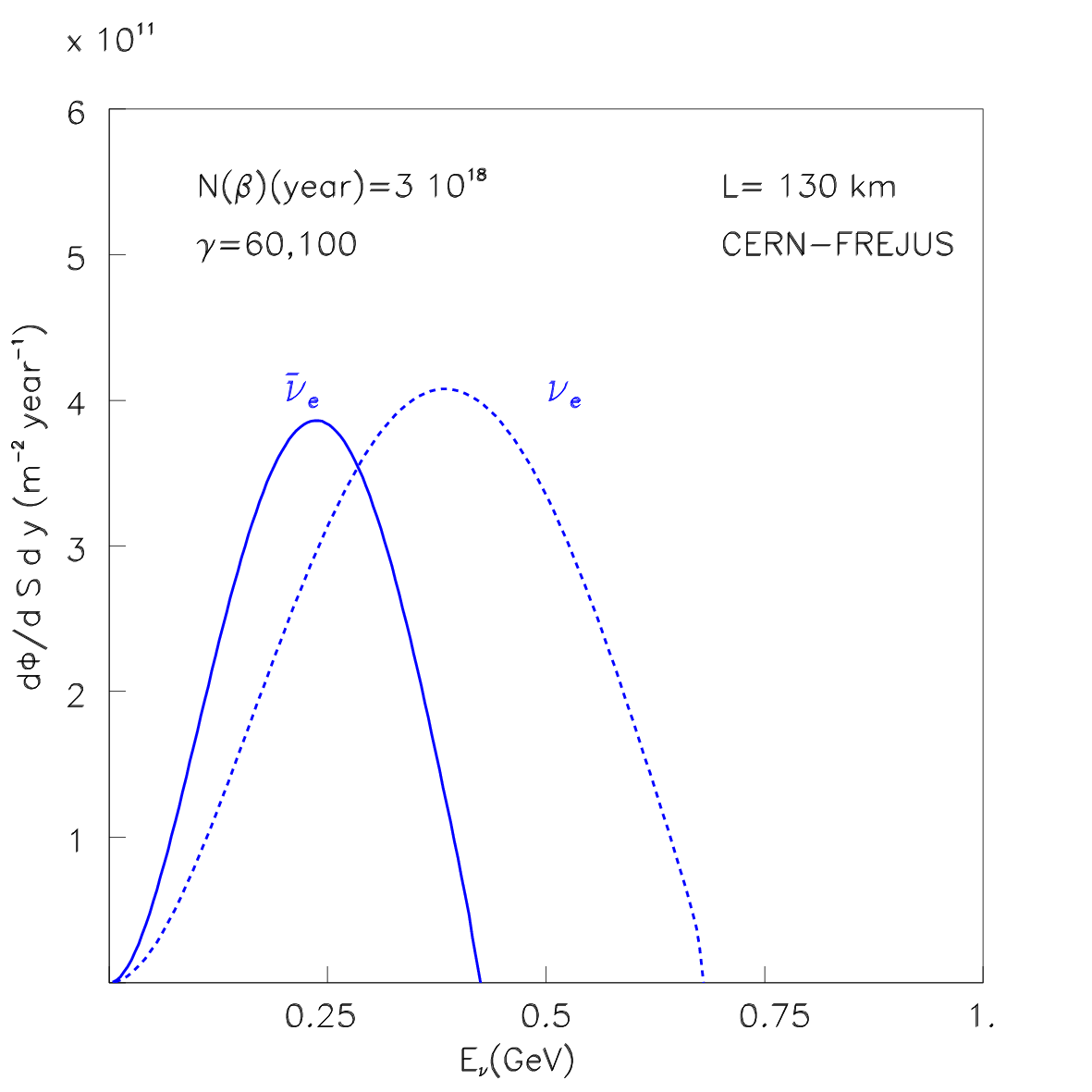}\hfill
\includegraphics[width=.47\linewidth]{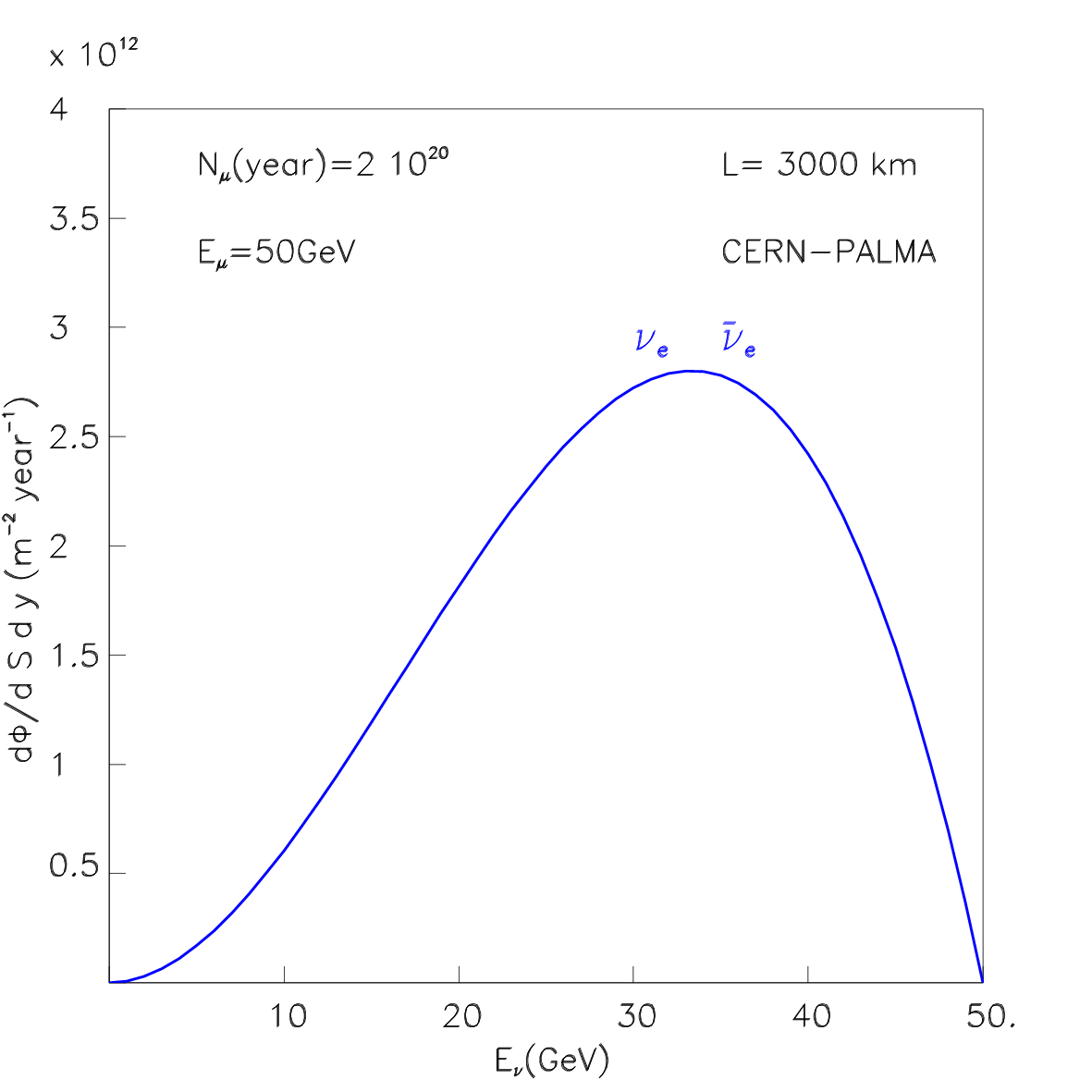}
\caption[]{Left: $\nu_e$ and $\bar{\nu}_e$ fluxes in the BB from
           $10^{18}~\Isotope[18]{Ne}/3\times 10^{18}~\Isotope[6]{He}$
           ion decays per year at $\gamma=100/60$ and
           $L=130\Ukm$. Right: $\nu_e$ and $\bar{\nu}_e$ fluxes at the
           NF from $2\times 10^{20}~50\UGeV \mu^-/\mu^+$ decays and
           $L=3000\Ukm$.}
\label{fig:fluxes} 
\end{figure}

These fluxes are shown in \Fref{fig:fluxes} for two standard setups
for the NF and the BB. Although the fluxes at the neutrino factory are
larger by at least one order of magnitude, the need to magnetize the
detector in the NF is a big limitation to how massive it can be in
practice. In the case of the $\beta$ beam no magnetization is needed,
which opens the possibility to use very massive water Cherenkov
detectors, like those that have been proposed to improve the limits on
proton decay and to study supernova neutrinos \cite{uno}.

In both the Neutrino Factory and the $\beta$-beam designs, the energy
of the parent muon or ion (which is proportional to the average neutrino
energy) can be optimized within a rather large range, since this is
fixed by the acceleration scheme that is part of the machine
design. Once the energy is fixed, the baseline is also fixed by the
atmospheric oscillation length. This optimization is, however, a complex
problem because there are often contradicting requirements in the
maximization of the intensity, the minimization of backgrounds, having
useful spectral information, measuring the \emph{silver} channel
in addition to the \emph{golden} one, having sizeable matter effects, etc. This
optimization was done for the NF some years ago and a muon energy of a
few tens of GeV and a baseline of a few thousand kilometres is
considered a reference setup \cite{golden}. For the BB, a scenario
with a neutrino beam of a few GeV and distances of a few hundred
kilometers is close to optimal.

\Fref[b]{fig:iss} shows a comparison of the physics reach for CP
violation and the neutrino hierarchy of the NF and BB complexes with
other second-generation superbeams that have also been proposed as
alternatives (SPL, T2HK, WBB). Even though this is probably not yet
the end of the story as regards optimization/comparison, these plots
show that reaching the realm of $\sin^2 2 \theta_{13} \sim 10^{-4}$
will be possible in the future, both for leptonic CP violation and the
neutrino hierarchy.

\begin{figure}
\centering
\includegraphics[width=.47\linewidth]{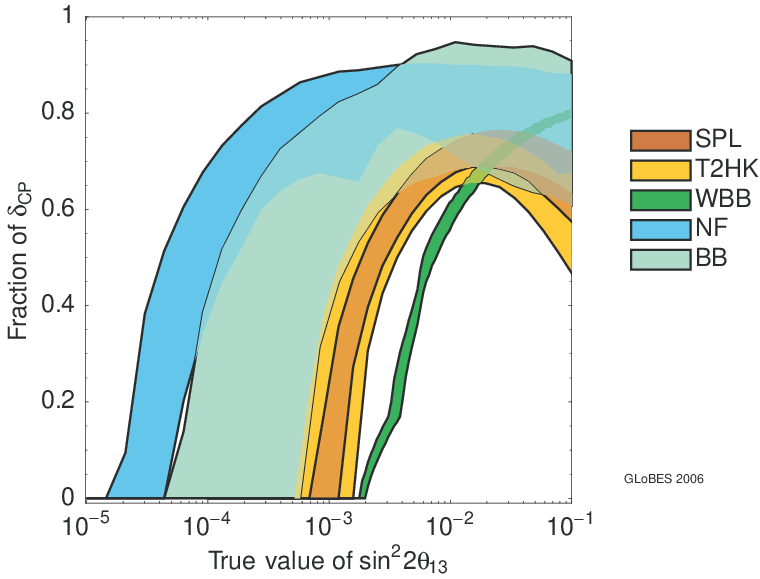}
\includegraphics[width=.47\linewidth]{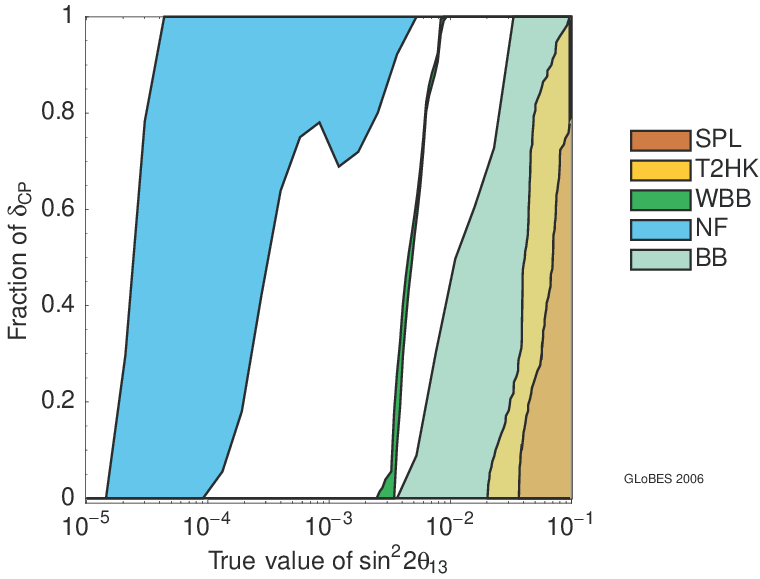}
\caption[]{Left: Sensitivity limit to leptonic CP violation in the
           plane $(\sin^2 2 \theta_{13}, \delta)$ of superbeams (SPL,
           T2KHK), the wide band beam (WBB), Neutrino Factory (NF) and
           $\beta$ beams (BB). The bands correspond to most/least
           conservative assumptions concerning the
           facility/detectors. Right: Sensitivity limits to the
           neutrino mass hierarchy in the same facilities.  Taken from Ref.~
           \cite{iss}.}
\label{fig:iss} 
\end{figure}

\section{Leptogenesis}
\label{sec:baryo}

The Universe is made of matter. The matter--antimatter asymmetry is
measured to be
\begin{equation}
\eta_B \equiv \frac{N_b - N_{\bar{b}}}{N_\gamma} \sim 6.15(25) \times 10^{-10}\SPp.
\end{equation}
It has been known for a long time that all the ingredients to generate dynamically 
such an asymmetry from a symmetric initial state are present in the 
laws of particle physics. These ingredients were first put forward by Sakharov: 

\emph{Baryon number violation}

\begin{figure}
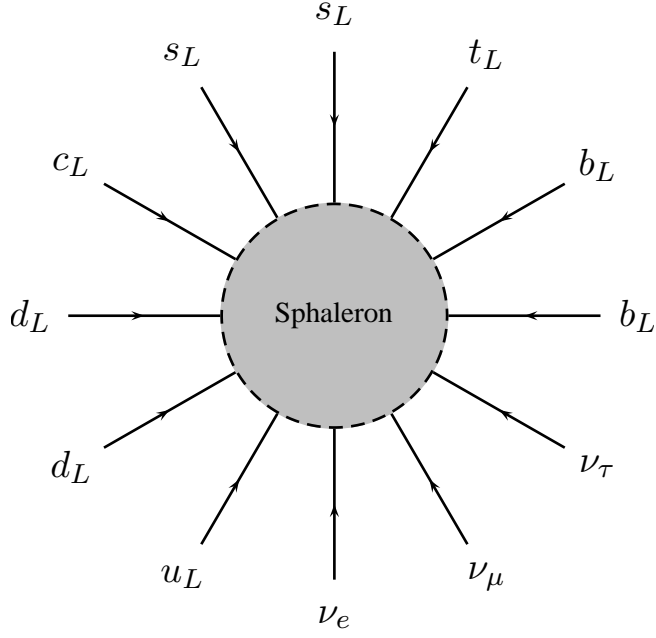

\centering
\pspicture*(-0.25,-2.2)(8.2,6.2)
     \psset{linecolor=lightgray}
     \qdisk(4,2){1.5cm}
     \psset{linecolor=black}
     \pscircle[linewidth=1pt,linestyle=dashed](4,2){1.5cm}
     \rput[cc]{0}(4,2){\scalebox{1.}{Sphaleron}}
     \psline[linewidth=1pt](5.50,2.00)(7.50,2.00)
     \psline[linewidth=1pt](5.30,2.75)(7.03,3.75)
     \psline[linewidth=1pt](4.75,3.30)(5.75,5.03)
     \psline[linewidth=1pt](4.00,3.50)(4.00,5.50)
     \psline[linewidth=1pt](3.25,3.30)(2.25,5.03)
     \psline[linewidth=1pt](2.70,2.75)(0.97,3.75)
     \psline[linewidth=1pt](2.50,2.00)(0.50,2.00)
     \psline[linewidth=1pt](2.70,1.25)(0.97,0.25)
     \psline[linewidth=1pt](3.25,0.70)(2.25,-1.03)
     \psline[linewidth=1pt](4.00,0.50)(4.00,-1.50)
     \psline[linewidth=1pt](4.75,0.70)(5.75,-1.03)
     \psline[linewidth=1pt](5.30,1.25)(7.03,0.25)
     \psline[linewidth=1pt]{<-}(6.50,2.00)(6.60,2.00)
     \psline[linewidth=1pt]{<-}(6.17,3.25)(6.25,3.30)
     \psline[linewidth=1pt]{<-}(5.25,4.17)(5.30,4.25)
     \psline[linewidth=1pt]{<-}(4.00,4.50)(4.00,4.60)
     \psline[linewidth=1pt]{<-}(2.75,4.17)(2.70,4.25)
     \psline[linewidth=1pt]{<-}(1.83,3.25)(1.75,3.30)
     \psline[linewidth=1pt]{<-}(1.50,2.00)(1.40,2.00)
     \psline[linewidth=1pt]{<-}(1.83,0.75)(1.75,0.70)
     \psline[linewidth=1pt]{<-}(2.75,-0.17)(2.70,-0.25)
     \psline[linewidth=1pt]{<-}(4.00,-0.50)(4.00,-0.60)
     \psline[linewidth=1pt]{<-}(5.25,-0.17)(5.30,-0.25)
     \psline[linewidth=1pt]{<-}(6.17,0.75)(6.25,0.70)
     \rput[cc]{0}(8.00,2.00){\scalebox{1.3}{$b_L$}}
     \rput[cc]{0}(7.46,4.00){\scalebox{1.3}{$b_L$}}
     \rput[cc]{0}(6.00,5.46){\scalebox{1.3}{$t_L$}}
     \rput[cc]{0}(4.00,6.00){\scalebox{1.3}{$s_L$}}
     \rput[cc]{0}(2.00,5.46){\scalebox{1.3}{$s_L$}}
     \rput[cc]{0}(0.54,4.00){\scalebox{1.3}{$c_L$}}
     \rput[cc]{0}(0.00,2.00){\scalebox{1.3}{$d_L$}}
     \rput[cc]{0}(0.54,0.00){\scalebox{1.3}{$d_L$}}
     \rput[cc]{0}(2.00,-1.46){\scalebox{1.3}{$u_L$}}
     \rput[cc]{0}(4.00,-2.00){\scalebox{1.3}{$\nu_e$}}
     \rput[cc]{0}(6.00,-1.46){\scalebox{1.3}{$\nu_{\mu}$}}
     \rput[cc]{0}(7.46,0.00){\scalebox{1.3}{$\nu_{\tau}$}}
\endpspicture
\caption[]{Artistic view of a sphaleron}
\label{fig:spha}
\end{figure}

$B+L$ is anomalous in the SM \cite{thooft} both with and without
massive neutrinos, while $B-L$ is preserved if the light neutrinos are
Dirac particles.  At high T in the early Universe, $B+L$ violating
transitions could be in thermal equilibrium \cite{krs} due to the
thermal excitation of configurations with topological charge called
sphalerons, see \Fref{fig:spha}.

These processes violate baryon and lepton numbers by the same amount:
\begin{eqnarray}
\Delta B = \Delta L. 
\end{eqnarray}
If there are heavy Majorana singlets, as in the see-saw models, there
is an additional source of $L$ violation (and $B-L$). If a lepton
charge is generated at temperatures where the sphalerons are still in
thermal equilibrium, a baryon charge can be generated.

\emph{Deviation from thermal equilibrium}

Sphalerons are in equilibrium for $T \geq 100\UGeV$ \cite{blr}, which
means that in order to get these processes out of equilibrium it is
necessary to go to the electroweak phase transition.

Electroweak baryogenesis which has been extensively studied both in the SM and
in the most popular extensions like the MSSM, is currently disfavoured
in the SM because the out-of-equilibrium condition is not well met:
the electroweak phase transition is not strongly first order.

A different out-of-equilibrium condition is met in the $L$ violation
processes associated to the heavy Majorana singlets
\cite{lepto0}. These singlets are in equilibrium until they decouple
at a temperature similar to their masses. Since their masses must be
significantly larger than the electroweak scale if we are to explain
the smallness of neutrino masses, sphalerons are still in equilibrium
when the heavy Majorana singlets decouple. Therefore if a lepton
number is generated in their decay, inducing a lepton number abundance
$Y_L$, the equilibrium of sphaleron processes implies that a baryon
abundance will also be present \cite{ht}:
\begin{equation}
Y_B = a Y_{B-L} = \frac{a}{a-1} Y_L \quad a = \frac{28}{79}\quad \text{in \;SM}\SPp.
\end{equation}

\emph{$C$ and $CP$ violation}

\begin{figure}
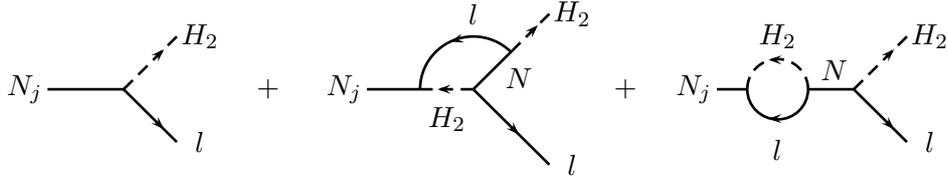

\centering
\pspicture(0,0)(3.7,2.6)
\psline[linewidth=1pt](0.6,1.3)(1.6,1.3)
\psline[linewidth=1pt](1.6,1.3)(2.3,0.6)
\psline[linewidth=1pt,linestyle=dashed](1.6,1.3)(2.3,2.0)
\psline[linewidth=1pt]{->}(2.03,0.87)(2.13,0.77)
\psline[linewidth=1pt]{->}(2.03,1.73)(2.13,1.83)
\rput[cc]{0}(0.3,1.3){$N_j$}
\rput[cc]{0}(2.6,0.6){$l$}
\rput[cc]{0}(2.6,2.0){$H_2$}
\rput[cc]{0}(3.5,1.3){$+$}
\endpspicture
\pspicture(-0.5,0)(4.2,2.6)
\psline[linewidth=1pt](0.6,1.3)(1.3,1.3)
\psline[linewidth=1pt,linestyle=dashed](1.3,1.3)(2.0,1.3)
\psline[linewidth=1pt](2,1.3)(2.5,1.8)
\psline[linewidth=1pt,linestyle=dashed](2.5,1.8)(3,2.3)
\psline[linewidth=1pt](2,1.3)(3,0.3)
\psarc[linewidth=1pt](2,1.3){0.7}{45}{180}
\psline[linewidth=1pt]{<-}(1.53,1.3)(1.63,1.3)
\psline[linewidth=1pt]{<-}(1.7,1.93)(1.8,1.96)
\psline[linewidth=1pt]{->}(2.75,2.05)(2.85,2.15)
\psline[linewidth=1pt]{->}(2.5,0.8)(2.6,0.7)
\rput[cc]{0}(0.3,1.3){$N_j$}
\rput[cc]{0}(1.65,0.9){$H_2$}
\rput[cc]{0}(2,2.3){$l$}
\rput[cc]{0}(2.6,1.45){$N$}
\rput[cc]{0}(3.3,2.3){$H_2$}
\rput[cc]{0}(3.3,0.3){$l$}
\rput[cc]{0}(4.0,1.3){$+$}
\endpspicture
\pspicture(-0.5,0)(3.5,2.6)
\psline[linewidth=1pt](0.5,1.3)(0.9,1.3)
\psline[linewidth=1pt](1.7,1.3)(2.3,1.3)
\psarc[linewidth=1pt](1.3,1.3){0.4}{-180}{0}
\psarc[linewidth=1pt,linestyle=dashed](1.3,1.3){0.4}{0}{180}
\psline[linewidth=1pt]{<-}(1.18,1.69)(1.28,1.69)
\psline[linewidth=1pt]{<-}(1.18,0.91)(1.28,0.91)
\psline[linewidth=1pt](2.3,1.3)(3.0,0.6)
\psline[linewidth=1pt,linestyle=dashed](2.3,1.3)(3.0,2.0)
\psline[linewidth=1pt]{->}(2.73,0.87)(2.83,0.77)
\psline[linewidth=1pt]{->}(2.73,1.73)(2.83,1.83)
\rput[cc]{0}(1.3,0.5){$l$}
\rput[cc]{0}(1.3,2){$H_2$}
\rput[cc]{0}(2.05,1.55){$N$}
\rput[cc]{0}(0.2,1.3){$N_j$}
\rput[cc]{0}(3.3,0.6){$l$}
\rput[cc]{0}(3.3,2.0){$H_2$}
\endpspicture
\caption[]{Tree-level and one-loop diagrams contributing to heavy
           neutrino decays}
\label{fig:decay}
\end{figure}

In order for lepton number to be generated in the decay of these
Majorana singlets, it is necessary that CP and C be violated in the
decays:
\begin{equation}
\epsilon_1 = \frac{\Gamma(N\rightarrow \Phi l)- \Gamma(N\rightarrow \Phi \bar{l})}
                  {\Gamma(N\rightarrow \Phi l)+ \Gamma(N\rightarrow \Phi \bar{l})} 
\neq 0\SPp.
\end{equation}
In fact this is generically the case since, as we have seen, there are
new CP-violating phases in the neutrino mixing matrices which induce
an asymmetry at the one-loop level (see \Fref{fig:decay}). 

\begin{figure}
\centering
\includegraphics[width=.6\linewidth]{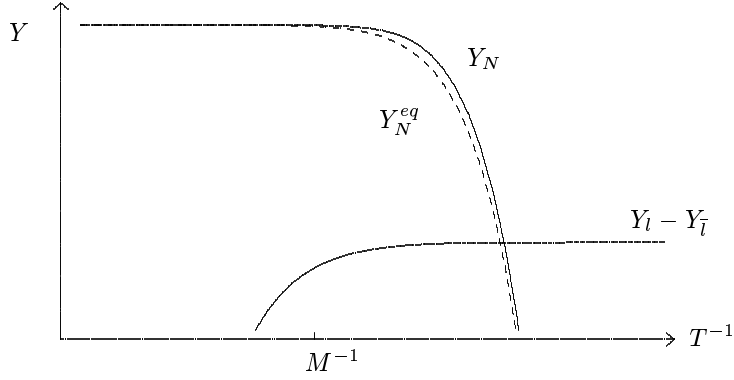}
\caption[]{Abundance of the heavy Majorana singlets at the decoupling
           temperature and the lepton number generated in the
           decay}
\label{fig:outofeq}
\end{figure}

These processes can then produce a net lepton asymmetry if the number
distributions of the Majorana singlets, $N_N$, differ from the thermal
ones. This can occur close to the decoupling temperature, when the
density of the heavy neutrinos gets exponentially suppressed, but they
are so weakly interacting that they cannot follow the fast depletion
(in other words if the decay rate is slower than the expansion of the
Universe close to the decoupling temperature) and
\begin{eqnarray}
N_N > N^\text{thermal}_N. 
\end{eqnarray}
This is shown in \Fref{fig:outofeq}. 
The final asymmetry is given by
\begin{equation}
Y_B = 10^{-2} \;\overbrace{\epsilon_1}^\text{CP-asym} \; 
                \overbrace{\kappa}^\text{eff.~factor}\SPp,
\end{equation}
where $\kappa$ is an efficiency factor which depends on the
non-equilibrium dynamics. Therefore a relation between the baryon
number of the Universe and the neutrino flavour parameters in
$\epsilon_1$ exists.

An interesting question is whether the baryon asymmetry can be
predicted quantitatively from the measurements at low energies of the
neutrino mass matrix.  Unfortunately this is not the case generically
because the asymmetry $\epsilon_1$ depends on more parameters than
those that are observable at low energies.

As we saw in \Sref{sec:seesaw}, at least three heavy Majorana
neutrinos of masses $M_i$ are needed to give masses to the three light
neutrinos. The asymmetry in the decay of the lighest of them in the
minimal model with $M_{2,3} \gg M_1$ is \cite{sasha}
\begin{equation}
\epsilon_1 = - \frac{3}{16 \pi} 
  \sum_i \frac{\IIm[(\tilde\lambda_\nu^\dagger \tilde\lambda_\nu)^2_{i1}]}
              {({\tilde\lambda}^\dagger \tilde\lambda)_{11}}
         \frac{M_1}{M_i}\SPp.
\end{equation}
Instead, at low energies, there is sensitivity only to the neutrino
mass matrix: 
\begin{equation} 
{\tilde\lambda}_\nu \frac{1}{M_R} {\tilde\lambda}^T_\nu,
\label{eq:le}
\end{equation}
where $M_R$ is the heavy Majorana mass matrix. The two combinations
are different and the measurement of the matrix in \Eref{eq:le} does
not allow one to compute $\epsilon_1$. This is because in general the
number of parameters measurable at high energies in the see-saw model
is larger than at low energies. The counting of parameters for $n$
generations before and after integrating out the heavy fields is shown
\Tref{tab:seesaw} (see Section~\ref{sec:mixing} for explanations).

\begin{table}
\caption[]{Number of physical parameters in the see-saw model with $n$
           families and the same number of right-handed Majorana
           neutrinos at high and low energies}
\label{tab:seesaw}
\[
\begin{array}{l|l|l|l|l|l}                        \hline\hline
 & \text{Yukawas} 
  & \text{Field redefinitons} 
   & No.~ m 
    &No.~ \theta 
     & No.~ \phi                                   \\\hline 
 &&&&&                                           \\
\text{see-saw} 
 & Y_l, Y_\nu, M_R=M^T_R 
  & U(n)^3 
   &&&                                           \\  
 E\geq M_i 
 & 5 n^2+n   
  & \frac{3 (n^2-n)}{2}, \frac{3(n^2+n)}{2} 
   & 3 n 
    & n^2-n 
     & n^2-n                                     \\
 &&&&&                                           \\ \hline
 &&&&&                                           \\ 
\text{see-saw}
 & Y_l, \alpha_\nu^T=\alpha_\nu 
  & U(n)^2
   &&&                                           \\  
E \ll M_i
 & 3 n^2 + n 
  & n^2-n, n^2+n 
   & 2 n 
    & \frac{n^2-n}{2}
     & \frac{n^2-n}{2}                           \\ 
 &&&&&                                           \\ \hline
\end{array}
\]
\end{table}

If the prediction of the lepton asymmetry is not possible, 
it should at least be possible to constrain the neutrino mass matrix, assuming
that the lepton asymmetry explains the measured baryon asymmetry.

Indeed, various upper bounds can be derived on the generated
asymmetry, through a bound on $\epsilon_1$ or on $\kappa$. In
particular $\epsilon_1$ has been shown to satisfy
\begin{equation}
|\epsilon_1| \leq \frac{8}{16 \pi} \frac{M_1}{v^2} |\Delta m^2_\text{atm}|^{1/2}\SPp,
\end{equation}
and therefore leptogenesis in this model requires that 
the lightest heavy neutrino is rather heavy:
\begin{equation}
M_1 \geq \mathcal{O}(10^9\UGeV)\SPp.
\end{equation}
A sufficiently large $\kappa$ implies an upper bound on the lightest neutrino mass:
\begin{equation}
m_i \leq \mathcal{O}(\UeVZ).
\end{equation} 
For further details and references see Ref.~\cite{sasha}.

\section{Outlook for theory}

One of the most important questions to resolve in neutrino physics is
whether the origin of neutrino masses is a new physics scale and if so
what this scale is.  One can envisage various possibilities for such
new physics, and the simplest is to assume that its associated energy
scale is above the electroweak scale. It is well known, since the
pioneering work of Weinberg \cite{eft}, that the appropriate language
to describe the low-energy effects of such new physics, no matter what
it is, is that of \emph{effective field theory}. The effects of
\emph{any} beyond-the-standard-model dynamics with a characteristic
energy scale, $\Lambda \gg v$, can be described at low-energies,
\ie $E < \Lambda$, by the SM Lagrangian plus a tower of operators with
mass dimension, $d>4$, constructed out of the SM fields and satisfying
all the gauge symmetries.  Even though the number of such operators is
infinite, they can be classified according to their dimension, $d$,
since an operator of dimension $d$ must be suppressed by the scale
$\Lambda^{d-4}$, and therefore higher dimensionality means stronger
suppression in the high-energy scale:
\begin{equation}
\mathcal{L} = \mathcal{L}_\text{SM} 
            + \sum_i \frac{\alpha_i}{\Lambda}  \mathcal{O}_i^{d=5} 
            + \sum_i \frac{\beta_i}{\Lambda^2} \mathcal{O}_i^{d=6} 
            + ...
\label{eq:eft}
\end{equation} 
Different fundamental theories correspond to different values for the
\emph{low-energy couplings} $\alpha_i, \beta_i,...$, but the structure
of the effective interactions is the same.

It turns out that the first operator in the list is the famous
Weinberg operator of \Eref{majo}:
\begin{equation}
\mathcal{O}^{d=5} = {\bar L}_L^c \tilde\Phi^T \tilde\Phi L_L\SPp,
\end{equation}
where $\tilde\Phi, L$ are the SM Higgs and lepton doublets,
respectively.  This operator is the only one with $d=5$ in the SM, and, 
as we have seen, brings in three essential new features to the minimal
SM:
\begin{itemize}
\item neutrino masses,
\item lepton mixing,
\item lepton number violation.
\end{itemize}
Upon spontaneous symmetry breaking, such an operator induces a neutrino
mass matrix of the form
\begin{equation}
m_\nu = \alpha \frac{v^2}{\Lambda}\SPp, 
\end{equation}
where $\alpha$ is generically a matrix in flavour space. Neutrino
masses are therefore expected to be naturally small if $\Lambda \gg
v$.

\begin{figure}[h]
\centering
\includegraphics[width=.6\linewidth]{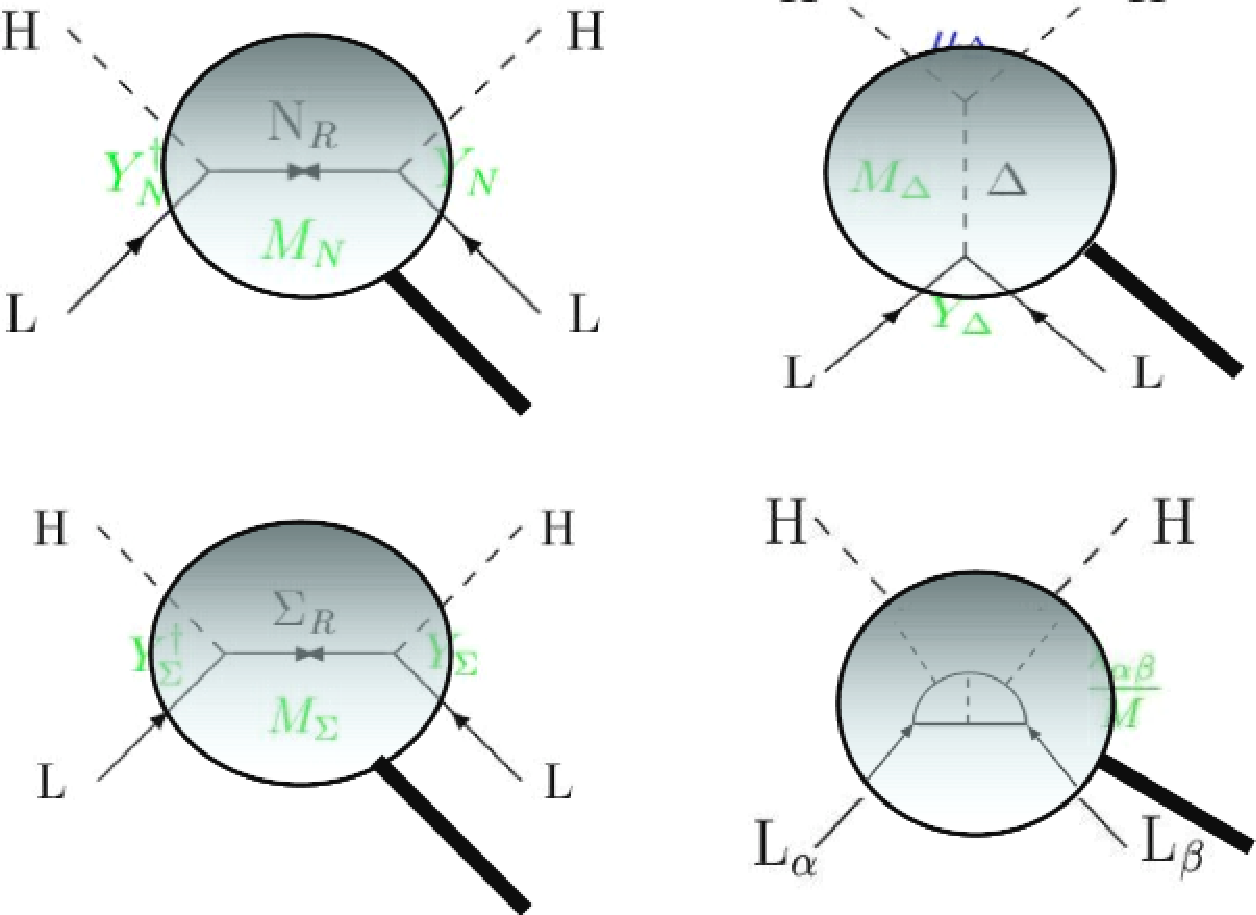}
\caption[]{Magnifying-glass view of Weinberg operator in see-saws
           Type I (top left), Type II (top right), Type III (bottom left)
           and Zee--Babu model (bottom right)}
\label{fig:weinberg}
\end{figure}

If we assume that the neutrino masses we have measured are the result
of this leading operator, one could ask the question: What type of new
physics would induce such an interaction ? In the same way that one can
conjecture the presence of a massive gauge boson from the Fermi
four-fermion interaction, one can classify the extra degrees of
freedom that can induce at tree-level Weinberg's interaction. It turns
out that there are the three well-known possibilities as depicted in
\Fref{fig:weinberg}:
\begin{itemize}
\item type I see-saw: SM+ heavy singlet fermions \cite{typeI},
\item type II see-saw: SM + heavy triplet scalar \cite{typeII},
\item type III see-saw: SM  + heavy triple fermions \cite{typeIII},
\end{itemize}
or combinations.  The masses of the extra states define the scale
$\Lambda$.

It is also possible that Weinberg's interaction is generated by new
physics at higher orders, such as in the famous Zee model \cite{zee}
and related ones \cite{2loop}. In this case, the coupling $\alpha$ in
\Eref{eq:eft} will be suppressed by loop factors $1/(16 \pi^2)$.

Unfortunately the measurement of neutrino masses alone will not tell
us which of these possibilities is the one chosen by Nature. In
particular, the measurement of Weinberg's interaction leaves behind
an unresolved \emph{$\alpha \leftrightarrow \Lambda$ degeneracy} that
makes it impossible to know what the scale of the new physics is, even
if we were to know the absolute value of neutrino masses.

Generically, however, the new physics will give other signals beyond
Weinberg's operator. The next in importance are the $d=6$ operators of
\Eref{eq:eft}~\cite{d6}. Recently the $d=6$ operators induced at
tree level in see-saw models of Types I to III have been worked out
\cite{d6_nu}.  They give rise to a rich phenomenology that could
discriminate between the models. In particular, they could induce
beyond-the-standard-model signals in $Z$ and $W$ decays, deviations in
the $\rho$ parameter or the $W$ mass, and mediate rare lepton decays,
as well as violations of universality and unitarity of the neutrino
mass matrix.  It would therefore be extremely important to search for
these effects. Whether they are large enough to be observed or not
depends strongly on how high the scale $\Lambda$ is, since all these
effects are suppressed by two powers of $\Lambda$.

As mentioned before, neutrino masses alone do not tell us what
$\Lambda$ is, but there are several theoretical prejudices of what
this scale should be. The most popular one is to relate $\Lambda$ to a
grand-unification scale, given the intriguing fact that the
seesaw-type ratio $\frac{v^2}{M_\text{GUT}} \sim 0.01--0.1\UeV$, in the
right ballpark of a neutrino mass scale. Recently, however, it has been
pointed out \cite{natural} that within see-saw models, and without
supersymmetry, this choice would destabilize the electroweak scale,
since the Higgs mass would receive quadratic loop corrections in
$\Lambda$. A naturalness argument would then imply that $\Lambda <
10^7\UGeV$, at least if there is no supersymmetry.

Another possibility is to consider $\Lambda$ to be related to the
electroweak scale, \ie not far from it. After all, the electroweak
scale is the only scale we are sure exits. The question is then if
such a choice would be testable via the measurement of the $d=6$
operators. The answer to this question is no in the simplest type I
see-saw model, because in order to get neutrino masses in the right
ballpark when $\Lambda \sim\UTeV$, it is necessary to have extremely
small Yukawa couplings, which suppress also the $d=6$ operators to an
unobservable level.  Several recent works have discussed the
possibility to have larger effects of the $d=6$ operators
\cite{mfv,d6vsd5,d6_nu}. One possibility is that realized in Zee-type
models where $d=5$ operators are forbidden at tree level and are
therefore suppressed by loop factors, while $d=6$ operators are
allowed at tree level and therefore unsuppressed. A more radical
possibility is the existence of two independent scales in
\Eref{eq:eft}, one that suppresses $d=6$ operators, $\Lambda_{6}$, and
another one, $\Lambda_5 \gg \Lambda_6$, that suppresses the $d=5$ one.
This possibility is not unnatural, because the $d=5$ and $d=6$
operators can be classified according to a a global symmetry: total
lepton number. If we therefore assume that the scale at which lepton
number is broken, $\Lambda_\text{LN}$, is much higher than the scale at
which lepton flavour violation, $\Lambda_\text{LFV}$, is relevant, we
can ensure that the $d=5$ operator, that breaks lepton number, is
suppressed by the former scale, $\Lambda_5 \sim \Lambda_\text{LN}$,
while the lepton-flavour effects induced by operators of $d=6$ would be
suppressed only by a lower scale $\Lambda_6 \sim \Lambda_\text{LFV} <<
\Lambda_\text{LN}$.  The effective field theory describing such a
possibility would look therefore like
\begin{equation}
\mathcal{L} = \mathcal{L}_\text{SM} 
            + \sum_i \frac{\alpha_i}{\Lambda_\text{LN}}   \mathcal{O}_i^{d=5} 
            + \sum_i \frac{\beta_i}{\Lambda^2_\text{LFV}} \mathcal{O}_i^{d=6} 
            + ..., 
\label{eq:eft2}
\end{equation} 
where the operators that break lepton number and those that preserve
this symmetry are generically suppressed by different scales. Such
a possibility has recently been considered in the context of the popular
Minimal Flavour Violation hypothesis~\cite{mfv}. The underlying
rationale for such an assumption is not completely ad hoc, since in
this context one could hope to explain two apparently contradictory
facts
\begin{itemize}
\item common origin of lepton and quark family mixing at a scale
      $\Lambda_\text{LFV}$,
\item large gap between neutrino masses and remaining fermions since
      neutrino masses would be suppressed by $\Lambda_\text{LN}$ .
\end{itemize}
In fact this separation of scales is built-in in several of the models
mentioned before. The simplest example being the type II see-saw
model, where the scalar-triplet mass, $M_\Delta$, is directly connected
with the $\Lambda_\text{LFV}$, while the scale of lepton number
violation is $M_\Delta^2/\mu$, where $\mu$ is a dimensionful coupling
in the scalar potential of the triplet. In fact, it is the separation
of scales that makes the phenomenology of this model much richer at
low energies than that of type I see-saw models in their simplest
version.

If this possibility is realized, there would be many interesting
consequences:
\begin{itemize}
\item lepton flavour violation could be measurable beyond neutrino
      oscillations, 
\item the scale of lepton flavour violation, $\Lambda_\text{LFV}$,
      could be reached at the LHC.
\end{itemize}
In recent years a lot of activity has been devoted to studying
possible signals of neutrino masses at the the LHC. Lepton number
violation could give rise to spectacular signals at LHC, like
same-charge lepton pairs \cite{lp}. This signal has been studied in
detail recently in various see-saw models. In one-scale models of type
I, neutrino masses restrict these processes to being highly suppressed
beyond detectable levels \cite{typeIlhc}. However, the separation of
scales mentioned before, allows  light enough triplets in the
type II see-saw to be pair-produced at LHC:
\begin{equation}
p p \rightarrow H^{++} H^{--} \rightarrow l^+ l^+ l^- l^- ,
\end{equation}
leading to the powerful signal of same-charge lepton pairs. Not only can
the invariant mass be reconstructed from the two leptons pairs,
but the flavour structure of the branching ratios to different leptons
is in one-to-one correspondence with the flavour structure of the
neutrino mass matrix. Therefore the putative measurement of these
processes would provide direct information on the neutrino mass matrix
\cite{typeIIlhc}.

Solving the flavour problem of the Standard Model is surely a
quixotic enterprise and we shall need to explore as many avenues as
we can. In recent years it has become increasingly clear that in addition to
quark flavour factories, we can obtain very valuable information on
different aspects of this puzzle also from LHC and lepton flavour
factories.
 
\section{Conclusions}

The results of many beautiful experiments in the last decade have
demonstrated beyond doubt that  neutrinos are massive and mix. The
standard $3\nu$ scenario can explain in terms of four fundamental
parameters all available data, except that of the unconfirmed signal
of LSND. The lepton flavour sector of the Standard Model is expected
to be at least as complex as the quark one, even though we know it
only partially.

The structure of the neutrino spectrum and mixing is quite different
from the one that has been observed for the quarks: there are large
leptonic mixing angles and the neutrino masses are much smaller than
those of the remaining leptons. These peculiar features of the lepton
sector strongly suggest that leptons and quarks constitute two
complementary approaches to understanding the origin of flavour in the
Standard Model.  In fact, the smallness of neutrino masses can be
naturally understood if there is new physics beyond the electroweak
scale.

Many fundamental questions remain to be answered in future neutrino
experiments, and these can have very important implications for our
understanding of the Standard Model and of what lies beyond: Are
neutrinos Majorana particles?  Are neutrino masses the result of a new
physics scale? Is CP violated in the lepton sector? Could neutrinos
be the seed of the matter--antimatter asymmetry in the Universe?

A rich experimental programme lies ahead where fundamental physics
discoveries are very likely (almost warrantied). We can only hope that
neutrinos will keep up with their old tradition and  provide a
window to what lies beyond the Standard Model.

\end{document}